\documentclass[runningheads,a4paper,12pt]{llncs}

\usepackage[utf8]{inputenc}
\usepackage{graphicx}
\usepackage{amsmath,amssymb}
\usepackage{natbib}
\usepackage[colorlinks=true,citecolor=blue]{hyperref}
\usepackage[a4paper,margin=3cm]{geometry}
\usepackage{subfigure}
\usepackage{color}
\usepackage{tikz}
\usetikzlibrary{trees,positioning,shapes,shadows,arrows}
\usepackage{newlfont}
\usepackage{multirow}
\usepackage{longtable} 
\usepackage{float}
\usepackage{url}
\urldef{\mailsa}\path|{alfred.hofmann, ursula.barth, ingrid.haas, frank.holzwarth,|
\urldef{\mailsb}\path|anna.kramer, leonie.kunz, christine.reiss, nicole.sator,|
\urldef{\mailsc}\path|erika.siebert-cole, peter.strasser, lncs}@springer.com|    
\newcommand{\keywords}[1]{\par\addvspace\baselineskip
\noindent\keywordname\enspace\ignorespaces#1}

\begin{document}

\mainmatter 

\title{Nowcasting of COVID-19 confirmed cases: Foundations, trends, and challenges}

\titlerunning{Short-term Forecasting of COVID-19}

\author{Tanujit Chakraborty \and Indrajit Ghosh \and Tirna Mahajan \and Tejasvi Arora}

\authorrunning{T. Chakraborty \and I. Ghosh \and T. Mahajan \and T. Arora}


\institute{Indian Statistical Institute, 203 B.T. Road, Kolkata - 700108, West Bengal, India \\
Corresponding Author's Email: indra7math@gmail.com}

\maketitle

\begin{abstract}
The coronavirus disease 2019 (COVID-19) has become a public health emergency of international concern affecting more than 200 countries and territories worldwide. As of September 30, 2020, it has caused a pandemic outbreak with more than 33 million confirmed infections and more than 1 million reported deaths worldwide. Several statistical, machine learning, and hybrid models have previously tried to forecast COVID-19 confirmed cases for profoundly affected countries. Due to extreme uncertainty and nonstationarity in the time series data, forecasting of COVID-19 confirmed cases has become a very challenging job. For univariate time series forecasting, there are various statistical and machine learning models available in the literature. But, epidemic forecasting has a dubious track record. Its failures became more prominent due to insufficient data input, flaws in modeling assumptions, high sensitivity of estimates, lack of incorporation of epidemiological features, inadequate past evidence on effects of available interventions, lack of transparency, errors, lack of determinacy, and lack of expertise in crucial disciplines. This chapter focuses on assessing different short-term forecasting models that can forecast the daily COVID-19 cases for various countries. In the form of an empirical study on forecasting accuracy, this chapter provides evidence to show that there is no universal method available that can accurately forecast pandemic data. Still, forecasters' predictions are useful for the effective allocation of healthcare resources and will act as an early-warning system for government policymakers.

\end{abstract}

\section{Introduction}
In December 2019, clusters of pneumonia cases caused by the novel Coronavirus (COVID-19) were identified at the Wuhan, Hubei province in China \cite{huang2020clinical, guan2020clinical} after almost hundred years of the 1918 Spanish flu \cite{trilla20081918}. Soon after the emergence of the novel beta coronavirus, World Health Organization (WHO) characterized this contagious disease as a ``global pandemic" due to its rapid spread worldwide \cite{roosa2020real}. Many scientists have attempted to make forecasts about its impact. However, despite involving many excellent modelers, best intentions, and highly sophisticated tools, forecasting COVID-19 pandemics is harder \cite{ioannidis2020forecasting}, and this is primarily due to the following major factors:
\begin{itemize}
    \item Very less amount of data is available;
    \item Less understanding of the factors that contribute to it;
    \item Model accuracy is constrained by our knowledge of the virus, however. With an emerging disease such as COVID-19, many transmission-related biologic features are hard to measure and remain unknown;
    \item The most obvious source of uncertainty affecting all models is that we don't know how many people are or have been infected;
    \item Ongoing issues with virologic testing mean that we are certainly missing a substantial number of cases, so models fitted to confirmed cases are likely to be highly uncertain \cite{holmdahl2020wrong};
    \item The problem of using confirmed cases to fit models is further complicated because the fraction of confirmed cases is spatially heterogeneous and time-varying \cite{weinberger2020estimating};
    \item Finally, many parameters associated with COVID-19 transmission are poorly understood.
\end{itemize}

Amid enormous uncertainty about the future of the COVID-19 pandemic, statistical, machine learning, and epidemiological models are critical forecasting tools for policymakers, clinicians, and public health practitioners \cite{chakraborty2020real, li2020trend,wu2020nowcasting, fanelli2020analysis,kucharski2020early, zhuang2020estimation}. COVID-19 modeling studies generally follow one of two general approaches that we will refer to as forecasting models and mechanistic models. Although there are hybrid approaches, these two model types tend to address different questions on different time scales, and they deal differently with uncertainty \cite{chakraborty2020integrated}. Compartmental epidemiological models have been developed over nearly a century and are well tested on data from past epidemics. These models are based on modeling the actual infection process and are useful for predicting long-term trajectories of the epidemic curves \cite{chakraborty2020integrated}. Short-term Forecasting models are often statistical, fitting a line or curve to data and extrapolating from there -- like seeing a pattern in a sequence of numbers and guessing the next number, without incorporating the process that produces the pattern \cite{chakraborty2020theta, chakraborty2019forecasting, chakraborty2020real}. Well constructed statistical frameworks can be used for short-term forecasts, using machine learning or regression. In statistical models, the uncertainty of the prediction is generally presented as statistically computed prediction intervals around an estimate \cite{hastie2009elements, james2013introduction}. Given that what happens a month from now will depend on what happens in the interim, the estimated uncertainty should increase as you look further into the future. These models yield quantitative projections that policymakers may need to allocate resources or plan interventions in the short-term. 

Forecasting time series datasets have been a traditional research topic for decades, and various models have been developed to improve forecasting accuracy \cite{chatfield2000time, armstrong2001principles, hanke2001business}. There are numerous methods available to forecast time series, including traditional statistical models and machine learning algorithms, providing many options for modelers working on epidemiological forecasting \cite{chakraborty2019forecasting, chakraborty2020integrated, brady2012refining, chakraborty2020real, messina2014global, buczak2018ensemble, ribeiro2020short}. Many research efforts have focused on developing a universal forecasting model but failed, which is also evident from the ``No Free Lunch Theorem" \cite{wolpert1997no}. This chapter focuses on assessing popularly used short-term forecasting (nowcasting) models for COVID-19 from an empirical perspective. The findings of this chapter will fill the gap in the literature of nowcasting of COVID-19 by comparing various forecasting methods, understanding global characteristics of pandemic data, and discovering real challenges for pandemic forecasters.    

The upcoming sections present a collection of recent findings on COVID-19 forecasting. Additionally, twenty nowcasting (statistical, machine learning, and hybrid) models are assessed for five countries of the United States of America (USA), India, Brazil, Russia, and Peru. Finally, some recommendations for policy-making decisions and limitations of these forecasting tools have been discussed. 

\section{Related works}
Researchers face unprecedented challenges during this global pandemic to forecast future real-time cases with traditional mathematical, statistical, forecasting, and machine learning tools \cite{li2020trend, wu2020nowcasting, fanelli2020analysis, kucharski2020early, zhuang2020estimation}. Studies in March with simple yet powerful forecasting methods like the exponential smoothing model predicted cases ten days ahead that, despite the positive bias, had reasonable forecast error \cite{petropoulos2020forecasting}. Early linear and exponential model forecasts for better preparation regarding hospital beds, ICU admission estimation, resource allocation, emergency funding, and proposing strong containment measures were conducted \cite{grasselli2020critical} that projected about 869 ICU and 14542 ICU admissions in Italy for March 20, 2020. Health-care workers had to go through the immense mental stress left with a formidable choice of prioritizing young and healthy adults over the elderly for allocation of life support, mostly unwanted ignoring of those who are extremely unlikely to survive \cite{emanuel2020fair,rosenbaum2020facing}. Real estimates of mortality with 14-days delay demonstrated underestimating of the COVID-19 outbreak and indicated a grave future with a global case fatality rate (CFR) of 5.7\% in March \cite{baud2020real}. The contact tracing, quarantine, and isolation efforts have a differential effect on the mortality due to COVID-19 among countries. Even though it seems that the CFR of COVID-19 is less compared to other deadly epidemics, there are concerns about it being eventually returning as the seasonal flu, causing a second wave or future pandemic \cite{petersen2020comparing, rajgor2020many}. 

Mechanistic models, like the Susceptible–Exposed–Infectious–Recovered (SEIR) frameworks, try to mimic the way COVID-19 spreads and are used to forecast or simulate future transmission scenarios under various assumptions about parameters governing the transmission, disease, and immunity \cite{hou2020effectiveness, he2020seir, annas2020stability, chen2020time, lopez2020end}. Mechanistic modeling is one of the only ways to explore possible long-term epidemiologic outcomes \cite{anderson1992infectious}. For example, the model from Ferguson et al. \cite{ferguson2020report} that has been used to guide policy responses in the United States and Britain examines how many COVID-19 deaths may occur over the next two years under various social distancing measures. Kissler et al. \cite{kissler2020projecting} ask whether we can expect seasonal, recurrent epidemics if immunity against novel coronavirus functions similarly to immunity against the milder coronaviruses that we transmit seasonally. In a detailed mechanistic model of Boston-area transmission, Aleta et al. \cite{aleta2020modeling} simulate various lockdown ``exit strategies". These models are a way to formalize what we know about the viral transmission and explore possible futures of a system that involves nonlinear interactions, which is almost impossible to do using intuition alone \cite{hellewell2020feasibility, mossong2008social}. Although these epidemiological models are useful for estimating the dynamics of transmission, targeting resources, and evaluating the impact of intervention strategies, the models require parameters and depend on many assumptions. 

Several statistical and machine learning methods for real-time forecasting of the new and cumulative confirmed cases of COVID-19 are developed to overcome limitations of the epidemiological model approaches and assist public health planning and policy-making \cite{chakraborty2020integrated, petropoulos2020forecasting, anastassopoulou2020data, chakraborty2020real, chakraborty2020theta}. Real-time forecasting with foretelling predictions is required to reach a statistically validated conjecture in this current health crisis. Some of the leading-edge research concerning real-time projections of COVID-19 confirmed cases, recovered cases, and mortality using statistical, machine learning, and mathematical time series modeling are given in Table \ref{Table1}.

\begin{table}[]
    \caption{Related works on nowcasting and forecasting of COVID-19 pandemic}
   \centering
   \normalsize
\resizebox{\columnwidth}{!}{
    \begin{tabular}{|p{3.5cm}|p{3.0cm}|p{2.5cm}|p{4cm}|p{4cm}|p{4cm}|}
\\\hline
\textbf{Research Topic}      
& \textbf{Date}   
& \textbf{Countries} 
&\textbf{Model}      
& \textbf{Results}   
& \textbf{Main Conclusion} 
\\\hline
Forecasting and risk assessment \cite{chakraborty2020real}
&January 30-31, 2020, to April 4, 2020
&Canada, France, India, South Korea, UK 
&ARIMA,Wavelet ARIMA(WBF),
Hybrid ARIMA-WBF
&MAE and RMSE least for Hybrid ARIMA-WBF
&Hybrid ARIMA-WBF performs better than traditional methods and important factors that impact on case fatality rates are estimated using regression tree.
\\\hline
Forecasting the confirmed and recovered cases \cite{maleki2020time}
& Jan 22,2020 to April 30, 2020       
& World data
& TP–SMN–AR time series (Autoregressive series based on two-piece scale mixture normal distributions)
& MAPE = 0.22 for confirmed cases; MAPE = 1.6 for Recovered cases 
& Provided reasonable forecasts in terms of error and model selection.
\\\hline
Short-term forecasting of cumulative confirmed cases \cite{ribeiro2020short}
&Inception - April 18-19, 2020
&Brazil
&ARIMA, Random forest, Ridge regression, Support vector regression, Ensemble learning
&Forecast errors lower than 6.9 percent
&SVR and stacking-ensemble learning model are suitable tools for forecasting COVID-19.
\\\hline
Modelling and forecasting daily cases \cite{anastassopoulou2020data}
& January 11 to February 10, 2020
&China
&Susceptible-Infectious-Recovered-Dead (SIRD) model
&Estimated average reproduction number $(R_0)\sim 2.6$ and $CFR\sim 0.15\%$
&simulations predicted a decline of the outbreak at the end of February.
\\\hline
Forecasting COVID-19 \cite{petropoulos2020forecasting}
& January 22, 2020 to March 11, 2020
& Global data
& Exponential smoothing models
& Ten-days-ahead forecasts have actual cases \ within $90\%$ CI
&Forecasts reflect the significant increase in the trend of global cases with growing uncertainty. 
\\\hline
Real-time forecasting \cite{roosa2020real}
&February 5to February 24, 2020
& China
&Generalized logistic growth model (GLM) and Sub-epidemic wave model
&Mean case estimates and 95\% prediction intervals emulsifies the global picture 15-days ahead
&All methods perform similarly and and increase in data inclusion decreases the width of prediction intervals. 
\\\hline
Predictions and role of interventions \cite{ray2020predictions}
&Live forecast
&India
&Extended state-space SIR epidemiological models
&Live forecasts with broad confidence intervals
& Lockdown has a high chance of reducing the number of COVID-19 cases.
\\\hline
Forecasting and nowcasting COVID-19 \cite{wu2020nowcasting}
&Dec 31, 2019, to Jan 28, 2020
&China
&Susceptible-exposed-infectious-recovered (SEIR) model
& $R_0$ = 2.68 (95\% CI 2.47, 2.86) ; Epidemic doubling time = 6.4 days (95\% CI 5.8, 7.1)
&COVID-19 is no longer contained within China, and human-human transmission became evident.
\\\hline
Forecast \cite{fanelli2020analysis}
&Jan, 22- March 15, 2020
&China, Italy and France
&Susceptible, infected, recovered, dead (SIRD) model
&The recovery rate is the same for Italy and China, while infection and death rate appear to be different.
&There is a certain universality in the time evolution of COVID-19.
\\\hline
AI-based forecasts \cite{hu2020artificial}
&Jan, 11 - February 27, 2020
&China
&Data driven AI-based methods
&Using the multiple-step forecasting, forecasts are given till April 19, 2020 for 34 provinces/cities.
&The accuracy of the AI-based methods for forecasting the trajectory of COVID-19 was high. 
\\\hline
Machine learning-based forecasts \cite{sujath2020machine}
&January 22, 2020, to April 10, 2020
&India
&Multi-layered perceptron (MLP) model
&Forecast of confirmed, deaths and recovered cases for 69 days
&MLP method is giving good prediction results than other methods.
\\\hline
Long-term trajectories of COVID-19 \cite{chakraborty2020integrated}
&Starting - June 17, 2020
&Spain and Italy
&Integrated stochastic-deterministic (ISA) approach
&Basic reproduction number and estimated future cases are computed.
&ISA model shows significant improvement in the long-term forecasting of COVID-19 cases. 
\\\hline
\end{tabular}}
\label{Table1}
\end{table}

\section{Global characteristics of pandemic time series}\label{global}
A univariate time series is the simplest form of temporal data and is a sequence of real numbers collected regularly over time, where each number represents a value \cite{chatfield2016analysis, box2015time}. There are broadly two major steps involved in univariate time series forecasting \cite{hyndman2018forecasting}: 
\begin{itemize}
    \item Studying the global characteristics of the time series data;
    \item Analysis of data with the `best-fitted' forecasting model.
\end{itemize}
Understanding the global characteristics of pandemic confirmed cases data can help forecasters determine what kind of forecasting method will be appropriate for the given situation \cite{tsay2000time}. As such, we aim to perform a meaningful data analysis, including the study of time series characteristics, to provide a suitable and comprehensive knowledge foundation for the future step of selecting an apt forecasting method. Thus, we take the path of using statistical measures to understand pandemic time series characteristics to assist method selection and data analysis. These characteristics will carry summarized information of the time series, capturing the `global picture' of the datasets. Based on the recommendation of \cite{de200625, wang2009rule, lemke2010meta, lemke2015metalearning}, we study several classical and advanced time series characteristics of COVID-19 data. This study considers eight global characteristics of the time series: periodicity, stationarity, serial correlation, skewness, kurtosis, nonlinearity, long-term dependence, and chaos. This collection of measures provides quantified descriptions and gives a rich portrait of the pandemic time-series' nature. A brief description of these statistical and advanced time-series measures are given below. 

\subsection{Periodicity}
A seasonal pattern exists when a time series is influenced by seasonal factors, such as the month of the year or day of the week. The seasonality of a time series is defined as a pattern that repeats itself over fixed intervals of time \cite{box2015time}. In general, the seasonality can be found by identifying a large autocorrelation coefficient or a large partial autocorrelation coefficient at the seasonal lag. Since the periodicity is very important for determining the seasonality and examining the cyclic pattern of the time series, the periodicity feature extraction becomes a necessity. Unfortunately, many time series available from the dataset in different domains do not always have known frequency or regular periodicity. Seasonal time series are sometimes also called cyclic series, although there is a significant distinction between them. Cyclic data have varying frequency lengths, but seasonality is of a fixed length over each period. For time series with no seasonal pattern, the frequency is set to 1. The seasonality is tested using the `stl' function within the ``stats" package in R statistical software \cite{hyndman2018forecasting}.  

\subsection{Stationarity}
Stationarity is the foremost fundamental statistical property tested for in time series analysis because most statistical models require that the underlying generating processes be stationary \cite{chatfield2000time}. Stationarity means that a time series (or rather the process rendering it) do not change over time. In statistics, a unit root test tests whether a time series variable is non-stationary and possesses a unit root \cite{phillips1988testing}. The null hypothesis is generally defined as the presence of a unit root, and the alternative hypothesis is either stationarity, trend stationarity, or explosive root depending on the test used. In econometrics, Kwiatkowski–Phillips–Schmidt–Shin (KPSS) tests are used for testing a null hypothesis that an observable time series is stationary around a deterministic trend (that is, trend-stationary) against the alternative of a unit root \cite{shin1992kpss}. The KPSS test is done using the `kpss.test' function within the ``tseries" package in R statistical software \cite{trapletti2007tseries}. 

\subsection{Serial correlation}
Serial correlation is the relationship between a variable and a lagged version of itself over various time intervals. Serial correlation occurs in time-series studies when the errors associated with a given time period carry over into future time periods \cite{box2015time}. We have used Box-Pierce statistics \cite{box1970distribution} in our approach to estimate the serial correlation measure and extract the measures from COVID-19 data. The Box-Pierce statistic was designed by Box and Pierce in 1970 for testing residuals from a forecast model \cite{wang2009rule}. It is a common portmanteau test for computing the measure. The mathematical formula of the Box-Pierce statistic is as follows: 
$$ Q_h = n \displaystyle \sum_{k=1}^{h} r_{k}^{2},$$
where $n$ is the length of the time series, $h$ is the maximum lag being considered (usually $h$ is chosen as 20), and $r_k$ is the autocorrelation function. The portmanteau test is done using the `Box.test' function within the ``stats" package in R statistical software \cite{hyndman2007automatic}. 

\subsection{Nonlinearity}
Nonlinear time series models have been used extensively to model complex dynamics not adequately represented by linear models \cite{kantz2004nonlinear}. Nonlinearity is one important time series characteristic to determine the selection of an appropriate forecasting method. \cite{tong2002nonlinear} There are many approaches to test the nonlinearity in time series models, including a nonparametric kernel test and a Neural Network test \cite{tsay1986nonlinearity}. In the comparative studies between these two approaches, the Neural Network test has been reported with better reliability \cite{wang2009rule}. In this research, we used Teräsvirta's neural network test \cite{terasvirta1993power} for measuring time series data nonlinearity. It has been widely accepted and reported that it can correctly model the nonlinear structure of the data \cite{terasvirta2005linear}. It is a test for neglected nonlinearity, likely to have power against a range of alternatives based on the NN model (augmented single-hidden-layer feedforward neural network model). This takes large values when the series is nonlinear and values near zero when the series is linear. The test is done using the `nonlinearityTest' function within the ``nonlinearTseries" package in R statistical software \cite{garcia2015nonlineartseries}. 

\subsection{Skewness}
Skewness is a measure of symmetry, or more precisely, the lack of symmetry. A distribution, or dataset, is symmetric if it looks the same to the left and the right of the center point \cite{wang2009rule}. A skewness
measure is used to characterize the degree of asymmetry of values around the mean value \cite{mood1950introduction}. For univariate data $Y_t$, the skewness coefficient is 
$$ S = \frac{1}{n \sigma^3} \sum_{t=1}^{n} \left( Y_t - \bar{Y} \right)^3, $$
where $\bar{Y}$ is the mean, $\sigma$ is the standard deviation, and $n$ is the number of data points. The skewness for a normal distribution is zero, and any symmetric data should have the skewness near zero. Negative values for the skewness indicate data that are skewed left, and positive values for the skewness indicate data that are skewed right. In other words, left skewness means that the left tail is heavier than the right tail. Similarly, right skewness means the right tail is heavier than the left tail \cite{kim2013statistical}. Skewness is calculated using the `skewness' function within the ``e1071" package in R statistical software \cite{meyer2019package}. 

\subsection{Kurtosis (heavy-tails)}
Kurtosis is a measure of whether the data are peaked or flat, relative to a normal distribution \cite{mood1950introduction}. A dataset with high kurtosis tends to have a distinct peak near the mean, decline rather rapidly, and have heavy tails. Datasets with low kurtosis tend to have a flat top near the mean rather than a sharp peak. For a univariate time series
$Y_t$, the kurtosis coefficient is $\frac{1}{n \sigma^4} \sum_{t=1}^{n} \left( Y_t - \bar{Y} \right)^4$. The kurtosis for a standard normal distribution is 3. Therefore, the excess
kurtosis is defined as 
$$ K = \frac{1}{n \sigma^4} \sum_{t=1}^{n} \left( Y_t - \bar{Y} \right)^4 - 3. $$
So, the standard normal distribution has an excess kurtosis of zero. Positive kurtosis indicates a `peaked' distribution and negative kurtosis indicates a `flat' distribution \cite{groeneveld1984measuring}. Kurtosis is calculated using the `kurtosis' function within the ``PerformanceAnalytics" package in R statistical software \cite{peterson2018package}. 

\subsection{Long-range Dependence}
Processes with long-range dependence have attracted a good deal of attention from a probabilistic perspective in time series analysis \cite{robinson1995log}. With such increasing importance of the `self-similarity' or `long-range dependence' as one of the time series characteristics, we study this feature into the group of pandemic data characteristics. The definition of self-similarity is most related to the self-similarity parameter, also called Hurst exponent (H) \cite{black1965long}. The class of autoregressive fractionally integrated moving average (ARFIMA) processes \cite{granger1980introduction} is a good estimation method for computing H.  In an ARIMA$(p,d,q)$, $p$ is the order of AR, $d$ is the degree first differencing involved, and $q$ is the order of MA. If the time series is suspected of exhibiting long-range dependency, parameter $d$ may be replaced by certain non-integer values in the ARFIMA model \cite{brockwell1991time}. We fit an ARFIMA$(0,d,0)$ to the maximum likelihood, which is approximated by using the fast and accurate method of Haslett and Raftery \cite{haslett1989space}. We then estimate the Hurst parameter using the relation $H = d + 0.5$. The self-similarity feature can only be detected in the RAW data of the time series. The value of H can be obtained using the `hurstexp' function within the ``pracma" package in R statistical software \cite{borchers2019package}. 

\subsection{Chaos (dynamic systems)}
Many systems in nature that were previously considered random processes are now categorized as chaotic systems. For several years, Lyapunov Characteristic Exponents are of interest in the study of dynamical systems to characterize quantitatively their stochasticity properties, related essentially to the exponential divergence of nearby orbits \cite{farmer1987predicting}. Nonlinear dynamical systems often exhibit chaos, characterized by sensitive dependence on initial values, or more precisely by a positive Lyapunov Exponent (LE) \cite{farmer1982chaotic}. Recognizing and quantifying chaos in time series are essential steps toward understanding the nature of random behavior and revealing the extent to which short-term forecasts may be improved \cite{hegger1999practical}. LE, as a measure of the divergence of nearby trajectories, has been used to qualifying chaos by giving a quantitative value \cite{benettin1980lyapunov}. The algorithm of computing LE from time-series is applied to continuous dynamical systems in an $n$-dimensional phase space \cite{rosenstein1993practical}. LE is calculated using the `Lyapunov exponent' function within the ``tseriesChaos" package in R statistical software \cite{antonio2013package}.

\section{Popular forecasting methods for pandemic nowcasting}
Time series forecasting models work by taking a series of historical observations and extrapolating future patterns. These are great when the data are accurate; the future is similar to the past. Forecasting tools are designed to predict possible future alternatives and help current planing and decision making \cite{armstrong2001principles}.

There are essentially three general approaches to forecasting a time series \cite{montero2020fforma}: 
\begin{enumerate}
    \item Generating forecasts from an individual model;
    \item Combining forecasts from many models (forecast model averaging);
    \item Hybrid experts for time series forecasting.
\end{enumerate}

Single (individual) forecasting models are either traditional statistical methods or modern machine learning tools. We study ten popularly used single forecasting models from classical time series, advanced statistics, and machine learning literature. There has been a vast literature on the forecast combinations motivated by the seminal work of Bates \& Granger \cite{bates1969combination} and followed by a plethora of empirical applications showing that combination forecasts are often superior to their counterparts (see, \cite{bordley1982combination, timmermann2006forecast}, for example). Combining forecasts using a weighted average is considered a successful way of hedging against the risk of selecting a misspecified model \cite{clemen1989combining}. A significant challenge is in choosing an appropriate set of weights, and many attempts to do this have been worse than simply using equal weights -- something that has become known as the ``forecast combination puzzle" (see, for example, \cite{smith2009simple}). To overcome this, hybrid models became popular with the seminal work of Zhang \cite{zhang2003time} and further extended for epidemic forecasting in \cite{chakraborty2019forecasting, chakraborty2020real, chakraborty2020theta}. The forecasting methods can be briefly reviewed and organized in the architecture shown in Figure \ref{fig:tsf_tools}.

\tikzset{
  basic/.style  = {draw, text width=2cm, drop shadow, font=\sffamily, rectangle},
  root/.style   = {basic, rounded corners=2pt, thin, align=center, fill=white},
  level-2/.style = {basic, rounded corners=5pt, thin,align=center, fill=white, text width=2cm},
  level-3/.style = {basic, thin, align=center, fill=white, text width=1.8cm}
}

\begin{figure}
    \centering
\begin{tikzpicture}[
  level 1/.style={sibling distance=8em, level distance=6em},
  edge from parent/.style={->,solid,black,thick,sloped,draw}, 
  edge from parent path={(\tikzparentnode.south) -- (\tikzchildnode.north)},
  >=latex, node distance=1.2cm, edge from parent fork down]

\node[root] {\textbf{Time series forecasting methods}}
  child {node[level-2] (c1) {\textbf{Classical}}}
  child {node[level-2] (c2) {\textbf{Smoothing}}}
  child {node[level-2] (c3) {\textbf{Advanced}}}
  child {node[level-2] (c4) {\textbf{ML}}}
  child {node[level-2] (c5) {\textbf{Hybrid}}}
  child {node[level-2] (c6) {\textbf{Ensemble}}};

\begin{scope}[every node/.style={level-3}]
\node [below of = c1, xshift=7pt] (c11) {ARIMA};
\node [below of = c11] (c12) {SETAR};
\node [below of = c12] (c13) {ARFIMA};

\node [below of = c2, xshift=7pt] (c21) {ETS};
\node [below of = c21] (c22) {TBATS};
\node [below of = c22] (c23) {Theta};

\node [below of = c3, xshift=7pt] (c31) {WARIMA};
\node [below of = c31] (c32) {BSTS};

\node [below of = c4, xshift=7pt] (c41) {ANN};
\node [below of = c41] (c42) {ARNN};

\node [below of = c5, xshift=7pt] (c51) {ARIMA-ANN};
\node [below of = c51] (c52) {ARIMA-ARNN};
\node [below of = c52] (c53) {ARIMA-WARIMA};
\node [below of = c53] (c54) {WARIMA-ANN};
\node [below of = c54] (c55) {WARIMA-ARNN};

\node [below of = c6, xshift=7pt] (c61) {ARIMA-ETS-Theta};
\node [below of = c61] (c62) {ARIMA-ETS-ARNN};
\node [below of = c62] (c63) {ARIMA-Theta-ARNN};
\node [below of = c63] (c64) {ETS-Theta-ARNN};
\node [below of = c64] (c65) {ANN-ARNN-WARIMA};
\end{scope}

\foreach \value in {1,2,3}
  \draw[->] (c1.195) |- (c1\value.west);

\foreach \value in {1,2,3}
  \draw[->] (c2.195) |- (c2\value.west);

\foreach \value in {1,2}
  \draw[->] (c3.195) |- (c3\value.west);
  
\foreach \value in {1,2}
  \draw[->] (c4.195) |- (c4\value.west);
  
\foreach \value in {1,...,5}
  \draw[->] (c5.195) |- (c5\value.west);
  
\foreach \value in {1,...,5}
  \draw[->] (c6.195) |- (c6\value.west);  
\end{tikzpicture}
    \caption{A systemic view of the various forecasting methods to be used in this study}
    \label{fig:tsf_tools}
\end{figure}
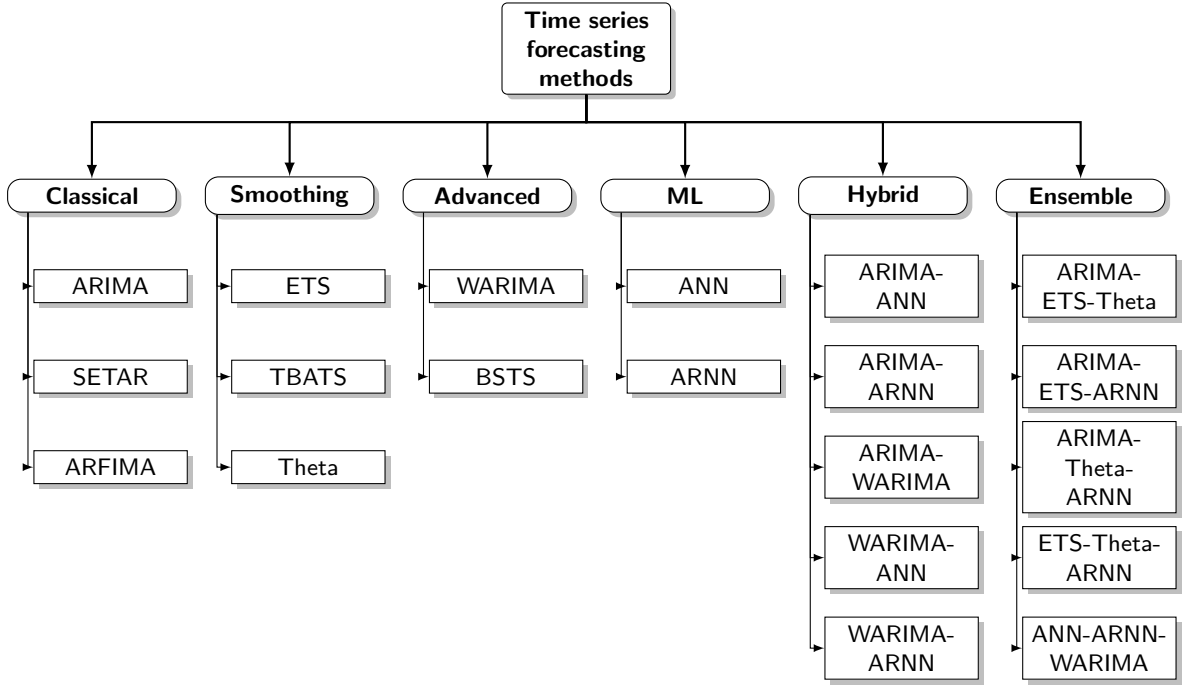

\subsection{Autoregressive integrated moving average (ARIMA) model}
The autoregressive integrated moving average (ARIMA) is one of the well-known linear models in time-series forecasting, developed in the early 1970s \cite{box2015time}. It is widely used to track linear tendencies in stationary time-series data. It is denoted by ARIMA($p,d,q$), where the three components have significant meanings. The parameters $p$ and $q$ represent the order of AR and MA models, respectively, and $d$ denotes the level of differencing to convert nonstationary data into stationary time series \cite{makridakis1997arma}. ARIMA model can be mathematically expressed as follows:
$$ y_t = \alpha_{0} + \sum_{i=1}^{p} \beta_i y_{t-i} + \epsilon_t - \sum_{j=1}^q \alpha_j \epsilon_{t-j}, $$
where  $y_t$ denotes the actual value of the variable at time $t$, $\epsilon_t$ denotes the random error at time $t$, $\beta_i$ and $\alpha_j$ are the coefficients of the model. Some necessary steps to be followed for any given time-series dataset to build an ARIMA model are as follows:
\begin{itemize}
\item Identification of the model (achieving stationarity).
\item Use autocorrelation function (ACF) and partial ACF plots to select the AR and MA model parameters, respectively, and finally estimate model parameters for the ARIMA model.
\item The `best-fitted' forecasting model can be found using the Akaike Information Criteria (AIC) or the Bayesian Information Criteria (BIC). Finally, one checks the model diagnostics to measure its performance. 
\end{itemize}
An implementation in R statistical software is available using the `auto.arima' function under the ``forecast" package, which returns the `best' ARIMA model according to either AIC or BIC values \cite{hyndman2020package}.

\subsection{Wavelet-based ARIMA (WARIMA) model}
Wavelet analysis is a mathematical tool that can reveal information within the signals in both the time and scale (frequency) domains. This property overcomes the primary drawback of Fourier analysis, and wavelet transforms the original signal data (especially in the time domain) into a different domain for data analysis and processing. Wavelet-based models are most suitable for nonstationary data, unlike standard ARIMA. Most epidemic time-series datasets are nonstationary; therefore, wavelet transforms are used as a forecasting model for these datasets \cite{chakraborty2020real}. When conducting wavelet analysis in the context of time series analysis \cite{aminghafari2007forecasting}, the selection of the optimal number of decomposition levels is vital to determine the performance of the model in the wavelet domain. The following formula for the number of decomposition levels, $WL=int[log(n)]$, is used to select the number of decomposition levels, where $n$ is the time-series length. 

The wavelet-based ARIMA (WARIMA) model transforms the time series data by using a hybrid maximal overlap discrete wavelet transform (MODWT) algorithm with a ‘haar’ filter \cite{percival2000wavelet}. Daubechies wavelets can produce identical events across the observed time series in so many fashions that most other time series prediction models cannot recognize. The necessary steps of a wavelet-based forecasting model, defined by \cite{aminghafari2007forecasting}, are as follows. Firstly, the Daubechies wavelet transformation and a decomposition level are applied to the nonstationary time series data. Secondly, the series is reconstructed by removing the high-frequency component, using the wavelet denoising method. Lastly, an appropriate ARIMA model is applied to the reconstructed series to generate out-of-sample forecasts of the given time series data. Wavelets were first considered as a family of functions by Morlet \cite{wang2002multiple}, constructed from the translations and dilation of a single function, which is called ``Mother Wavelet". These wavelets are defined as follows:
$$ \phi_{m,n}(t) = \frac{1}{\sqrt{|m|}} \phi\left(\frac{t-n}{m}\right); \; \; m, n \in \mathcal{R},$$
where the parameter $m \; (\neq 0)$ is denoted as the scaling parameter or scale, and it measures the degree of compression. The parameter $n$ is used to determine the time location of the wavelet, and it is called the translation parameter. If the value $|m| < 1$, then the wavelet in $m$ is a compressed version (smaller support is the time domain) of the mother wavelet and primarily corresponds to higher frequencies, and when $|m| > 1$, then $\phi_(m,n) (t)$ has larger time width than $\phi(t)$ and corresponds to lower frequencies. Hence wavelets have time width adopted to their frequencies, which is the main reason behind the success of the Morlet wavelets in signal processing and time-frequency signal analysis \cite{nury2017comparative}. An implementation of the WARIMA model is available using the `WaveletFittingarma’ function under the ``WaveletArima" package in R statistical software \cite{paul2017package}.

\subsection{Autoregressive fractionally integrated moving average (ARFIMA) model}
Fractionally autoregressive integrated moving average or autoregressive fractionally integrated moving average models are the generalized version ARIMA model in time series forecasting, which allow non-integer values of the differencing parameter \cite{granger1980introduction}. It may sometimes happen that our time-series data is not stationary, but when we try differencing with parameter $d$ taking the value to be an integer, it may over difference it. To overcome this problem, it is necessary to difference the time series data using a fractional value. These models are useful in modeling time series, which has deviations from the long-run mean decay more slowly than an exponential decay; these models can deal with time-series data having long memory \cite{pumi2019beta}. ARFIMA models can be mathematically expressed as follows:
$$ \left( 1 - \sum_{i=1}^{p} \Phi_i B^i \right) (1 - B)^d X_t = \left( 1 + \sum_{i=1}^{q} \theta_i B^i \right)\epsilon_t,$$
where $B$ is is the backshift operator, $p, q$ are ARIMA parameters, and $d$ is the differencing term (allowed to take non-integer values).  
An R implementation of ARFIMA model can be done with `arfima' function under the ``forecast"package \cite{hyndman2020package}. An ARFIMA$(p,d,q)$ model is selected and estimated automatically using the Hyndman-Khandakar (2008) \cite{hyndman2008forecasting} algorithm to select $p$ and $q$ and the Haslett and Raftery (1989) \cite{haslett1989space} algorithm to estimate the parameters including $d$.

\subsection{Exponential smoothing state space (ETS) model}
Exponential smoothing state space methods are very effective methods in case of time series forecasting. Exponential smoothing was proposed in the late 1950s \cite{winters1960forecasting} and has motivated some of the most successful forecasting methods. Forecasts produced using exponential smoothing methods are weighted averages of past observations, with the weights decaying exponentially as the observations get older. The ETS models belong to the family of state-space models, consisting of three-level components such as an error component (E), a trend component (T), and a seasonal component(S). This method is used to forecast univariate time series data. Each model consists of a measurement equation that describes the observed data, and some state equations that describe how the unobserved components or states (level, trend, seasonal) change over time \cite{hyndman2018forecasting}. Hence, these are referred to as state-space models. The flexibility of the ETS model lies in its ability to trend and seasonal components of different traits. Errors can be of two types: Additive and Multiplicative. Trend Component can be any of the following: None, Additive, Additive Damped, Multiplicative and Multiplicative Damped. Seasonal Component can be of three types: None, Additive, and Multiplicative. Thus, there are 15 models with additive errors and 15 models with multiplicative errors. To determine the best model of 30 ETS models, several criteria such as Akaike's Information Criterion (AIC), Akaike's Information Criterion correction (AICc), and Bayesian Information Criterion (BIC) can be used \cite{hyndman2008forecasting}. An R implementation of the model is available in the `ets' function under ``forecast" package \cite{hyndman2020package}.

\subsection{Self-exciting threshold autoregressive (SETAR) model}
As an extension of autoregressive model, Self-exciting threshold autoregressive (SETAR) model is used to model time series data, in order to allow for higher degree of flexibility in model parameters through a regime switching behaviour \cite{tong1990non}. Given a time-series data $y_t$, the SETAR model is used to predict future values, assuming that the behavior of the time series changes once the series enters a different regime. This switch from one to another regime depends on the past values of the series. The model consists of $k$ autoregressive (AR) parts, each for a different regime. The model is usually denoted as SETAR $(k,p)$ where $k$ is the number of threshold, there are $k+1$ number of regime in the model and $p$ is the order of the autoregressive part. For example, suppose an AR(1) model is assumed in both regimes, then a 2-regime SETAR model is given by \cite{franses2000non}:
\begin{equation} 
\begin{split}
y_t & = \phi_{0,1} + \phi_{1,1}y_{t-1} + \epsilon_t \; \; \text{if} \; \; y_{t-1} \leq c,  \\
 & = \phi_{0,2} + \phi_{1,2}y_{t-1} + \epsilon_t \; \; \text{if} \; \; y_{t-1} > c,
\end{split}
\end{equation}
where for the moment the $\epsilon_t$ are assumed to be an i.i.d. white noise sequence conditional upon the history of the time series and $c$ is the threshold value. The SETAR model assumes that the border between the two regimes is given by a specific value of the threshold variable $y_{t-1}$. The model can be implemented using `setar' function under the ``tsDyn" package in R \cite{di2020package}. 

\subsection{Bayesian structural time series (BSTS) model}
Bayesian Statistics has many applications in the field of statistical techniques such as regression, classification, clustering, and time series analysis. Scott and Varian \cite{scott2014predicting} used structural time series models to show how Google search data can be used to improve short-term forecasts of economic time series. In the structural time series model, the observation in time $t$, $y_t$ is defined as follows:
$$ y_t = X_{t}^{T}\beta_t + \epsilon_t$$
where $\beta_t$ is the vector of latent variables, $X_t$ is the vector of model parameters, and $\epsilon_t$ are assumed follow Normal distributions with zero mean and $H_t$ as the variance. In addition, $\beta_t$ is represented as follows:   
$$ \beta_{t+1} = S_t \beta_t + R_t \delta_t, $$
where $\delta_t$ are assumed to follow Normal distributions with zero mean and $Q_t$ as the variance. Gaussian distribution is selected as the prior of the BSTS model since we use the occurred frequency values ranging from 0 to $\infty$ \cite{jammalamadaka2018multivariate}. An R implementation is available under the ``bsts" package \cite{scott2020package}, where one can add local linear trend and seasonal components as required. The state specification is passed as an argument to `bsts' function, along with the data and the desired number of Markov chain Monte Carlo (MCMC) iterations, and the model is fit using an MCMC algorithm \cite{scott2013bayesian}.

\subsection{Theta model}
The `Theta method' or `Theta model' is a univariate time series forecasting technique that performed particularly well in M3 forecasting competition and of interest to forecasters \cite{assimakopoulos2000theta}. The method decomposes the original data into two or more lines, called theta lines, and extrapolates them using forecasting models. Finally, the predictions are combined to obtain the final forecasts. The theta lines can be estimated by simply modifying the `curvatures' of the original time series \cite{spiliotis2020generalizing}. This change is obtained from a coefficient, called $\theta$ coefficient, which is directly applied to the second differences of the time series:
\begin{equation}
	Y^{"}_{new}(\theta) = \theta  Y^{"}_{data},  
	\label{eqn1}
\end{equation}
where $ Y^{"}_{data}= Y_t - 2 Y_{t-1} + Y_{t-2}$ at time $t$ for $t=3,4,\cdots,n$ and $\{Y_1,Y_2,\cdots,Y_n\}$ denote the observed univariate time series. In practice, coefficient $\theta$ can be considered as a transformation parameter which creates a series of the same mean and slope with that of the original data but having different variances. Now, Eqn. (\ref{eqn1}) is a second-order difference equation and has solution of the following form \cite{hyndman2003unmasking}:
\begin{equation}
	Y_{new}(\theta) = a_{\theta} + b_{\theta}(t-1) + \theta  Y_{t},  
	\label{eqn2}
\end{equation}
where $a_{\theta}$ and $b_{\theta}$ are constants and $t=1,2,\cdots,n$. Thus, $Y_{new}(\theta)$ is equivalent to a linear function of $Y_t$ with a linear trend added. The values of $a_{\theta}$ and $b_{\theta}$ are computed by minimizing the sum of squared differences:
\begin{equation}
	\displaystyle\sum_{i=1}^{t} \left[ Y_t - Y_{new}(\theta) \right]^2 = \displaystyle\sum_{i=1}^{t}\left[(1-\theta) Y_{t} - a_{\theta} - b_{\theta}(t-1) \right]^2.
	\label{eqn3}
\end{equation}
Forecasts from the Theta model are obtained by a weighted average of forecasts of $Y_{new}(\theta)$ for different values of $\theta$. Also, the prediction intervals and likelihood-based estimation of the parameters can be obtained based on a state-space model, demonstrated in \cite{hyndman2003unmasking}. An R implementation of the Theta model is possible with `thetaf' function in ``forecast" package \cite{hyndman2020package}. 

\subsection{TBATS model}
The main objective of TBATS model is to deal with complex seasonal patterns using exponential smoothing \cite{de2011forecasting}. The name is acronyms for key features of the models: Trigonometric seasonality (T), Box-Cox Transformation (B), ARMA errors (A), Trend (T) and Seasonal (S) components. TBATS makes it easy for users to handle data with multiple seasonal patterns. This model is preferable when the seasonality changes over time \cite{hyndman2018forecasting}. TBATS models can be described as follows:
$$ y_{t}^{(\mu)} = l_{t-1} + \phi b_{t-1} + \sum_{i=1}^{T} s_{t-m_i}^{(i)} + d_t $$
$$ l_{t} = l_{t-1} + \phi b_{t-1} + \alpha d_t $$
$$ b_{t} = \phi b_{t-1} + \beta d_t $$
$$ d_t = \sum_{i=1}^{p} \psi_i d_{t-i} + \sum_{j=1}^{q} \theta_j e_{t-j} +  e_{t} ; $$
where $y_{t}^{(\mu)}$ is the time series at time point $t$ (Box-Cox Transformed), $s_{t}^{(i)}$ is the $i$-th seasonal component, $l_t$ is the local level, $b_t$ is the trend with damping, $d_t$ is the ARMA$(p,q)$ process for residuals and $e_t$ as the Gaussian white noise. TBATS model can be implemented using `tbats' function under the ``forecast" package in R statistical software \cite{hyndman2020package}.

\subsection{Artificial neural networks (ANN) model}
Forecasting with artificial neural networks (ANN) has received increasing interest in various research and applied domains in the late 1990s. It has been given special attention in epidemiological forecasting \cite{philemon2019review}. Multi-layered feed-forward neural networks with back-propagation learning rules are the most widely used models with applications in classification and prediction problems \cite{zhang1998forecasting}. There is a single hidden layer between the input and output layers in a simple feed-forward neural net, and where weights connect the layers. Denoting by $\omega_{ji}$ the weights between the input layer and hidden layer and $\nu_{kj}$ denotes the weights between the hidden and output layers. Based on the given inputs $x_i$, the neuron's net input is calculated as the weighted sum of its inputs. The output layer of the neuron, $y_j$, is based on a sigmoidal function indicating the magnitude of this net-input \cite{goodfellow2016deep}. For the $j^{th}$ hidden neuron, the calculation for the net input and output are:
$$net_j^h= \displaystyle\sum_{i=1}^{n} \omega_{ji} x_i \; \; \; \text{and} \; \; \; y_j=f(net_j^h).$$
For the $k^{th}$ output neuron:
$$net_k^o=\displaystyle\sum_{j=1}^{J+1} \nu_{kj} y_j \; \; \; \text{and} \; \; \; o_k=f(net_k^o), \; \; \; 
\text{where} \; f(net)= \frac{1}{1+e^{-\lambda net}}$$ 
with $\lambda \in (0,1)$ is a parameter used to control the gradient of the function and $J$ is the number of neurons in the hidden layer. The back-propagation \cite{rumelhart1985learning} learning algorithm is the most commonly used technique in ANN. In the error back-propagation step, the weights in ANN are updated by minimizing 
$$E=\frac{1}{2P} \displaystyle\sum_{p=1}^{P} \displaystyle\sum_{k=1}^{K} (d_{pk}-O_{pk})^2,$$
where, $d_{pk}$ is the desired output of neuron $k$ and for input pattern $p$. The common formula for number of neurons in the hidden layer is $h=\frac{(i+j)}{2} + \sqrt{d}$, for selecting the number of hidden neurons, where $i$ is the number of output $y_j$ and $d$ denotes the number of $i$ training patterns in the input $x_i$ \cite{zhang2005neural}. The application of ANN for time series data is possible with `mlp' function under "nnfor" package in R \cite{kourentzes2017nnfor}.

\subsection{Autoregressive neural network (ARNN) model}
Autoregressive neural network (ARNN) received attention in time series literature in late 1990s \cite{faraway1998time}. The architecture of a simple feedforward neural network can be described as a network of neurons arranged in input layer, hidden layer, and output layer in a prescribed order. Each layer passes the information to the next layer using weights that are obtained using a learning algorithm \cite{zhang2005neural}. ARNN model is a modification to the simple ANN model especially designed for prediction problems of time series datasets \cite{faraway1998time}. ARNN model uses a pre-specified number of lagged values of the time series as inputs and number of hidden neurons in its architecture is also fixed \cite{hyndman2018forecasting}. ARNN($p,k$) model uses $p$ lagged inputs of the time series data in a one hidden layered feedforward neural network with $k$ hidden units in the hidden layer. Let $\underline{x}$ denotes a $p$-lagged inputs and $f$ is a neural network of the following architecture:
\begin{equation}
	f(\underline{x}) = c_{0} + \displaystyle \sum_{j=1}^{k} w_j \phi \left( a_j + b_{j} '\underline{x} \right);
	\label{eqn4}
\end{equation}
where $c_0, a_j, w_j$ are connecting weights, $b_j$ are $p$-dimensional weight vector and $\phi$ is a bounded nonlinear sigmoidal function (e.g., logistic squasher function or tangent hyperbolic activation function). These Weights are trained using a gradient descent backpropagation \cite{rumelhart1985learning}. Standard ANN faces the dilemma to choose the number of hidden neurons in the hidden layer and optimal choice is unknown. But for ARNN model, we adopt the formula $k = [(p+1)/2]$ for non-seasonal time series data where $p$ is the number of lagged inputs in an autoregressive model \cite{hyndman2018forecasting}. ARNN model can be applied using the `nnetar' function available in the R statistical package ``forecast" \cite{hyndman2020package}. 

\subsection{Ensemble forecasting models}
\begin{figure}[H]
	\centering
	\includegraphics[scale=0.75]{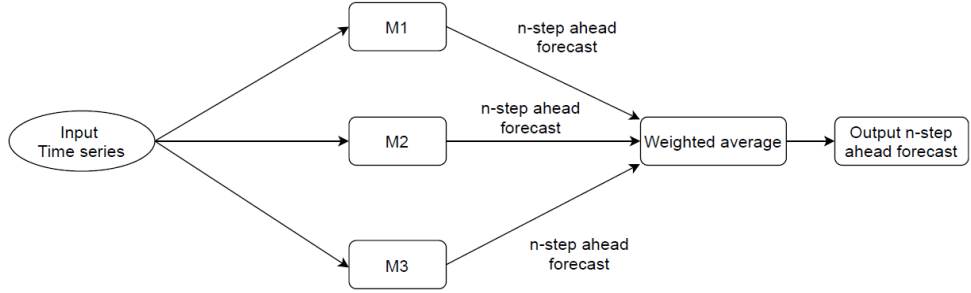}
	\caption{Flow diagram of the ensemble model where M1, M2, and M3 are three different univariate time series models} \label{flow_chart_ensemble}
\end{figure}

The idea of ensemble time series forecasts was given by Bates and Granger (1969) in their seminal work \cite{bates1969combination}. Forecasts generated from ARIMA, ETS, Theta, ARNN, WARIMA can be combined with equal weights, weights based on in-sample errors, or cross-validated weights. In the ensemble framework, cross-validation for time series data with user-supplied models and forecasting functions is also possible to evaluate model accuracy \cite{shaub2020fast}. Combining several candidate models can hedge against an incorrect model specification. Bates and Granger(1969) \cite{bates1969combination} suggested such an approach and observed, somewhat surprisingly, that the combined forecast can even outperform the single best component forecast. While combination weights selected equally or proportionally to past model errors are possible approaches, many more sophisticated combination schemes, have been suggested. For example, rather than normalizing weights to sum to unity, unconstrained and even negative weights could be possible \cite{granger1984improved}. The simple equal-weights combination might appear woefully obsolete and probably non-competitive compared to the multitude of sophisticated combination approaches or advanced machine learning and neural network forecasting models, especially in the age of big data. However, such simple combinations can still be competitive, particularly for pandemic time series \cite{shaub2020fast}. A flow diagram of the ensemble method is presented in Figure \ref{flow_chart_ensemble}.

The ensemble method by \cite{bates1969combination} produces forecasts out to a horizon $h$ by applying a weight $w_m$ to each $m$ of the $n$ model forecasts in the ensemble. The ensemble forecast $f(i)$ for time horizon $1 \leq i \leq h$ and with individual component model forecasts $f_m(i)$ is then 
$$ f(i) = \displaystyle\sum_{m=1}^{n} w_m f_m(i). $$
The weights can be determined in several ways (for example, supplied by the user, set equally, determined by in-sample errors, or determined by cross-validation).  The ``forecastHybrid" package in R includes these component models in order to enhance the ``forecast" package base models with easy ensembling (e.g., `hybridModel' function in R statistical software) \cite{shaub4forecasthybrid}.

\subsection{Hybrid forecasting models}
\begin{figure}[H]
	\centering
	\includegraphics[scale=0.75]{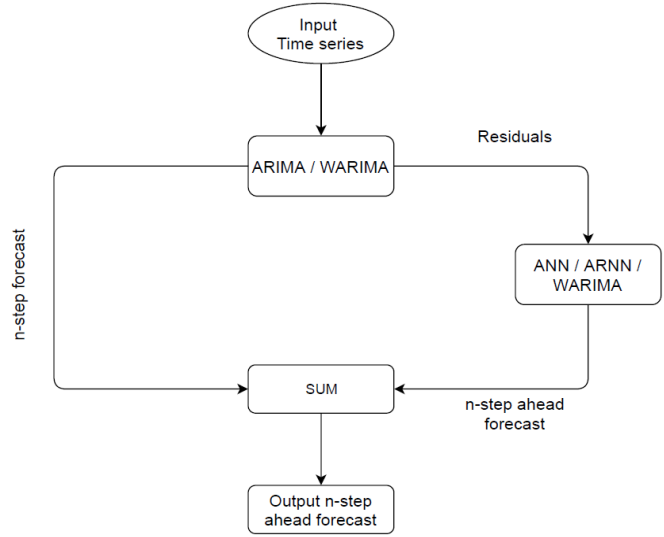}
	\caption{Flow diagram of the hybrid forecasting model} \label{flow_chart_hybrid}
\end{figure}

The idea of hybridizing time series models and combining different forecasts was first introduced by Zhang \cite{zhang2003time} and further extended by \cite{khashei2010artificial, chakraborty2019forecasting, chakraborty2020real, chakraborty2020theta}. The hybrid forecasting models are based on an error re-modeling approach, and there are broadly two types of error calculations popular in the literature, which are given below \cite{mosleh1986assessment, chowdhury2020multiplicative}:
\begin{definition}
In the additive error model, the forecaster treats the expert's
estimate as a variable, $\hat{Y_t}$, and thinks of it as the sum of two terms:
$$\hat{Y_t}=Y_t + e_t,$$ where $Y_t$ is the true value and $e_t$ be the additive error term. \label{def1}
\end{definition}
\begin{definition}
In the multiplicative error model, the forecaster treats the
expert's estimate $\hat{Y_t}$ as the product of two terms:
$$\hat{Y_t}=Y_t \times e_t,$$ where $Y_t$ is the true value and $e_t$ be the multiplicative error term. \label{def2}
\end{definition}
Now, even if the relationship is of product type, in the log-log scale it becomes additive. Hence, without loss of generality, we may assume the relationship to be additive and expect errors (additive) of a forecasting model to be random shocks \cite{chakraborty2020theta}. These hybrid models are useful for complex correlation structures where less amount of knowledge is available about the data generating process. A simple example is the daily confirmed cases of the COVID-19 cases for various countries where very little is known about the structural properties of the current pandemic. The mathematical formulation of the proposed hybrid model ($Z_t$) is as follows:
\begin{eqnarray*}
	Z_t &=& L_t + N_t,
\end{eqnarray*}
where $L_t$ is the linear part and $N_t$ is the nonlinear part of the hybrid model. We can estimate both $L_t$ and $N_t$ from the available time series data. Let $\hat{L_t}$ be the forecast value of the linear model (e.g., ARIMA) at time $t$ and $\epsilon_{t}$ represent the error
residuals at time $t$, obtained from the linear model. Then, we write
\begin{eqnarray*}
	\epsilon_{t} & = & Z_t - \hat{L_t}.
\end{eqnarray*}
These left-out residuals are further modeled by a nonlinear model (e.g., ANN or ARNN) and can be represented as follows:
\begin{eqnarray*}
	\epsilon_{t} &=& f(\epsilon_{t-1}, \epsilon_{t-2}, . . . , \epsilon_{t-p}) + \varepsilon_t,
\end{eqnarray*}
where $f$ is a nonlinear function, and the modeling is done by the nonlinear ANN or ARNN model as defined in Eqn. (\ref{eqn4}) and $\varepsilon_t$ is supposed to be the random shocks. Therefore, the combined forecast can be obtained as follows:
\begin{eqnarray*}
	\hat{Z_t} &=& \hat{L_t} + \hat{N_t},
\end{eqnarray*}
where $\hat{N_t}$ is the forecasted value of the nonlinear time series model. An overall flow diagram of the proposed hybrid model is given in Figure \ref{flow_chart_hybrid}. In the hybrid model, a nonlinear model is applied in the second stage to re-model the left-over autocorrelations in the residuals, which the linear model could not model. Thus, this can be considered as an error re-modeling approach. This is important because due to model misspecification and disturbances in the pandemic rate time series, the linear models may fail to generate white noise behavior for the forecast residuals. Thus, hybrid approaches eventually can improve the predictions for the epidemiological forecasting problems, as shown in \cite{chakraborty2019forecasting, chakraborty2020real, chakraborty2020theta}. These hybrid models only assume that the linear and nonlinear components of the epidemic time series can be separated individually. The implementation of the hybrid models used in this study are available in \cite{github2020csf}.

\section{Experimental analysis}
Five time series COVID-19 datasets for the USA, India, Russia, Brazil, and Peru UK are considered for assessing twenty forecasting models (individual, ensemble, and hybrid). The datasets are mostly nonlinear, nonstationary, and non-gaussian in nature. We have used root mean square error (RMSE), mean absolute error (MAE), mean absolute percentage error (MAPE), and symmetric MAPE (SMAPE) to evaluate the predictive performance of the models used in this study. Since the number of data points in both the datasets is limited, advanced deep learning techniques will over-fit the datasets \citep{hastie2009elements}.

\subsection{Datasets}
We use publicly available datasets to compare various forecasting frameworks. COVID-19 cases of five countries with the highest number of cases were collected \cite{owd2020, wom2020}. The datasets and their description is presented in Table \ref{tab:data_descrip}.

\begin{table}
	\centering
	\caption{Description of COVID-19 datasets}
	\begin{tabular}{|p{2cm}|p{3cm}|p{3cm}|p{2cm}|}
		\hline
		\textbf{Countries} & \textbf{Start date} & \textbf{End date} & \textbf{Length} \\
		\hline
		USA & 20/01/2020 & 15/09/2020 & 240\\
		\hline
		India & 29/01/2020 & 15/09/2020 & 231\\
		\hline
		Brazil & 25/02/2020 & 15/09/2020 & 204\\
		\hline
		Russia & 31/01/2020 & 15/09/2020 & 229 \\
		\hline
		Peru & 06/03/2020 & 15/09/2020 & 194  \\
		\hline
	\end{tabular}
	\label{tab:data_descrip}
\end{table}

\subsection{Global characteristics}
Characteristics of these five time series were examined using Hurst exponent, KPSS test and Terasvirta test and other measures as described in Section \ref{global}. Hurst exponent (denoted by H), which ranges between zero to one, is calculated to measure the long-range dependency in a time series and provides a measure of long-term nonlinearity. For values of H near zero, the time series under consideration is mean-reverting. An increase in the value will be followed by a decrease in the series and vice versa. When H is close to 0.5, the series has no autocorrelation with past values. These types of series are often called Brownian motion. When H is near one, an increase or decrease in the value is most likely to be followed by a similar movement in the future. All the five COVID-19 datasets in this study possess the Hurst exponent value near one, which indicates that these time series datasets have a strong trend of increase followed by an increase or decrease followed by another decline.

KPSS tests are performed to examine the stationarity of a given time series. The null hypothesis for the KPSS test is that the time series is stationary. Thus, the series is nonstationary when the p-value less than a threshold. From Table \ref{tab:data_tests}, all the five datasets can be characterized as non-stationary as the p-value $<$ 0.01 in each instances. Terasvirta test examines the linearity of a time series against the alternative that a nonlinear process has generated the series. It is observed that the USA, Russia, and Peru COVID-19 datasets are likely to follow a nonlinear trend. On the other hand, India and Brazil datasets have some linear trends.

\begin{table}
	\centering
	\caption{Test results on COVID-19 datasets}
	\begin{tabular}{|p{2cm}|p{3cm}|p{3cm}|p{3cm}|}
		\hline
		\textbf{Countries} & \textbf{Hurst exponent} & \textbf{KPSS test} & \textbf{Terasvirta test} \\
		\hline
		USA & 0.9996 & p-value $<$ 0.01 & p-value = 0.0181\\
		\hline
		India & 0.9997 & p-value $<$ 0.01 & p-value $<$ 0.01 \\
		\hline
		Brazil & 0.9974 & p-value $<$ 0.01 & p-value $<$ 0.01\\
		\hline
		Russia & 0.9992 & p-value $<$ 0.01 & p-value = 0.0566 \\
		\hline
		Peru & 0.9983 & p-value $<$ 0.01 & p-value = 0.8471  \\
		\hline
	\end{tabular}
	\label{tab:data_tests}
\end{table}

Further, we examine serial correlation, skewness, kurtosis, and maximum Lyapunov exponent for the five COVID-19 datasets. The results are reported in Table \ref{tab:data_chars}. The serial correlation of the datasets is computed using the Box-Pierce test statistic for the null hypothesis of independence in a given time series. The p-values related to each of the datasets were found to be below the significant level (see Table \ref{tab:data_chars}). This indicates that these COVID-19 datasets have no serial correlation when lag equals one. Skewness for Russia COVID-19 dataset is found to be negative, whereas the other four datasets are positively skewed. This means for the Russia dataset; the left tail is heavier than the right tail. For the other four datasets, the right tail is heavier than the left tail. The Kurtosis values for the India dataset are found positive while the other four datasets have negative kurtosis values. Therefore, the COVID-19 dataset of India tends to have a peaked distribution, and the other four datasets may have a flat distribution. We observe that each of the five datasets is non-chaotic in nature, i.e., the maximum Lyapunov exponents are less than unity. A summary of the implementation tools is presented in Table \ref{r}.

\begin{table}
	\centering
	\caption{Characteristics of COVID-19 datasets}
	\begin{tabular}{|p{2cm}|p{3cm}|p{2cm}|p{2cm}|p{4cm}|}
		\hline
		\textbf{Countries} & \textbf{Box test} & \textbf{Skewness} & \textbf{Kurtosis} & \textbf{Chaotic/Non-chaotic} \\
		\hline
		USA &  p-value $<$ 0.01 & 0.4971 & - 0.7465 & Non-chaotic \\
		\hline
		India & p-value $<$ 0.01 & 1.4981 & 0.9422 & Non-chaotic \\
		\hline
		Brazil & p-value $<$ 0.01 & 0.6897 & -0.7124 & Non-chaotic \\
		\hline
		Russia & p-value $<$ 0.01 & - 0.0544 & -1.4439 & Non-chaotic \\
		\hline
		Peru & p-value $<$ 0.01 & 0.4421 & -0.2142 & Non-chaotic \\
		\hline
	\end{tabular}
	\label{tab:data_chars}
\end{table}

\begin{table}
	\centering \caption{R functions and packages for implementation.}\label{R_functions_packages}
	\begin{tabular}{|c|c|c|c|}
		\hline
	    Model & R function & R package & Reference \\ \hline
		ARIMA & auto.arima & forecast & \cite{hyndman2007automatic} \\ \hline
		ETS & ets & forecast & \cite{hyndman2007automatic}  \\ \hline
		SETAR & setar & tsDyn & \cite{di2020package} \\ \hline
		TBATS & tbats & forecast & \cite{hyndman2007automatic} \\ \hline
		Theta & thetaf & forecast  & \cite{hyndman2007automatic} \\ \hline
		ANN & mlp & nnfor & \cite{kourentzes9nnfor} \\ \hline
		ARNN & nnetar & forecast & \cite{hyndman2007automatic} \\ \hline
		WARIMA & WaveletFittingarma & WaveletArima & \cite{paul2017package} \\ \hline
		BSTS & bsts & bsts & \cite{scott2020package} \\ \hline
		ARFIMA & arfima & forecast & \cite{hyndman2007automatic} \\ \hline
		Ensemble models & hybridModel & forecastHybrid & \cite{shaub4forecasthybrid} \\ \hline
		Hybrid models & - & - & \cite{github2020csf}  \\ \hline
	\end{tabular}
	\label{r}
\end{table}

\subsection{Accuracy metrics}
We used four popular accuracy metrics to evaluate the performance of different time series forecasting models. The expressions of these metrics are given below.
$$ RMSE = \sqrt{\frac{1}{n} \sum_{i=1}^n (y_i - \hat{y}_i)^2} ; \;
MAE = \frac{\displaystyle 1}{\displaystyle n} \sum_{i=1}^n |y_i - \hat{y}_i| ; $$
$$ MAPE = \frac{1}{n} \sum_{i=1}^n |\frac{\hat{y}_i - y_i}{y_i}| \times 100 ; \; SMAPE =\frac{1}{n} \sum_{i=1}^n \frac{|\hat{y}_i - y_i|}{(|\hat{y}_i|+|y_i|)/2} \times 100 ; $$
where $y_i$ are actual series values, $\hat{y}_i$ are the predictions by different models and $n$ represent the number of data points of the time series. The models with least accuracy metrics is the best forecasting model.

\subsection{Analysis of results}
This subsection is devoted to the experimental analysis of confirmed COVID-19 cases using different time series forecasting models. The test period is chosen to be 15 days and 30 days, whereas the rest of the data is used as training data (see Table \ref{tab:data_descrip}). In first columns of Tables \ref{table_15_days} and \ref{table_30_days}, we present training data and test data for USA, India, Brazil, Russia and Peru. The autocorrelation function (ACF) and partial autocorrelation function (PACF) plots are also depicted for the training period of each of the five countries in Tables \ref{table_15_days} and \ref{table_30_days}. ACF and PACF plots are generated after applying the required number of differencing of each training data using the r function `diff'. The required order of differencing is obtained by using the R function `ndiffs' which estimate the number of differences required to make a given time series stationary. The integer-valued order of differencing is then used as the value of '$d$' in the ARIMA$(p,d,q)$ model. Other two parameters `$p$' and `$q$' of the model are obtained from ACF and PACF plots respectively (see Tables \ref{table_15_days} and \ref{table_30_days}). However, we choose the `best' fitted ARIMA model using AIC value for each training dataset. Table \ref{table_15_days} presents the training data (black colored) and test data (red-colored) and corresponding ACF and PACF plots for the five time-series datasets. 

\begin{table}[H]
    \centering
    \caption{Pandemic datasets and corresponding ACF, PACF plots with 15-days test data}\label{table_15_days} \vspace{1cm}
    \begin{tabular}{ | c | p{4cm} | p{4cm} | p{4cm} |}
        \hline
        Country & Data & ACF plot & PACF plot  \\ \hline
        USA
        &
        \begin{minipage}{.30\textwidth}
            \includegraphics[width=40mm, height=30mm]{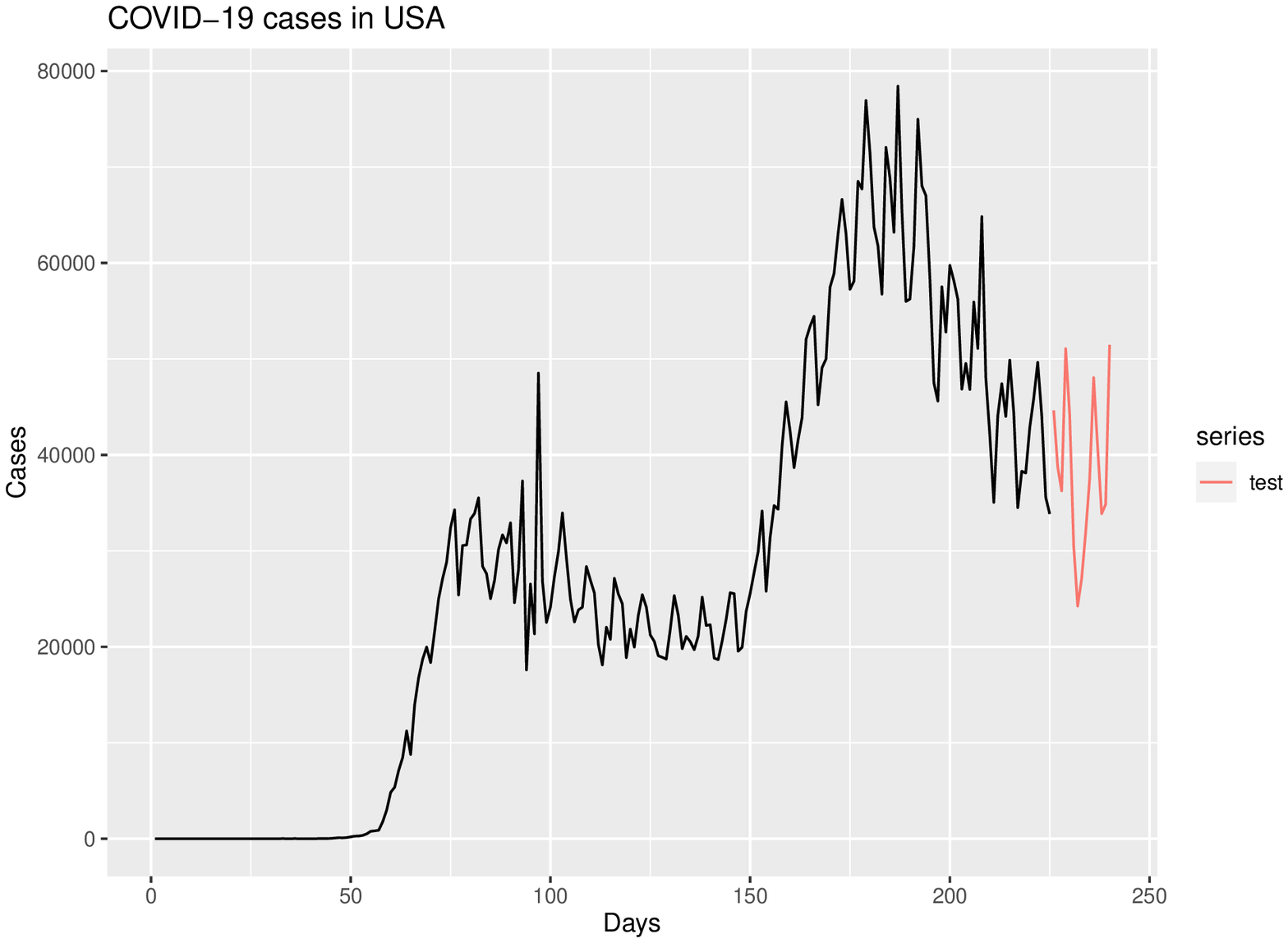}
        \end{minipage}
        &
        \begin{minipage}{.30\textwidth}
            \includegraphics[width=40mm, height=30mm]{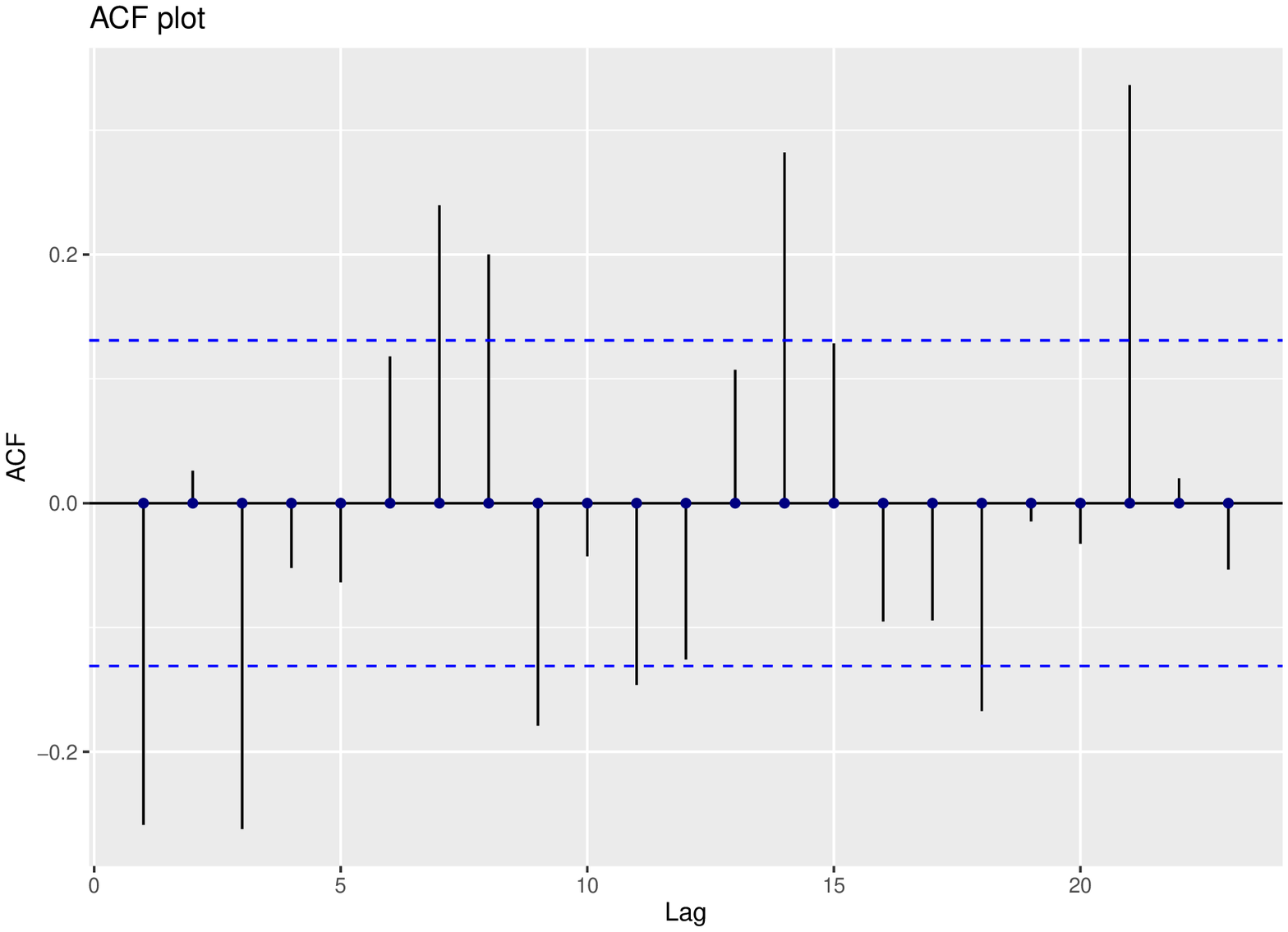}
        \end{minipage}
        &
        \begin{minipage}{.30\textwidth}
            \includegraphics[width=40mm, height=30mm]{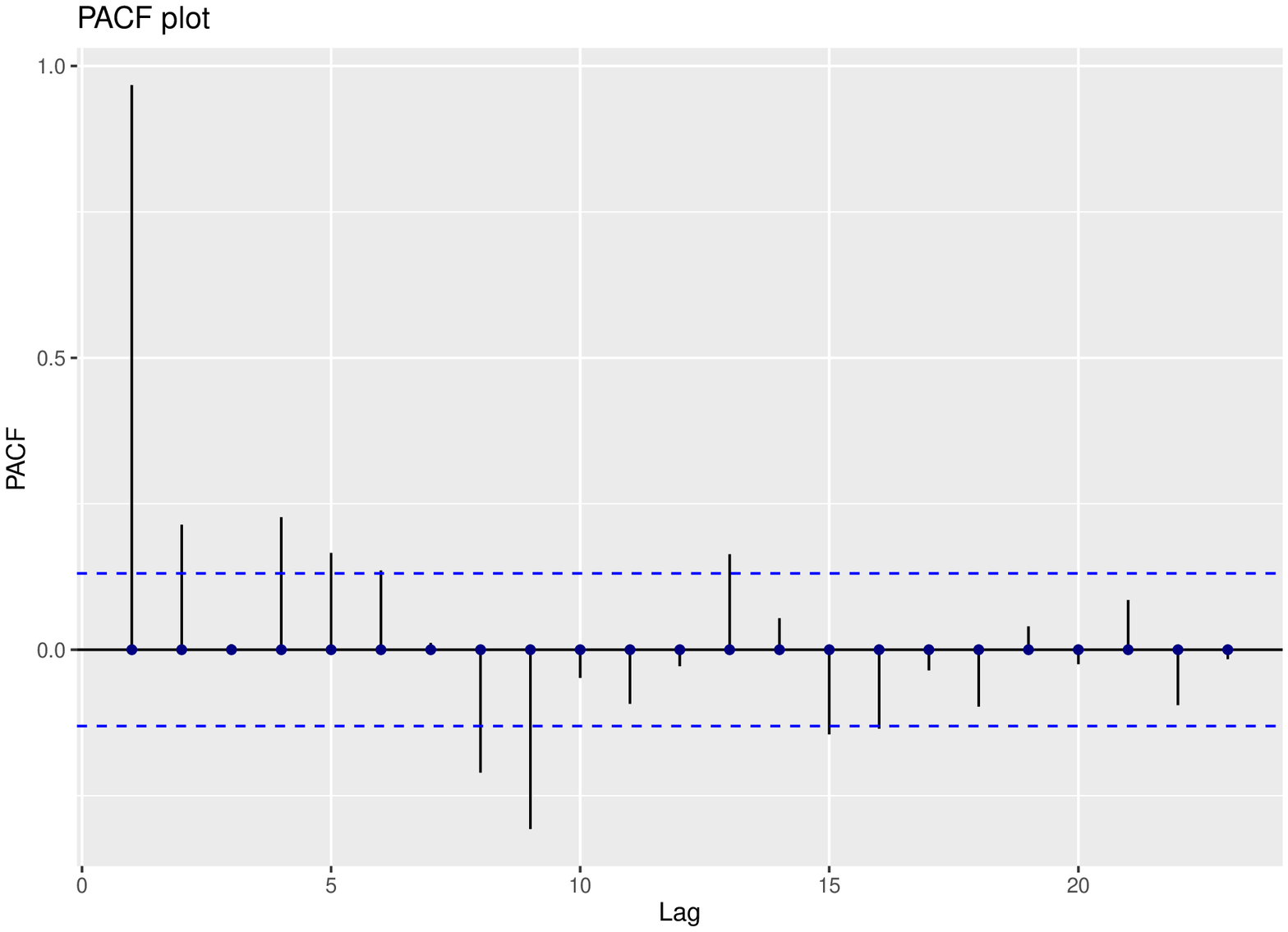}
        \end{minipage} \\ \hline
        India
        &
        \begin{minipage}{.3\textwidth}
            \includegraphics[width=40mm, height=30mm]{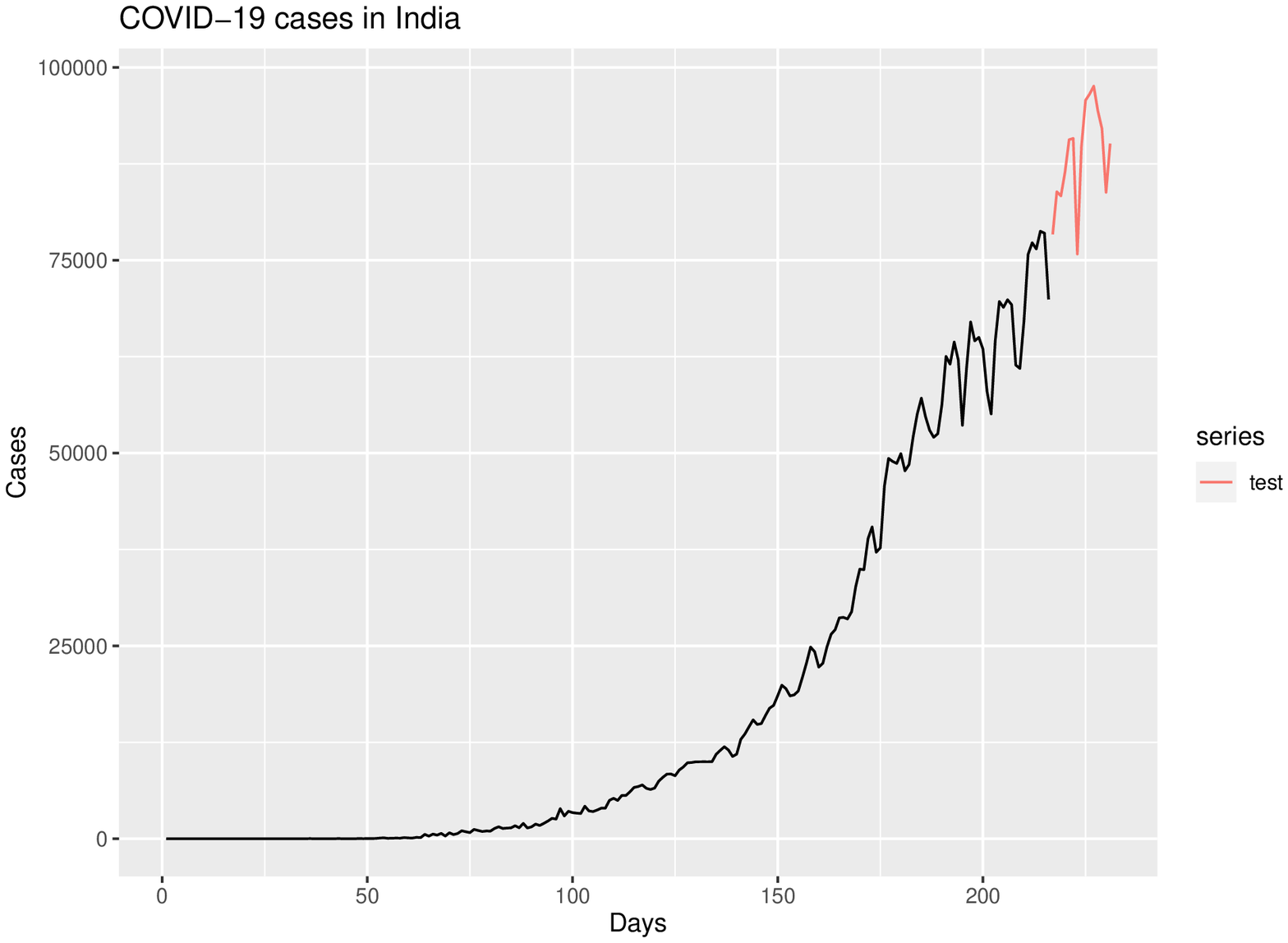}
        \end{minipage}
        &
        \begin{minipage}{.3\textwidth}
            \includegraphics[width=40mm, height=30mm]{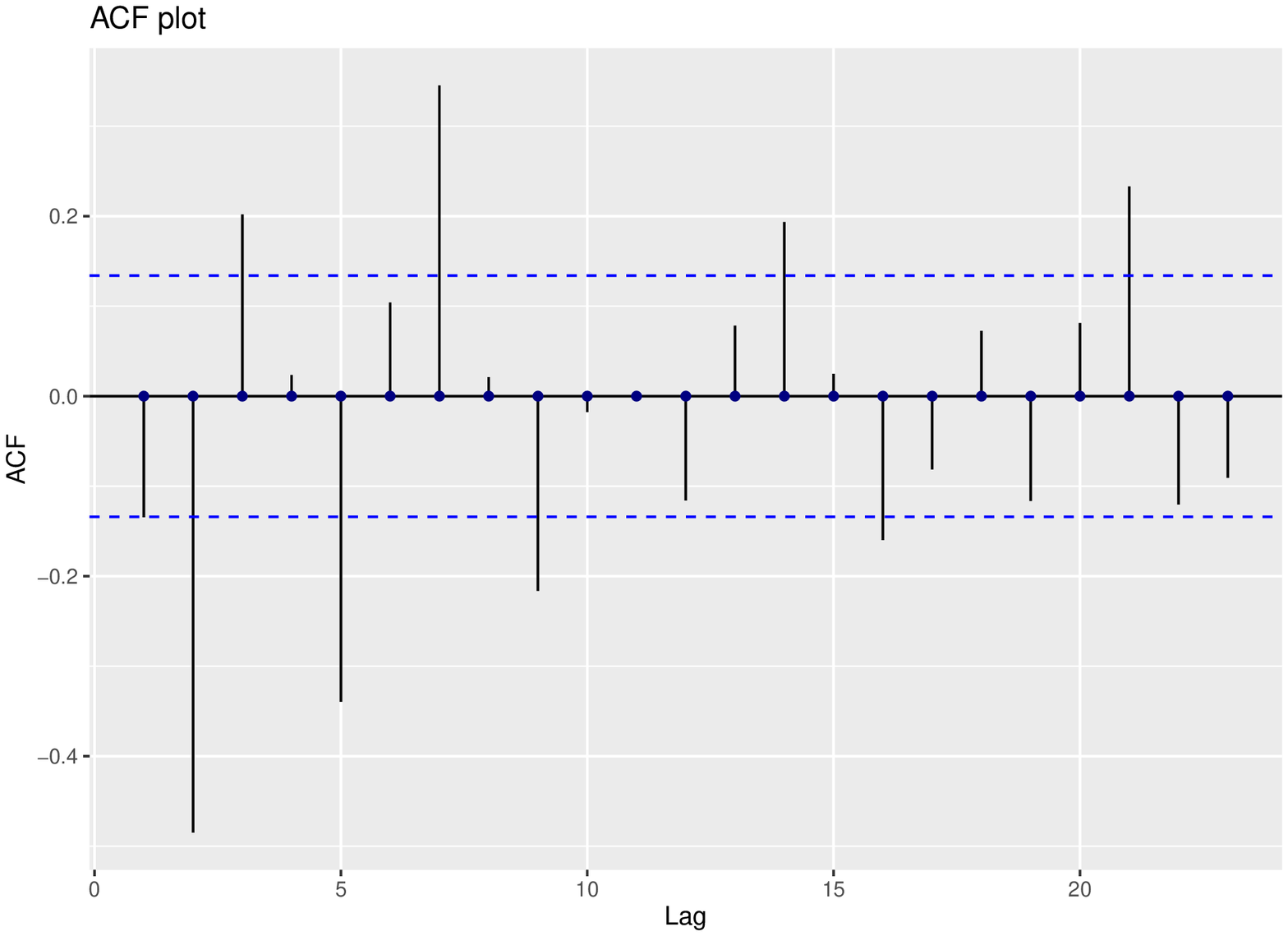}
        \end{minipage}
        &
        \begin{minipage}{.3\textwidth}
            \includegraphics[width=40mm, height=30mm]{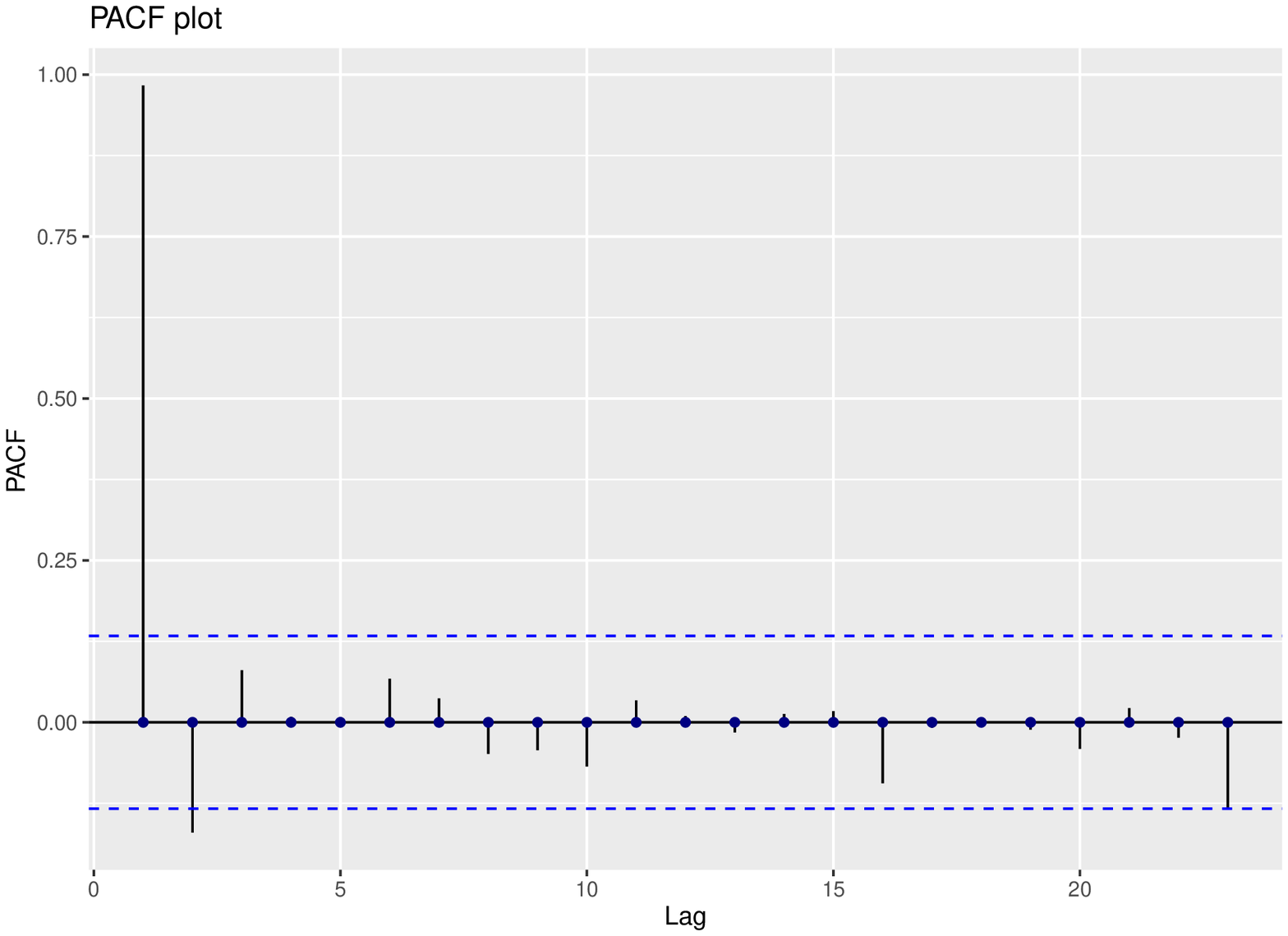}
        \end{minipage}
        \\ \hline
        Brazil
        &
        \begin{minipage}{.3\textwidth}
            \includegraphics[width=40mm, height=30mm]{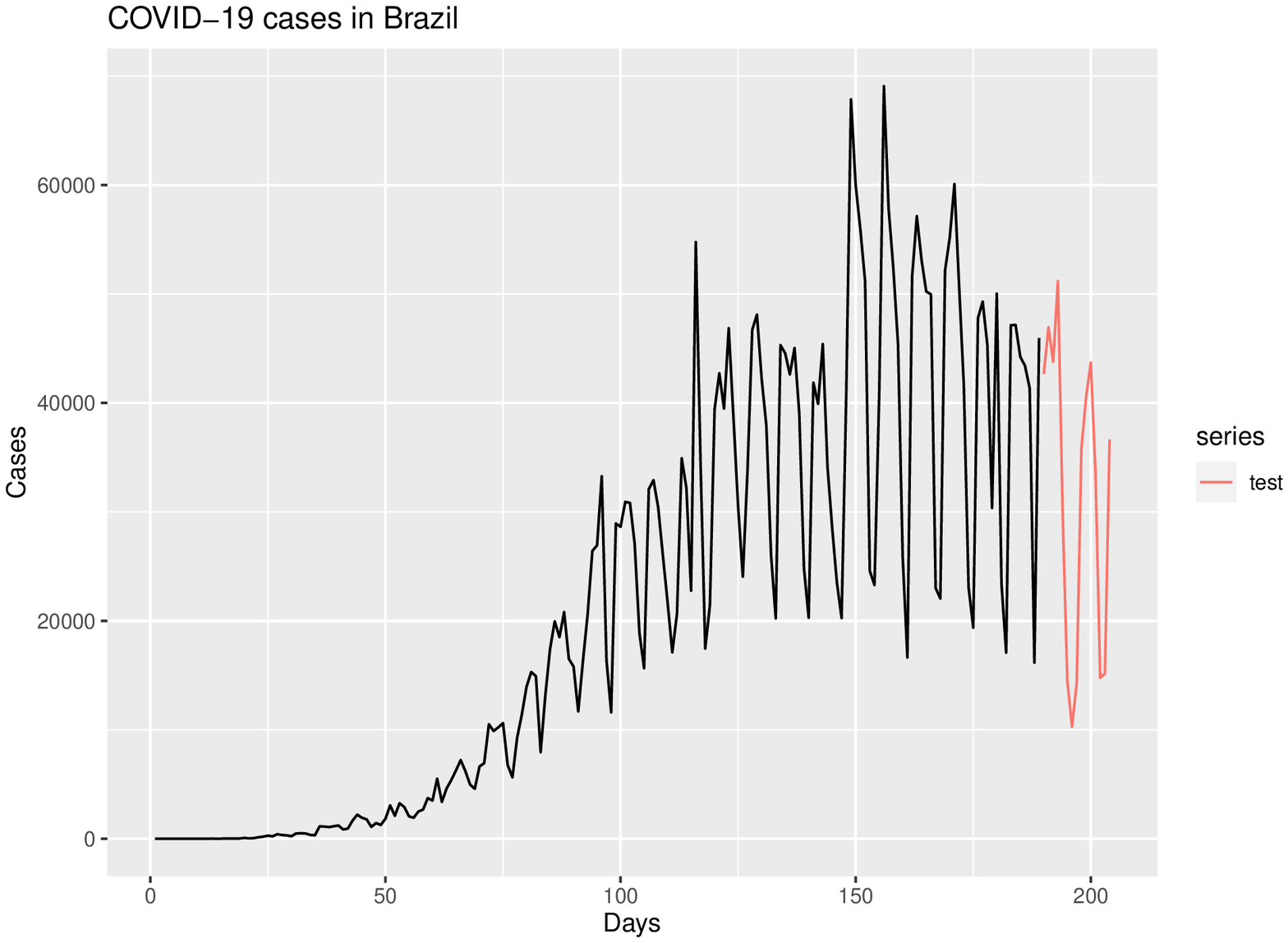}
        \end{minipage}
        &
        \begin{minipage}{.3\textwidth}
            \includegraphics[width=40mm, height=30mm]{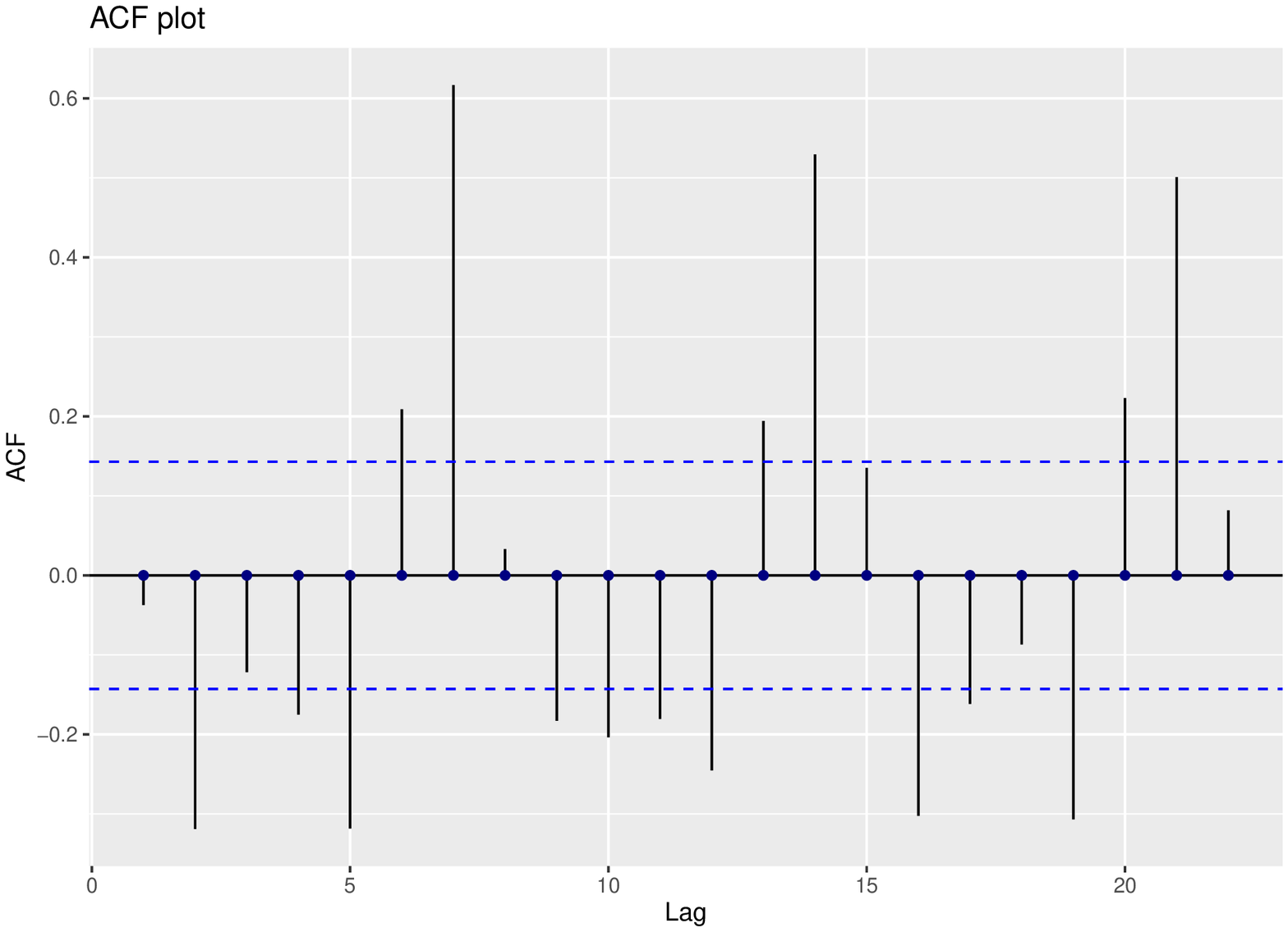}
        \end{minipage}
        &
        \begin{minipage}{.3\textwidth}
            \includegraphics[width=40mm, height=30mm]{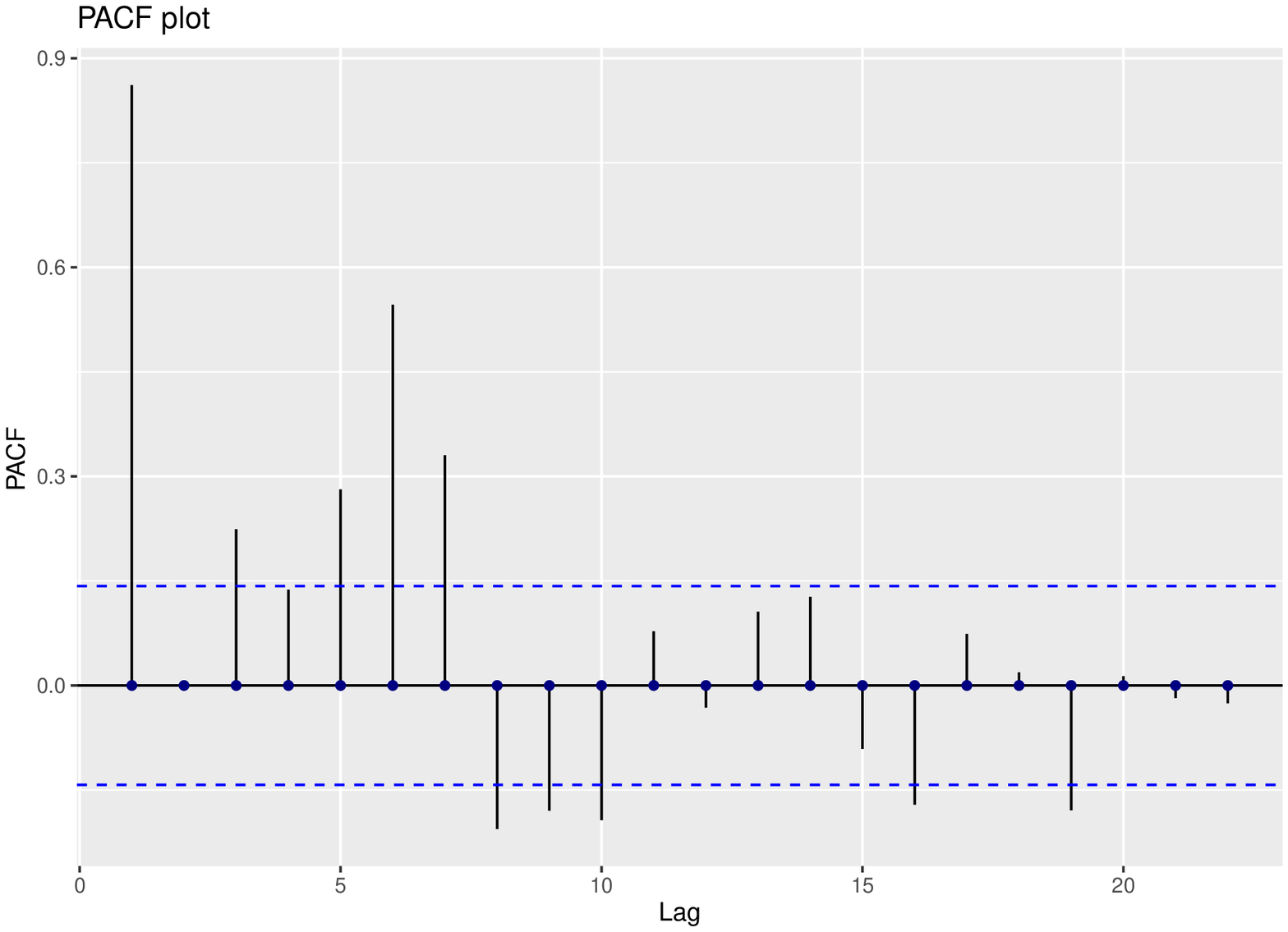}
        \end{minipage}
        \\ \hline
        Russia
        &
        \begin{minipage}{.3\textwidth}
        \includegraphics[width=40mm, height=30mm]{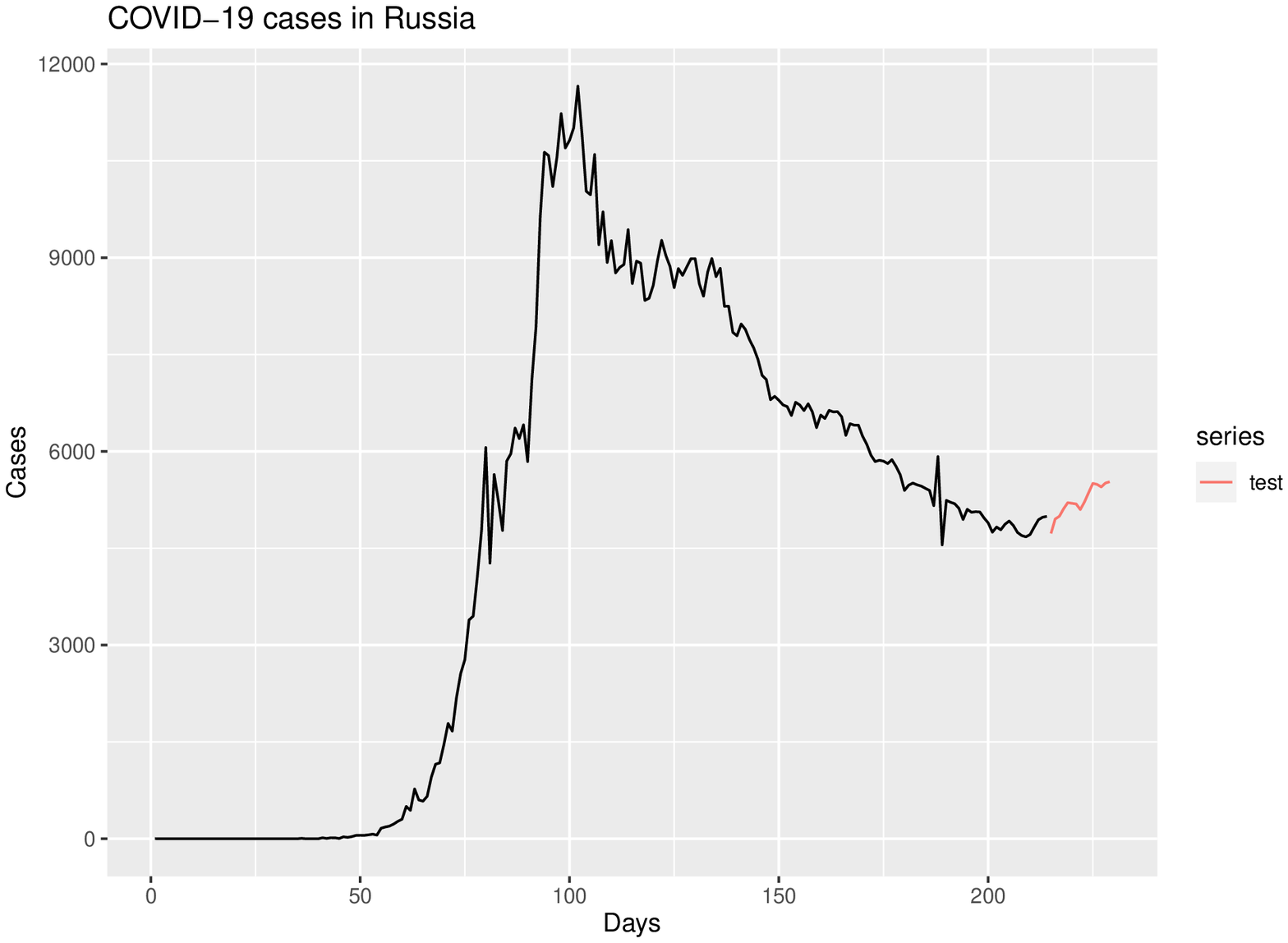}
        \end{minipage}
         &
        \begin{minipage}{.3\textwidth}
        \includegraphics[width=40mm, height=30mm]{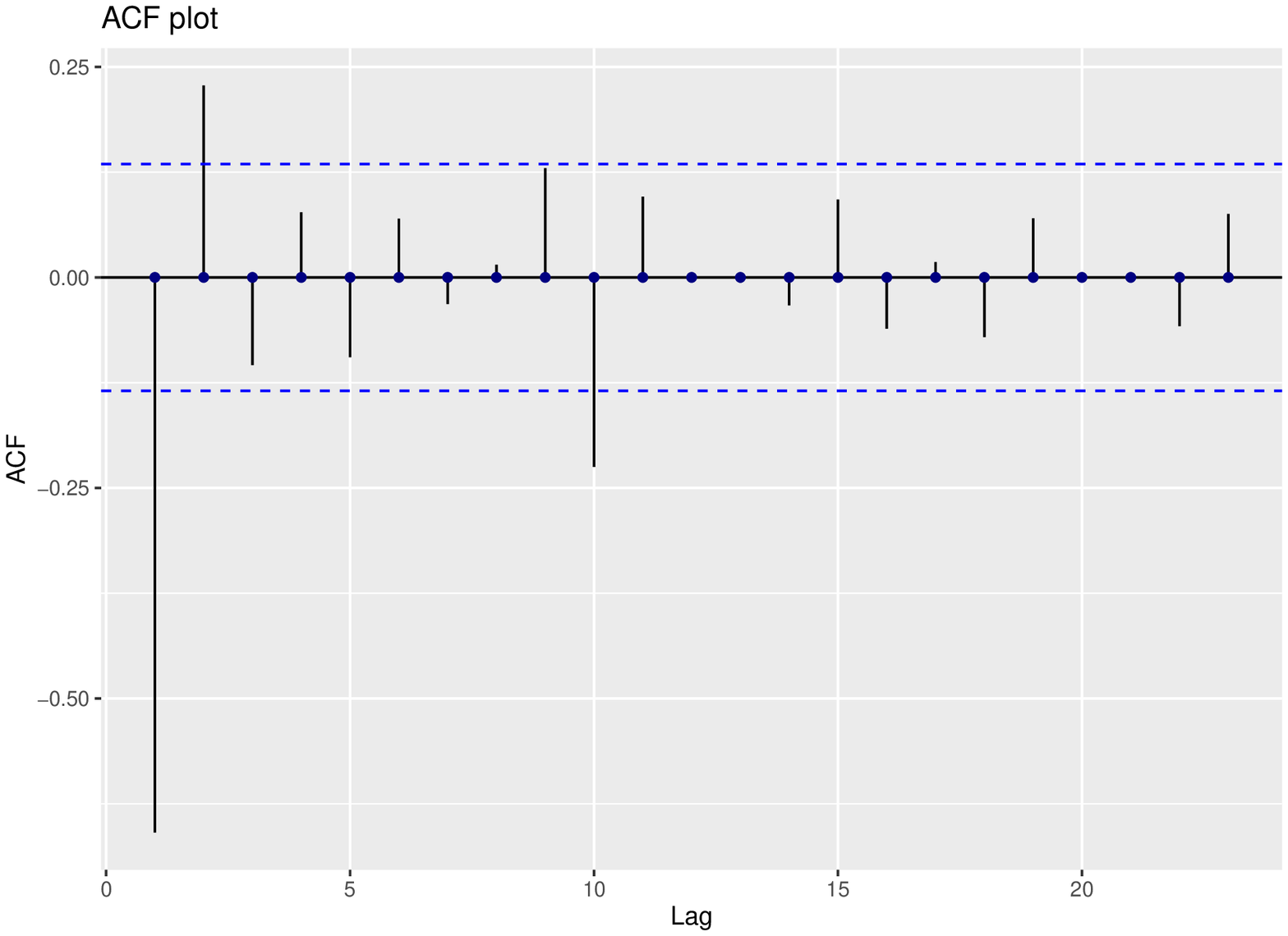}
         \end{minipage}
         &
         \begin{minipage}{.3\textwidth}
        \includegraphics[width=40mm, height=30mm]{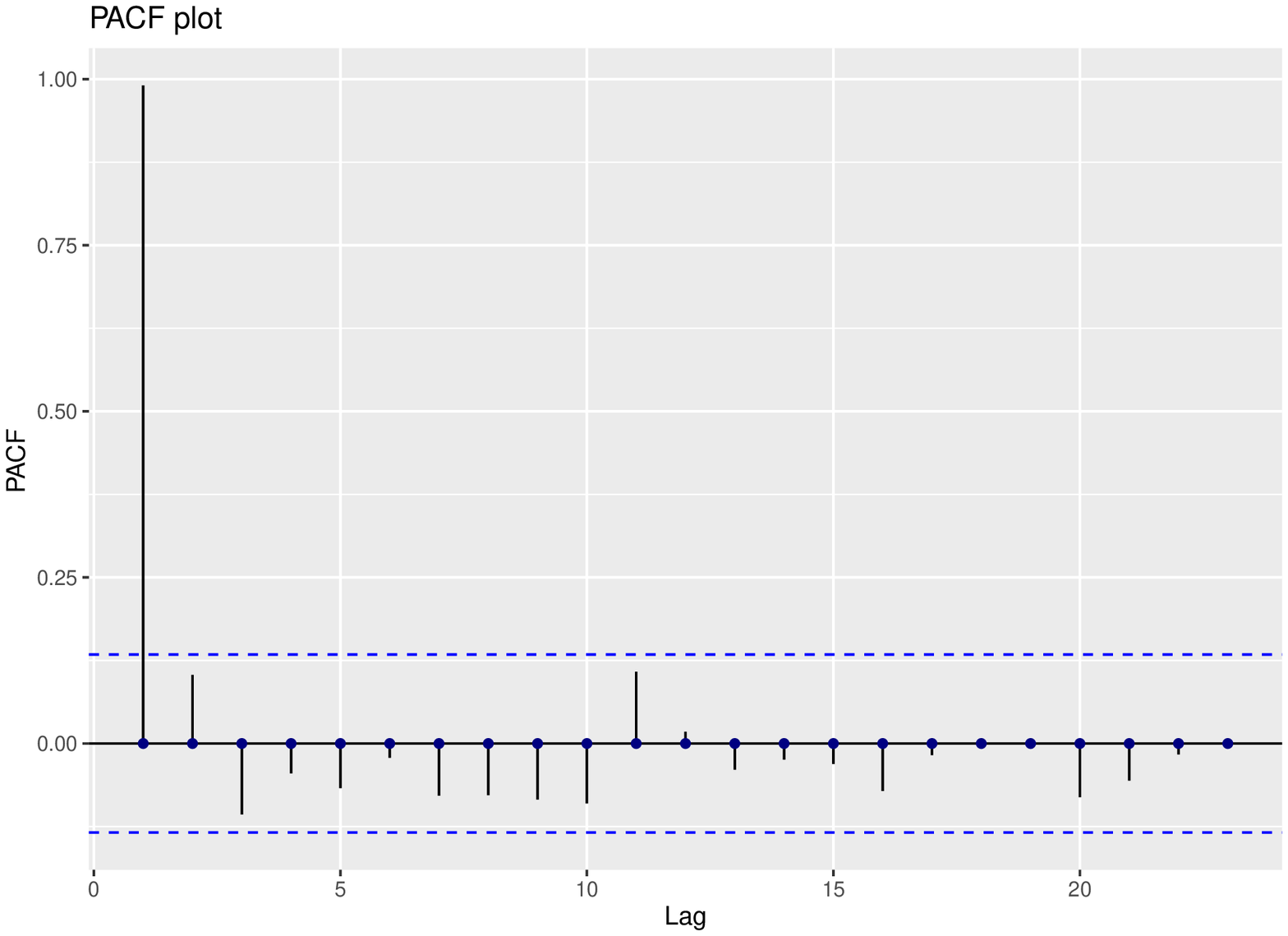}
         \end{minipage}
         \\ \hline
         Peru
         &
         \begin{minipage}{.3\textwidth}
         \includegraphics[width=40mm, height=30mm]{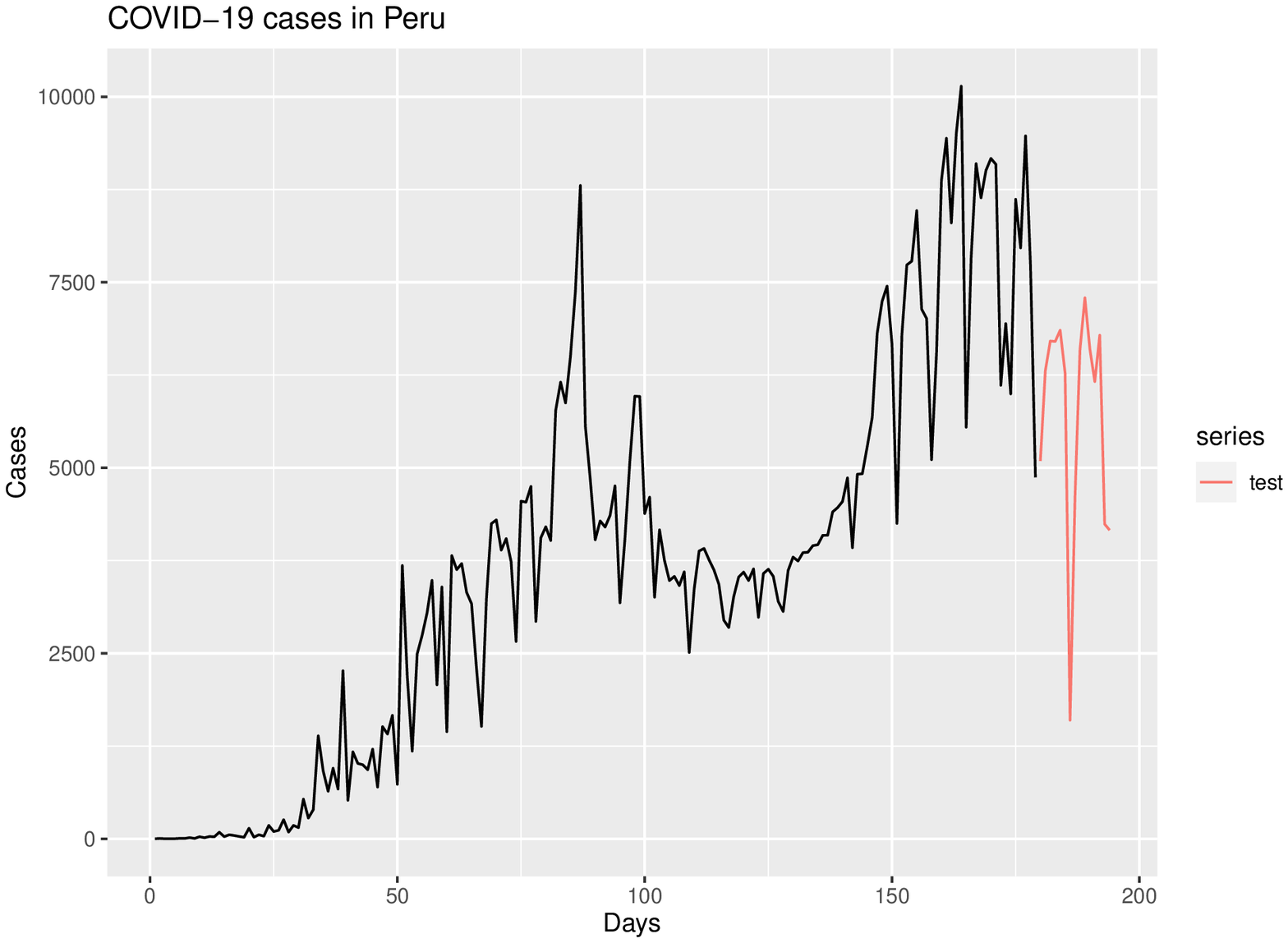}
         \end{minipage}
          &
         \begin{minipage}{.3\textwidth}
         \includegraphics[width=40mm, height=30mm]{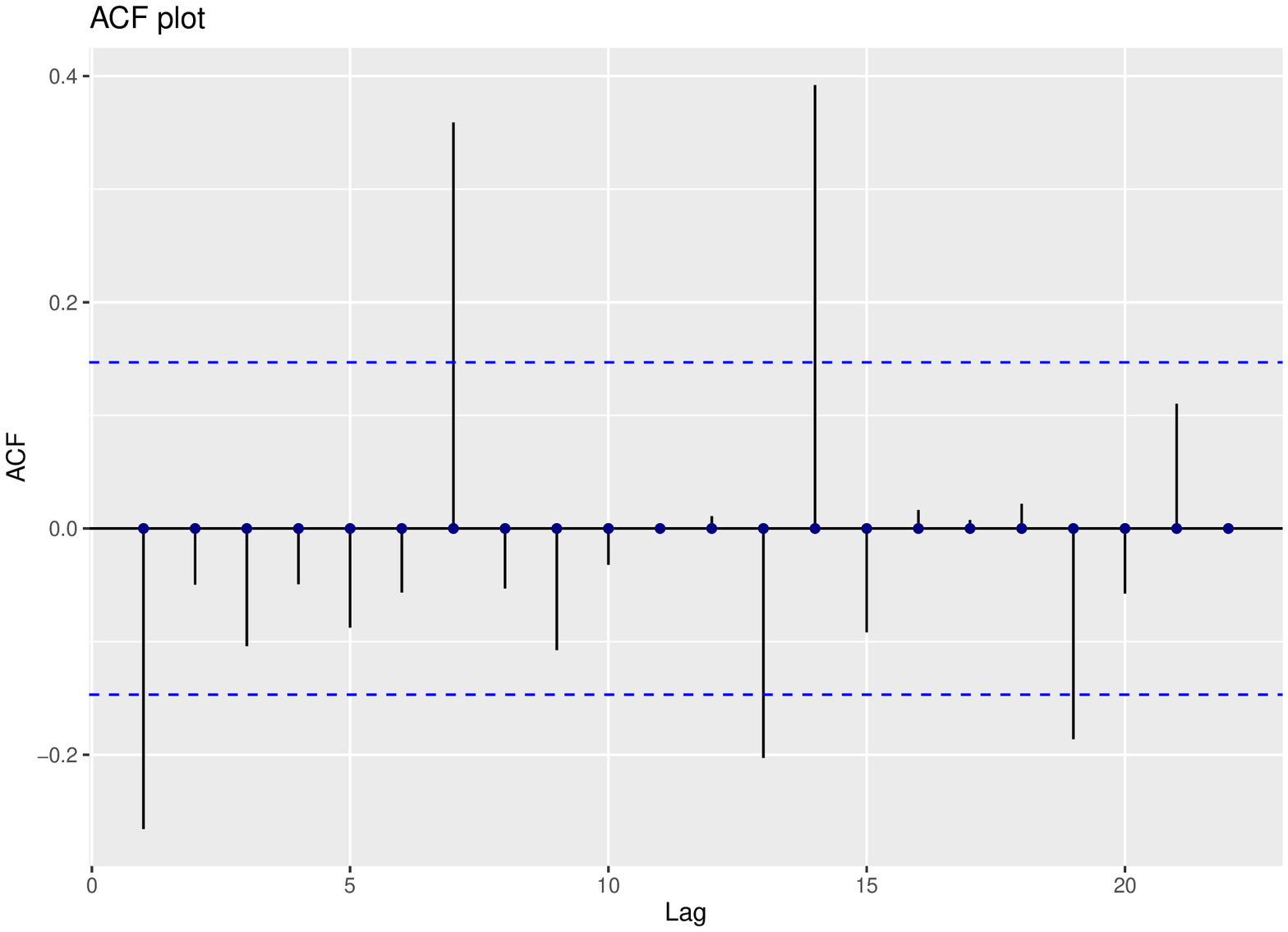}
         \end{minipage}
          &
          \begin{minipage}{.3\textwidth}
          \includegraphics[width=40mm, height=30mm]{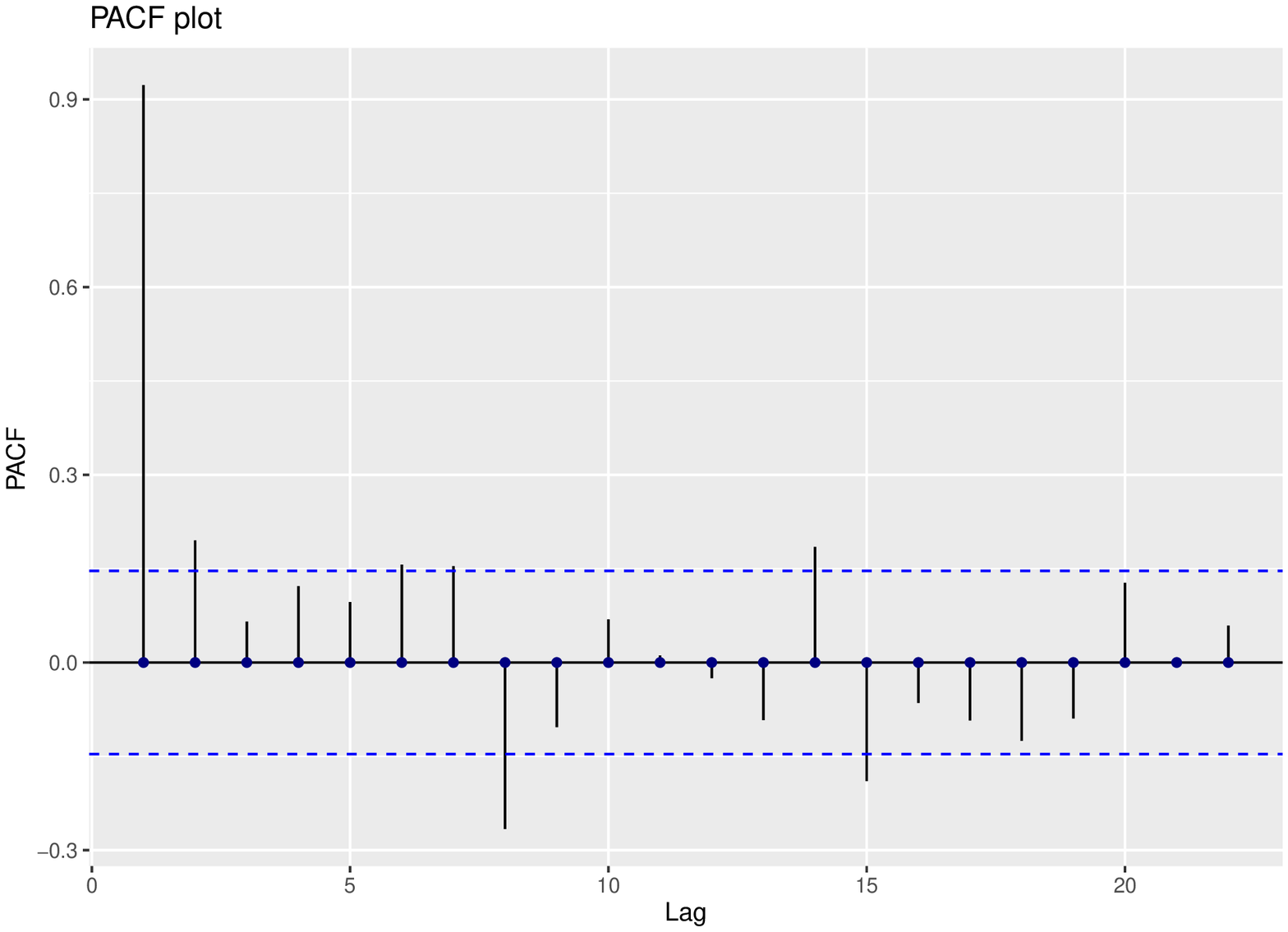}
          \end{minipage}
        \\ \hline
    \end{tabular}
\end{table}

Further, we checked twenty different forecasting models as competitors for the short-term forecasting of COVID-19 confirmed cases in five countries. 15-days and 30-days ahead forecasts were generated for each model, and accuracy metrics were computed to determine the best predictive models. From the ten popular single models, we choose the best one based on the accuracy metrics. On the other hand, one hybrid/ensemble model is selected from the rest of the ten models. The best-fitted ARIMA parameters, ETS, ARNN, and ARFIMA models for each country are reported in the respective tables. Table \ref{table_30_days} presents the training data (black colored) and test data (red-colored) and corresponding plots for the five datasets. Twenty forecasting models are implemented on these pandemic time-series datasets. Table \ref{r} gives the essential details about the functions and packages required for implementation. 

\begin{table}[H]
    \centering
    \caption{Pandemic datasets and corresponding ACF, PACF plots with 30-days test data}\label{table_30_days} \vspace{1cm}
    \begin{tabular}{ | c | p{4cm} | p{4cm} | p{4cm} |}
        \hline
        Country & Data & ACF plot & PACF plot  \\ \hline
        USA
        &
        \begin{minipage}{.30\textwidth}
            \includegraphics[width=40mm, height=30mm]{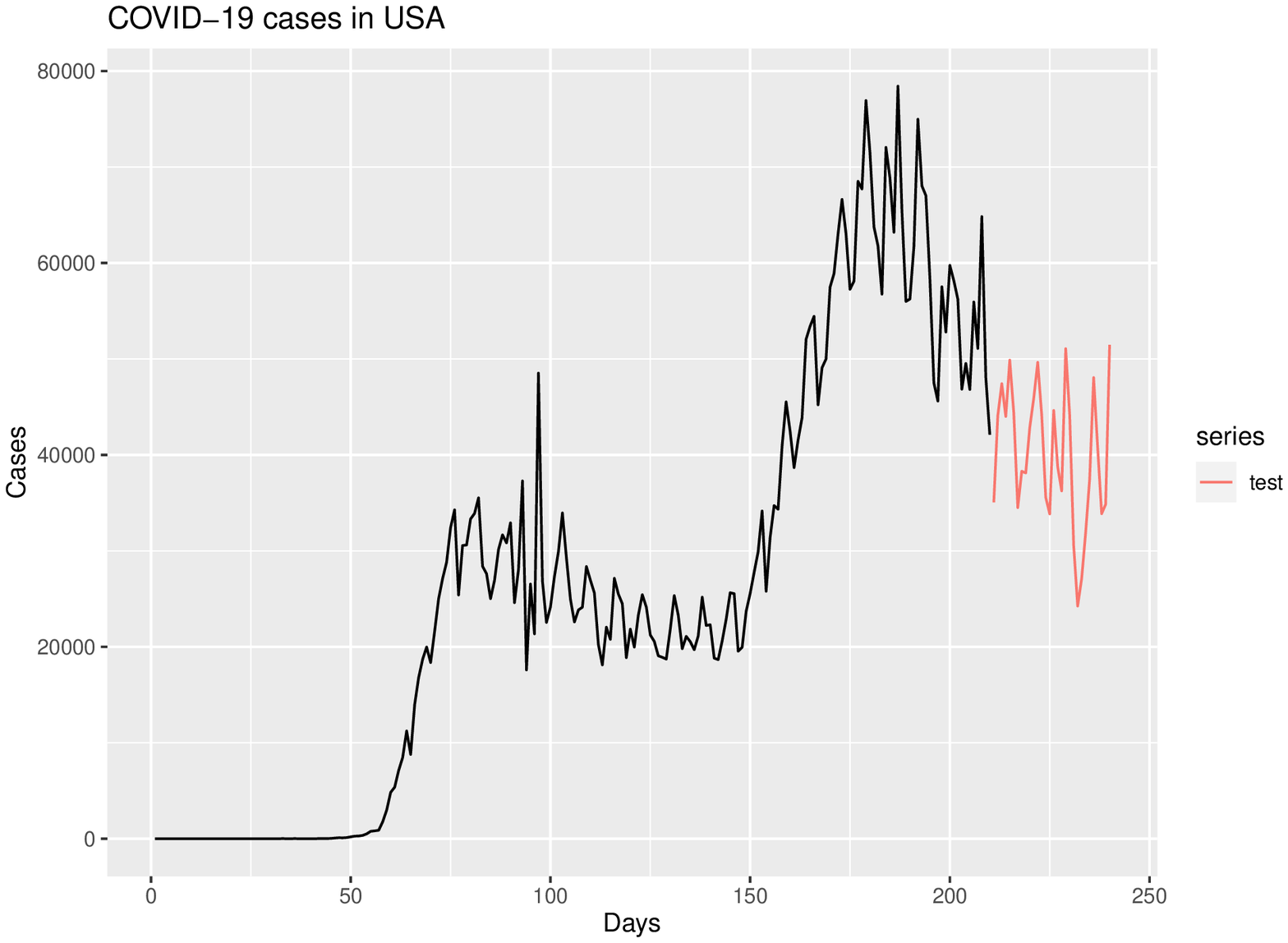}
        \end{minipage}
        &
        \begin{minipage}{.30\textwidth}
            \includegraphics[width=40mm, height=30mm]{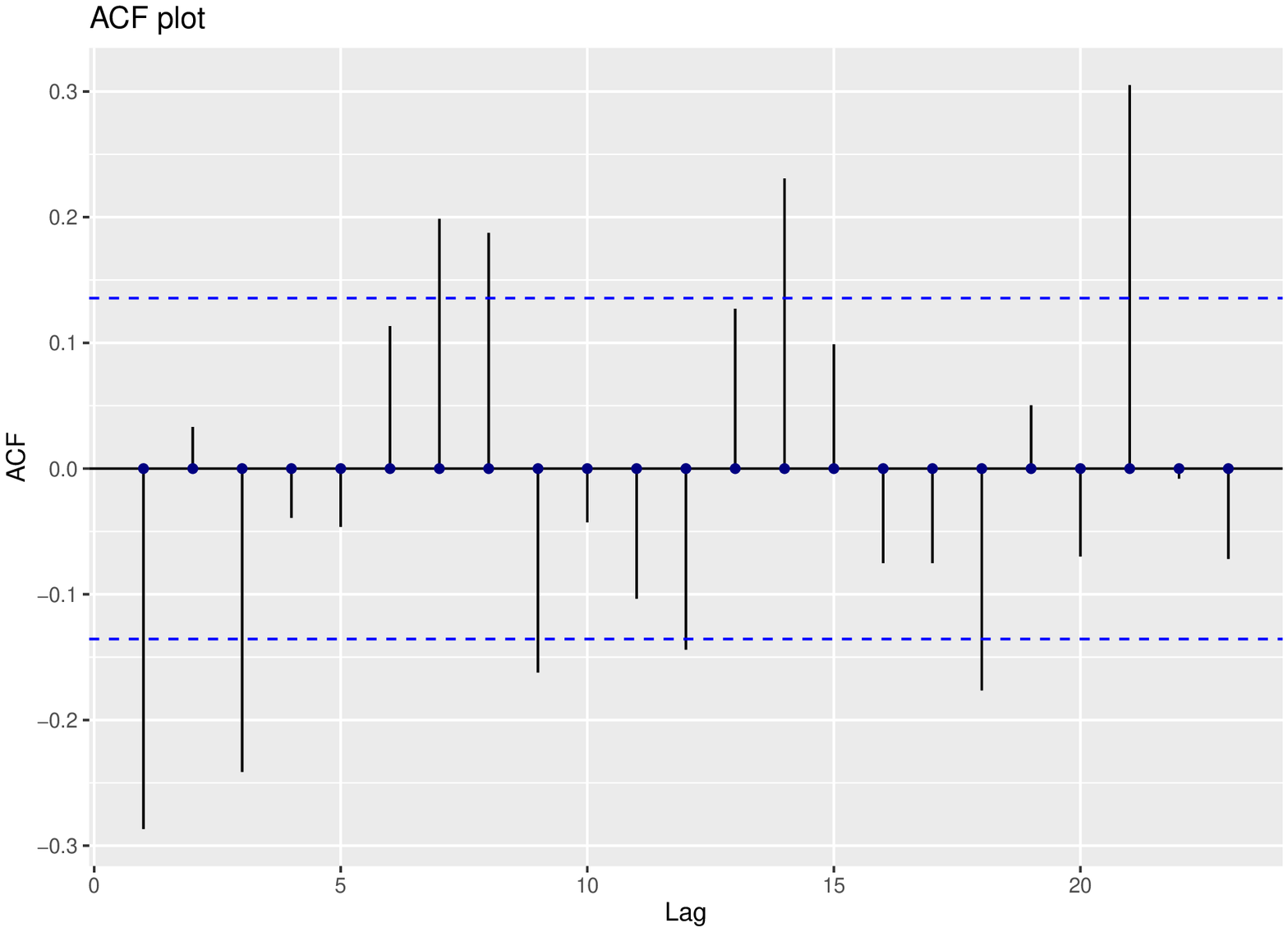}
        \end{minipage}
        &
        \begin{minipage}{.30\textwidth}
            \includegraphics[width=40mm, height=30mm]{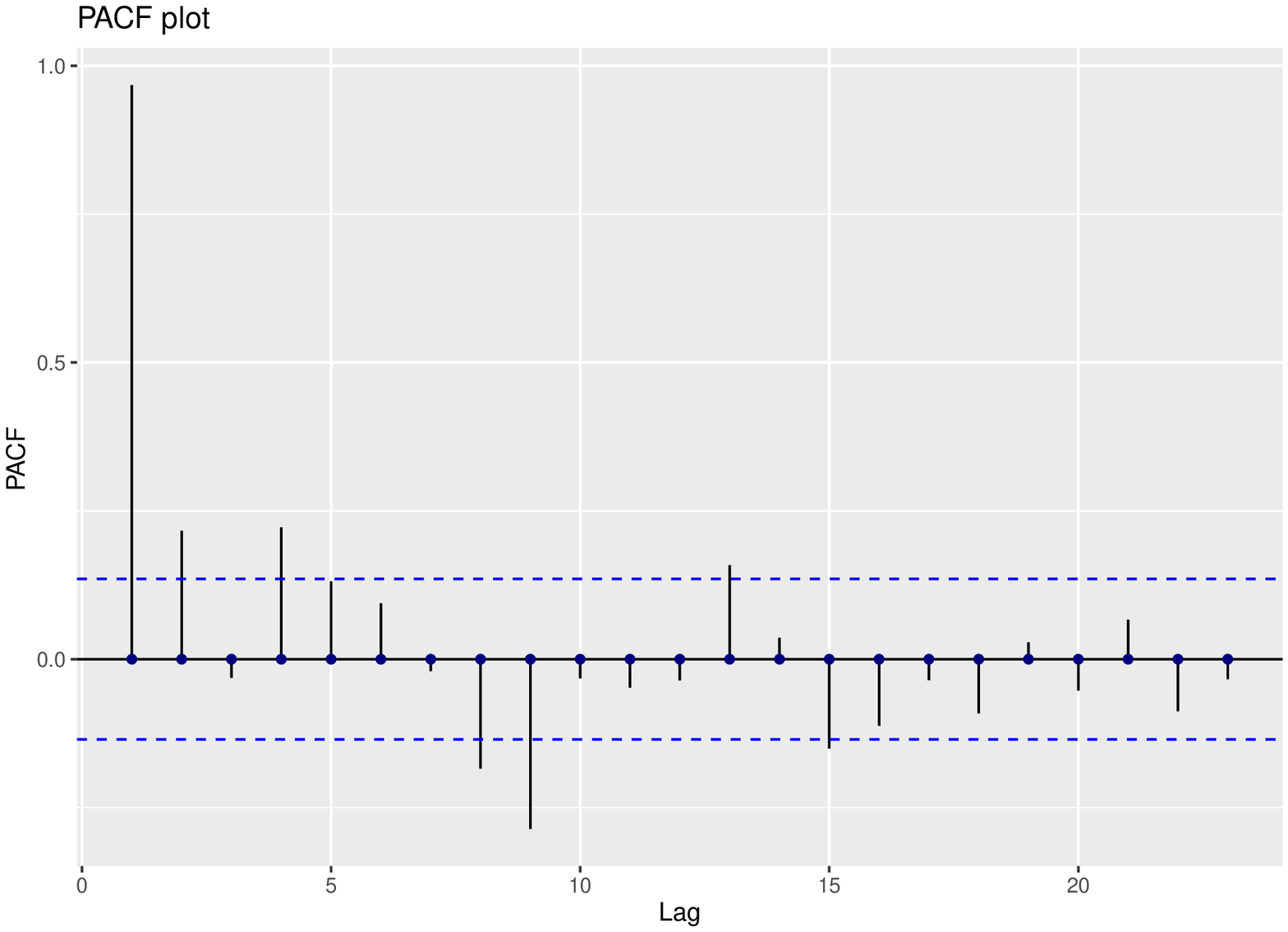}
        \end{minipage} \\ \hline
        India
        &
        \begin{minipage}{.3\textwidth}
            \includegraphics[width=40mm, height=30mm]{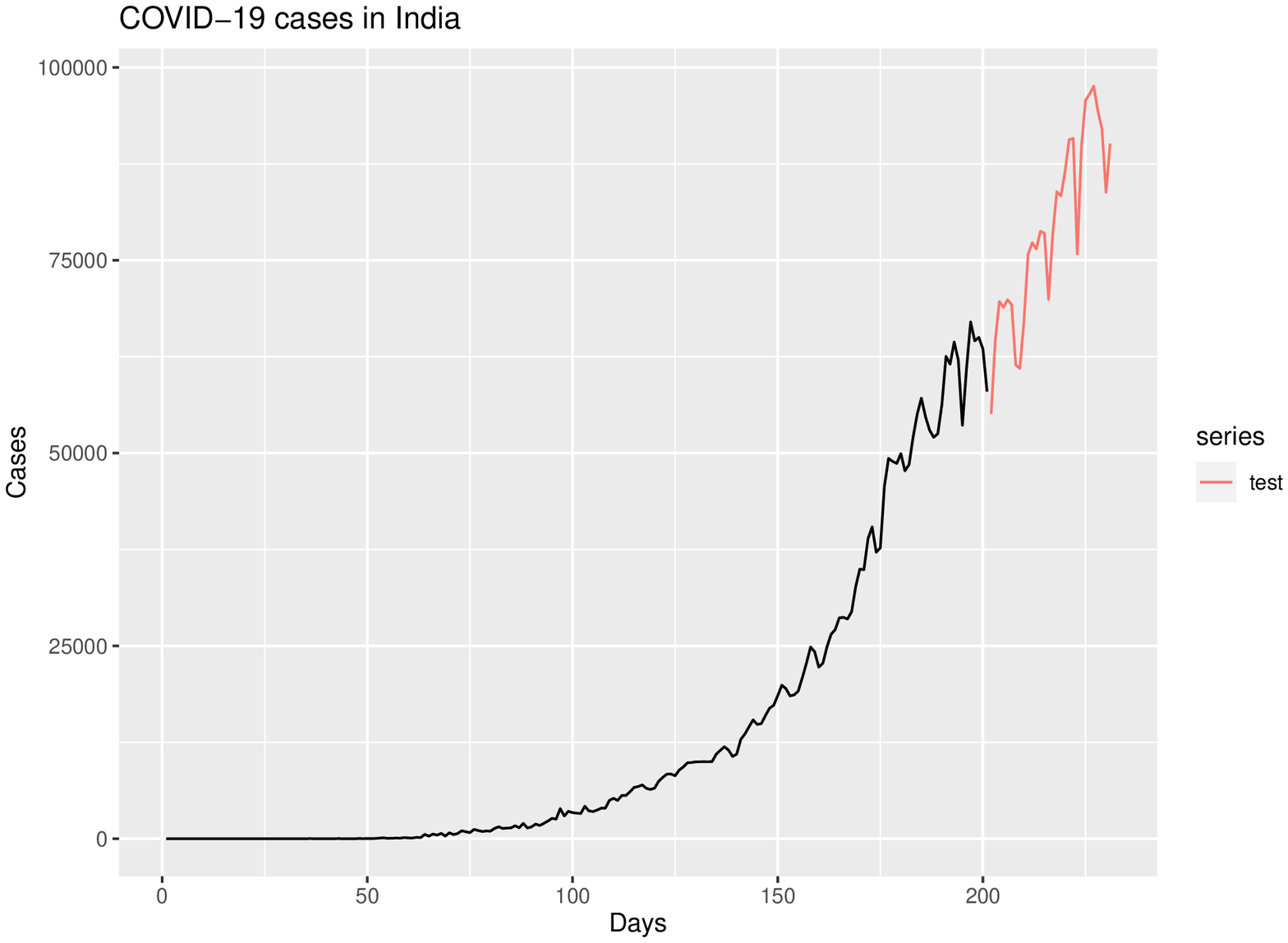}
        \end{minipage}
        &
        \begin{minipage}{.3\textwidth}
            \includegraphics[width=40mm, height=30mm]{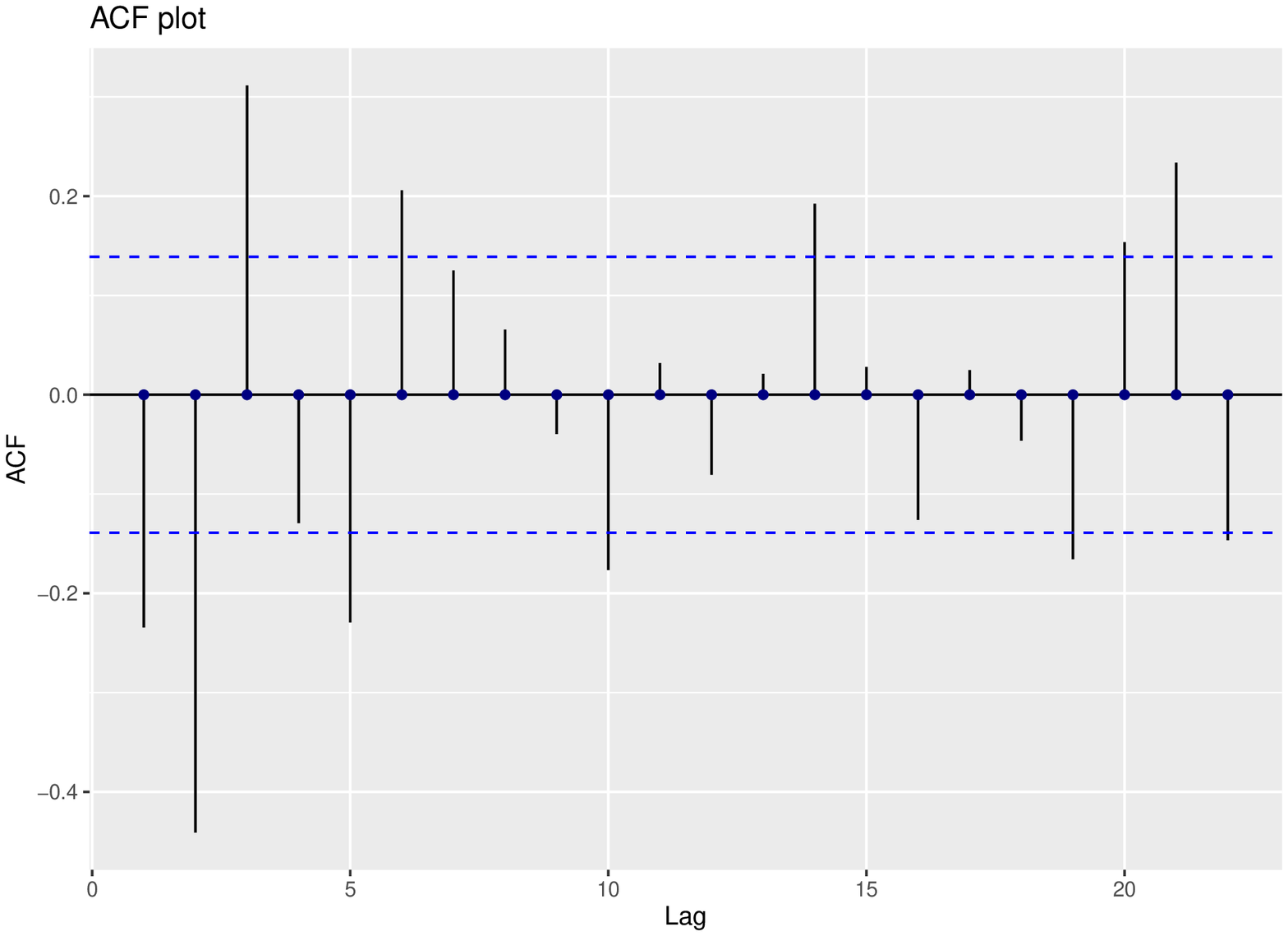}
        \end{minipage}
        &
        \begin{minipage}{.3\textwidth}
            \includegraphics[width=40mm, height=30mm]{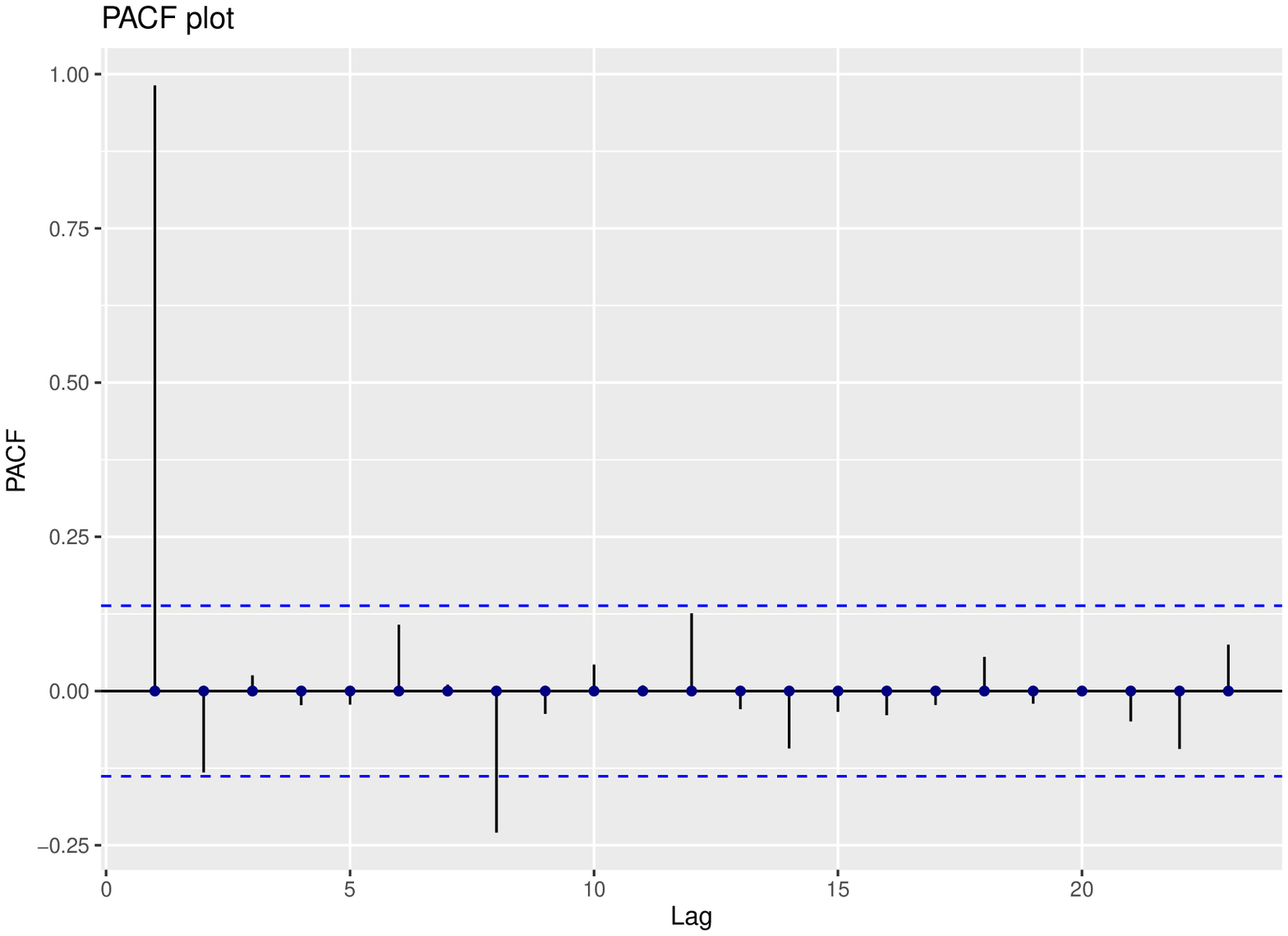}
        \end{minipage}
        \\ \hline
        Brazil
        &
        \begin{minipage}{.3\textwidth}
            \includegraphics[width=40mm, height=30mm]{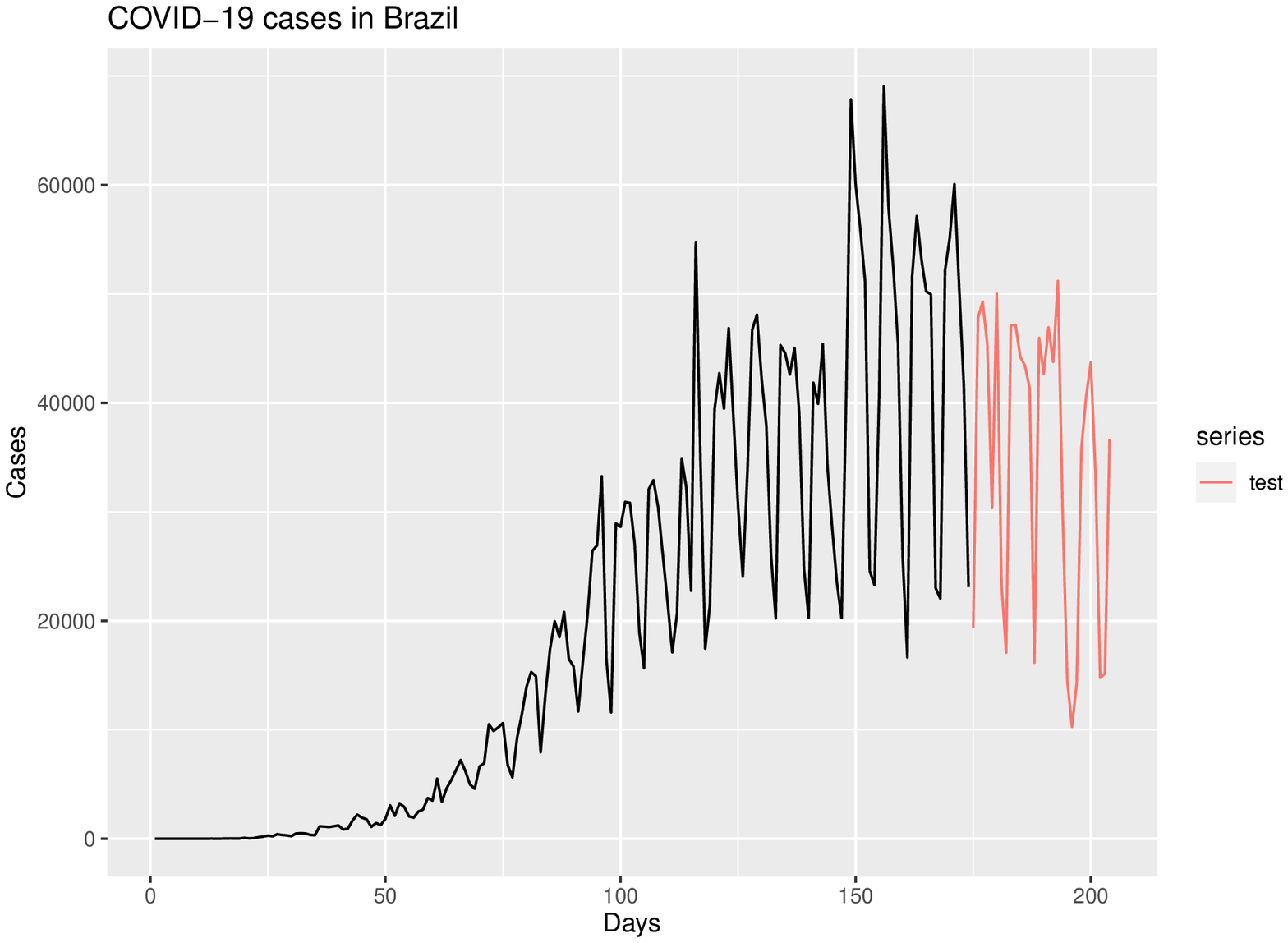}
        \end{minipage}
        &
        \begin{minipage}{.3\textwidth}
            \includegraphics[width=40mm, height=30mm]{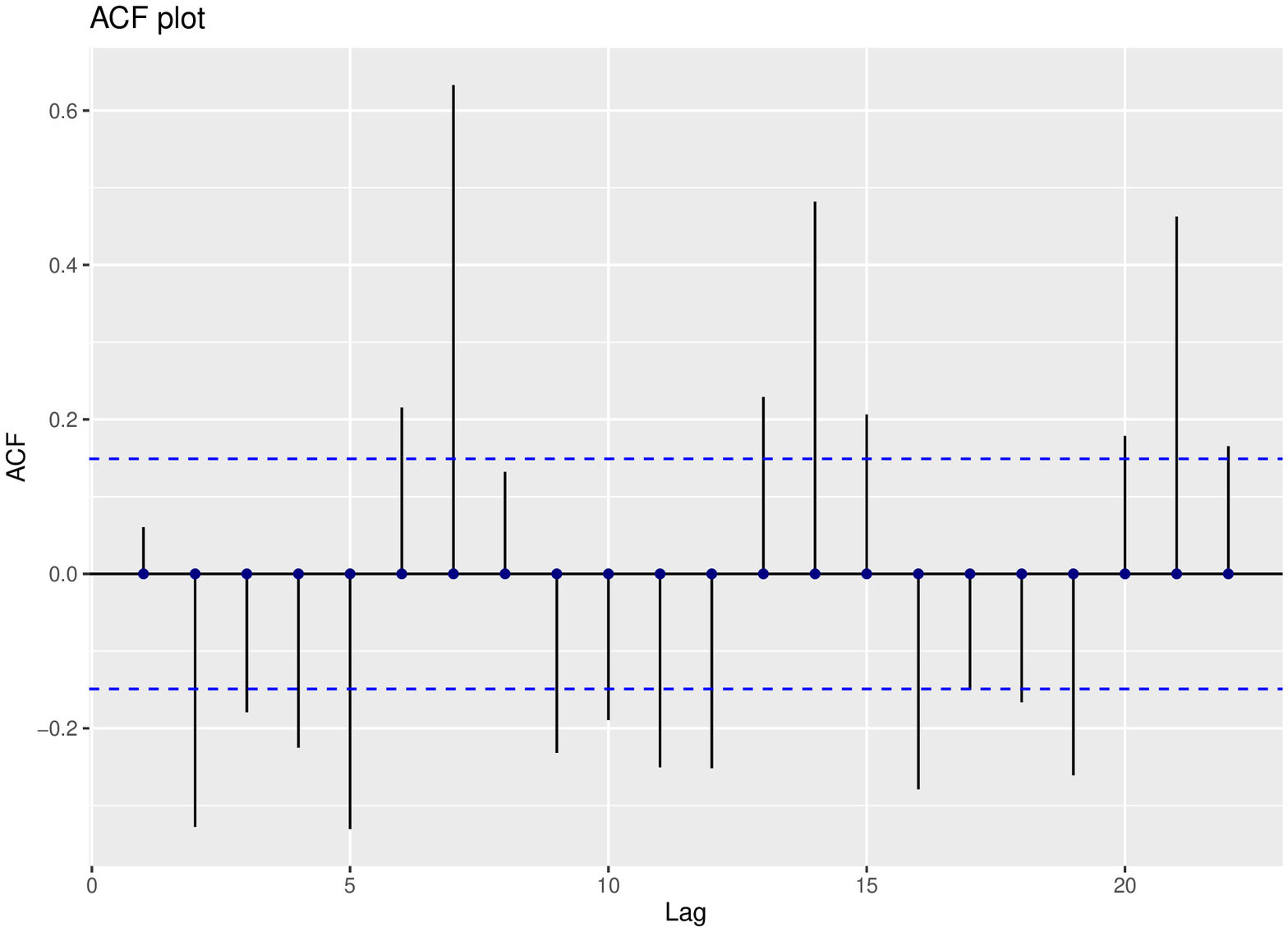}
        \end{minipage}
        &
        \begin{minipage}{.3\textwidth}
            \includegraphics[width=40mm, height=30mm]{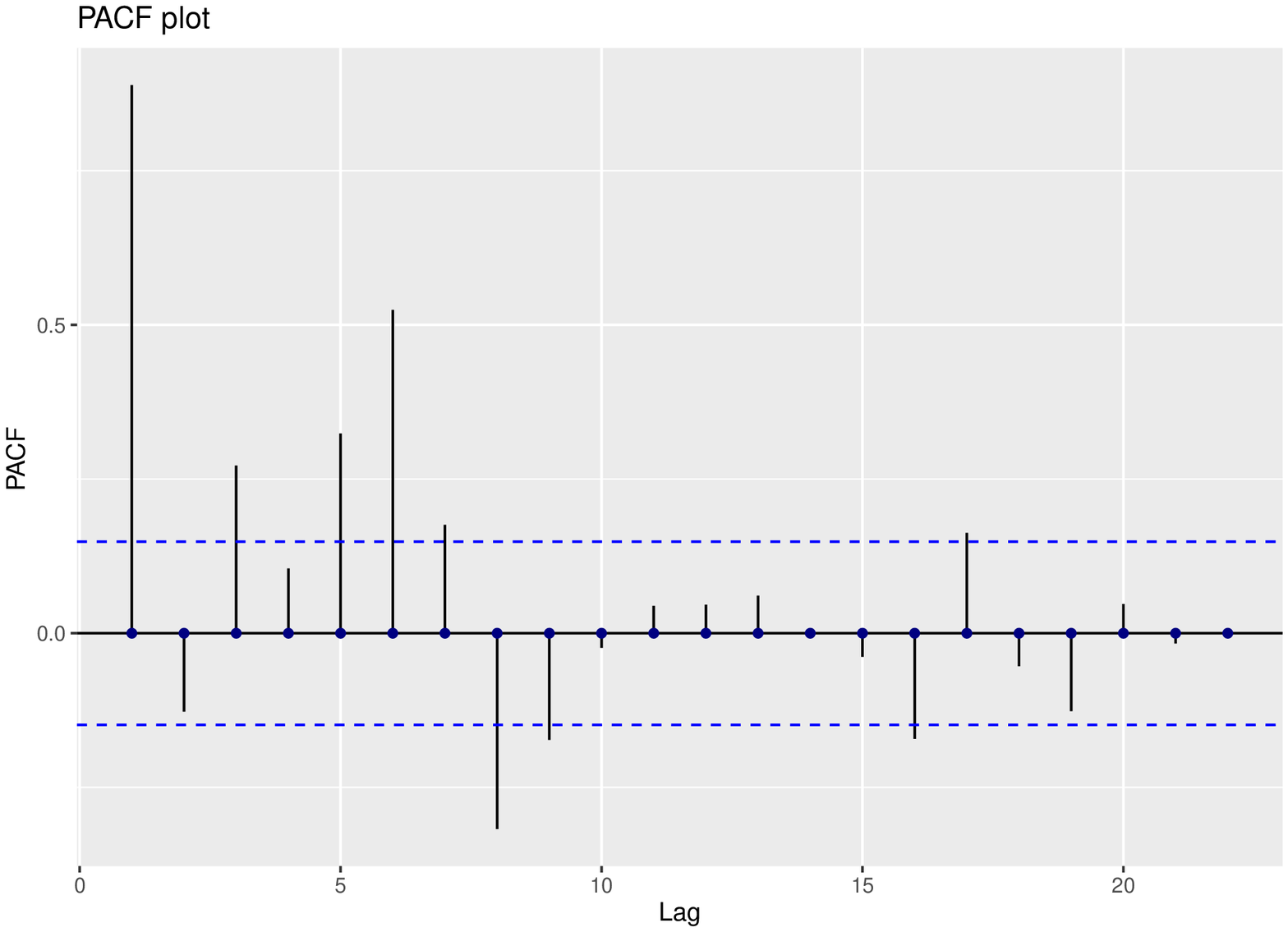}
        \end{minipage}
        \\ \hline
        Russia
        &
        \begin{minipage}{.3\textwidth}
        \includegraphics[width=40mm, height=30mm]{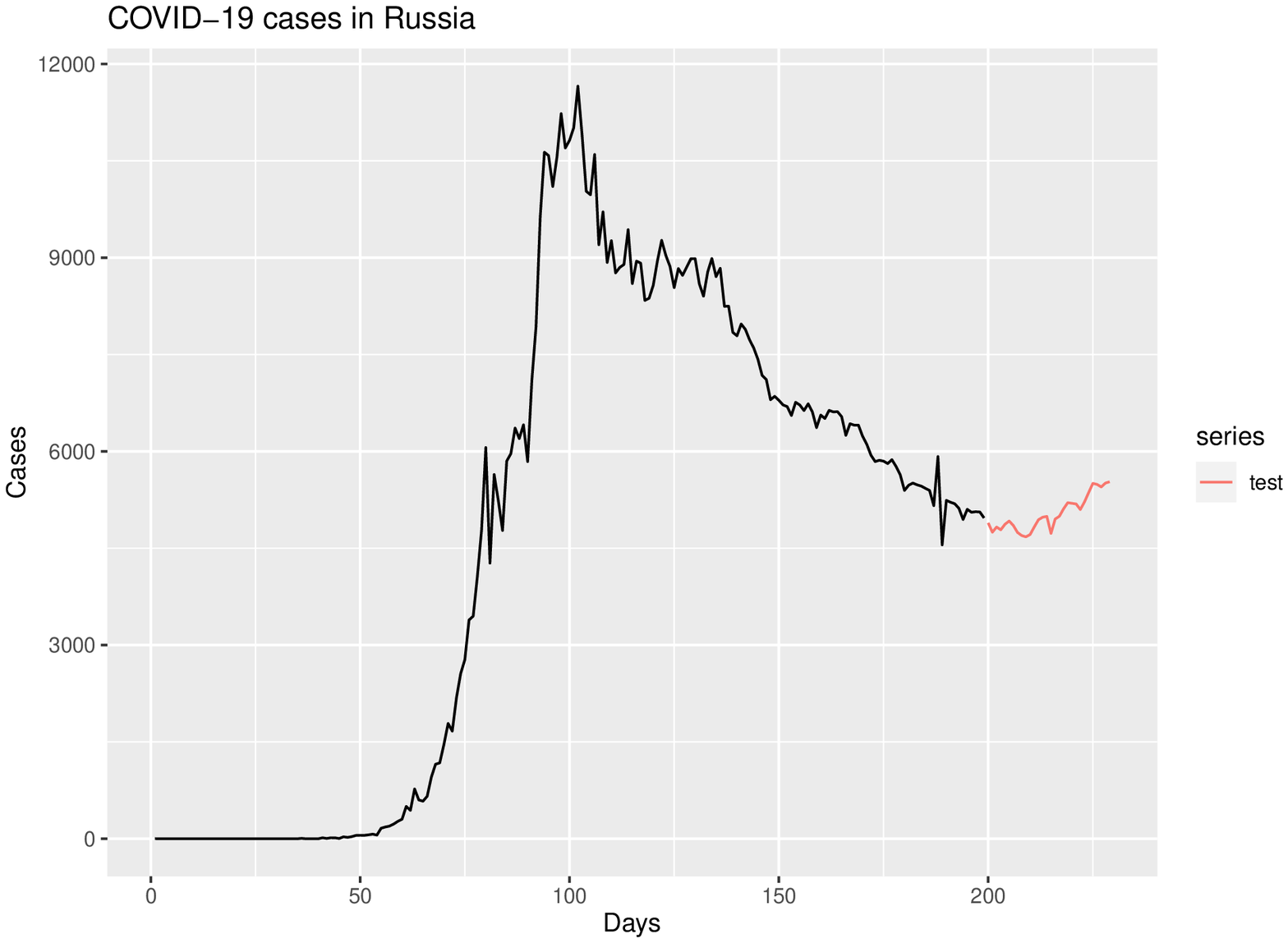}
        \end{minipage}
         &
        \begin{minipage}{.3\textwidth}
        \includegraphics[width=40mm, height=30mm]{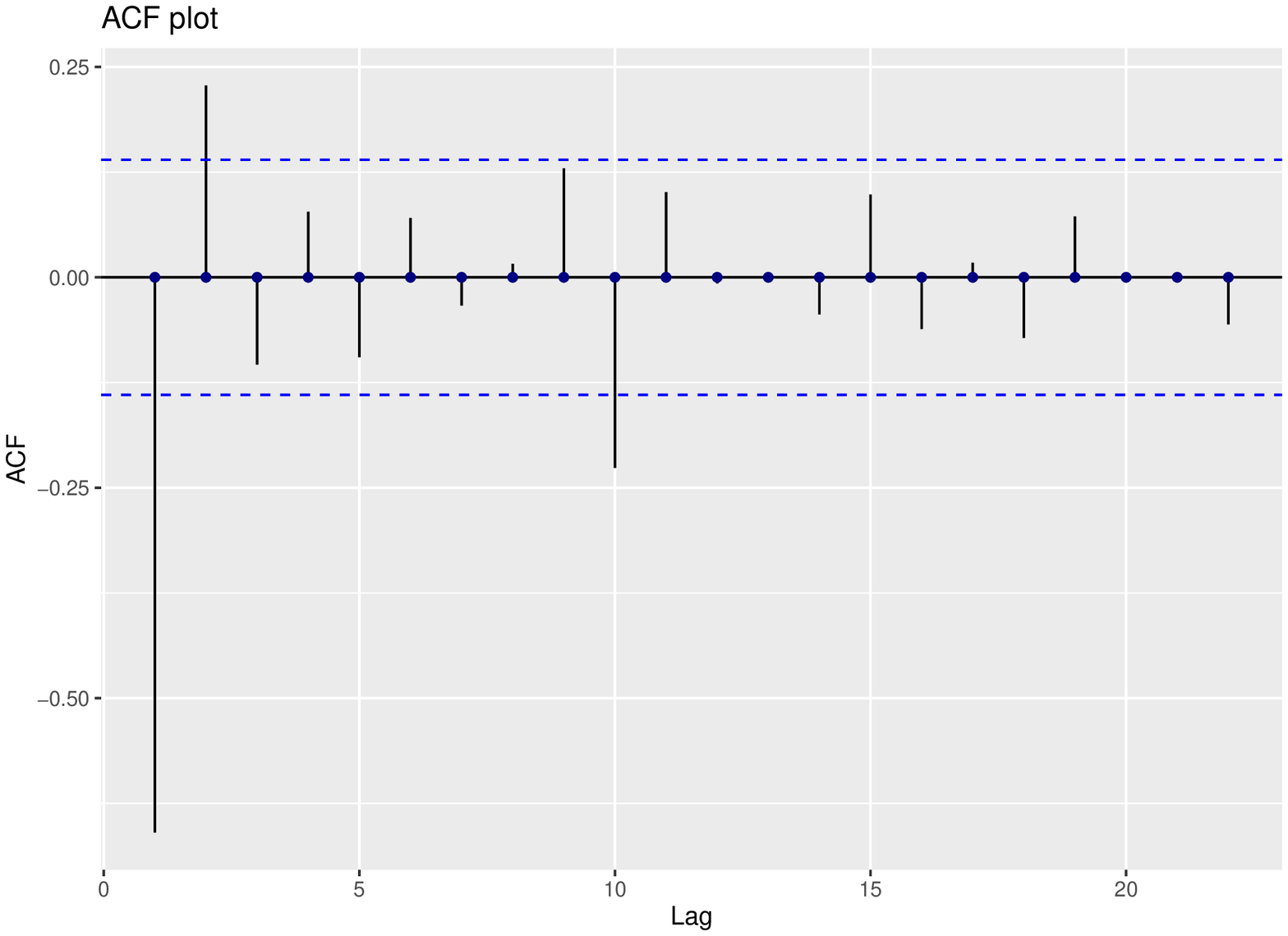}
         \end{minipage}
         &
         \begin{minipage}{.3\textwidth}
        \includegraphics[width=40mm, height=30mm]{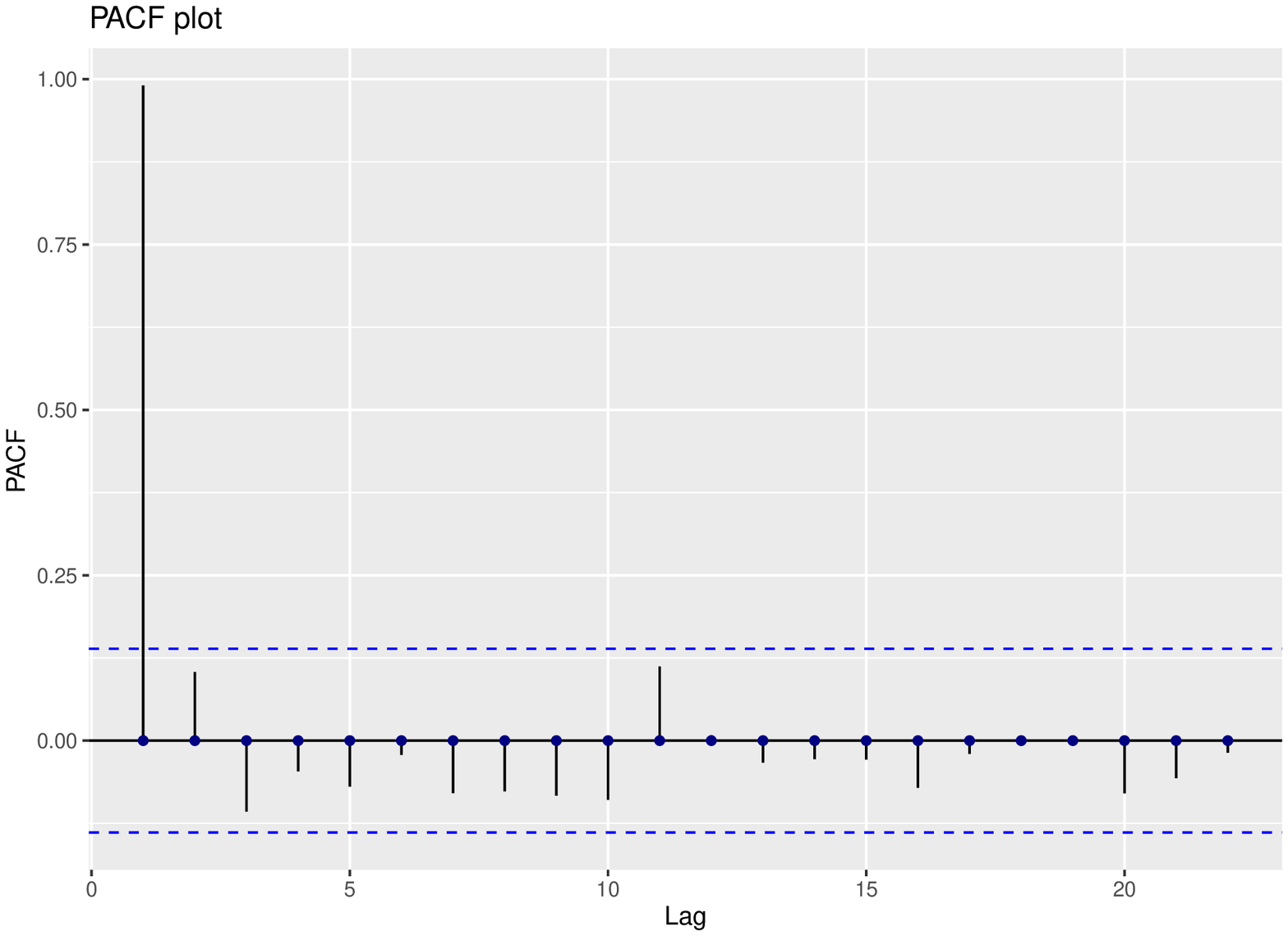}
         \end{minipage}
         \\ \hline
         Peru
         &
         \begin{minipage}{.3\textwidth}
         \includegraphics[width=40mm, height=30mm]{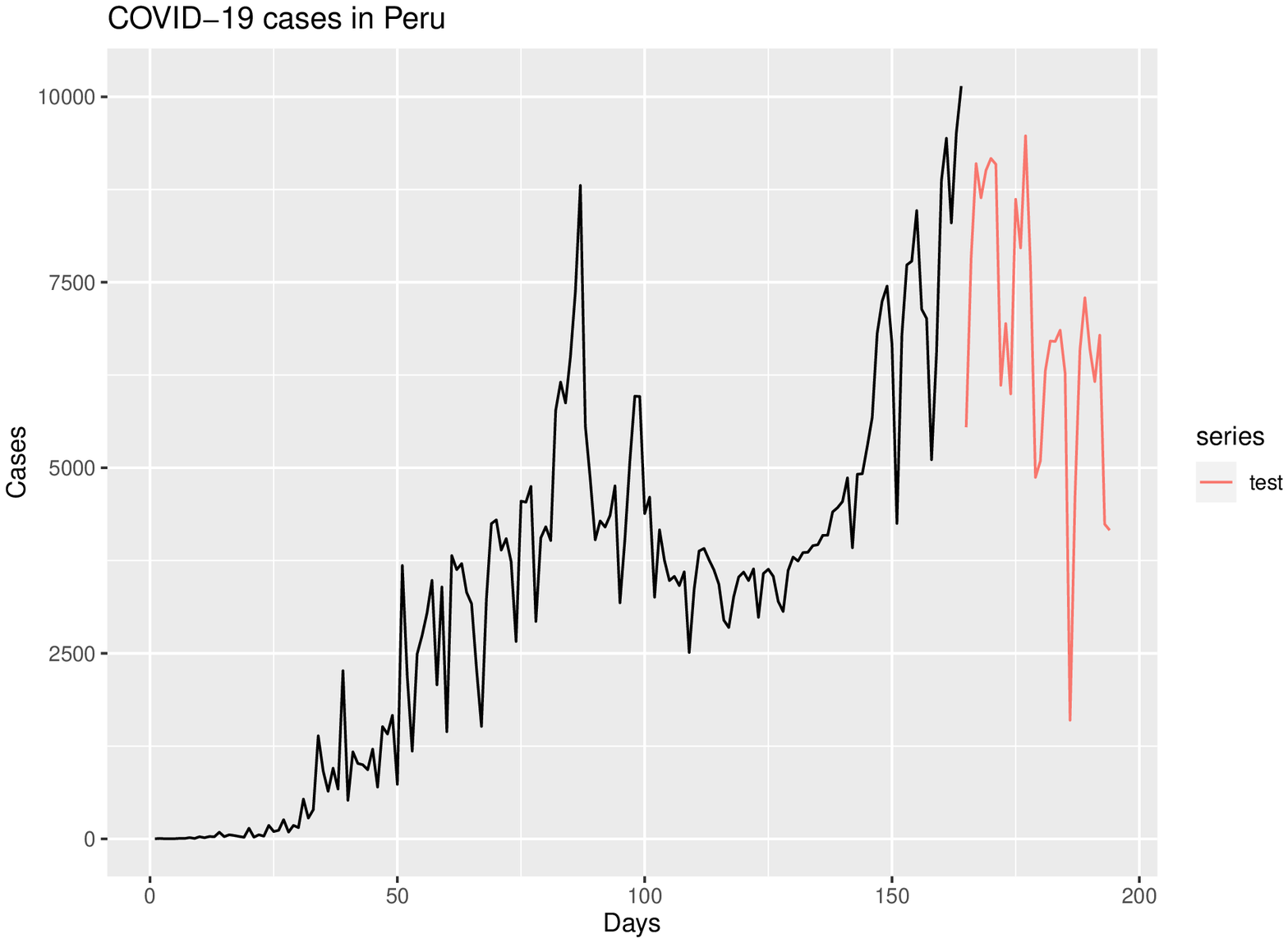}
         \end{minipage}
          &
         \begin{minipage}{.3\textwidth}
         \includegraphics[width=40mm, height=30mm]{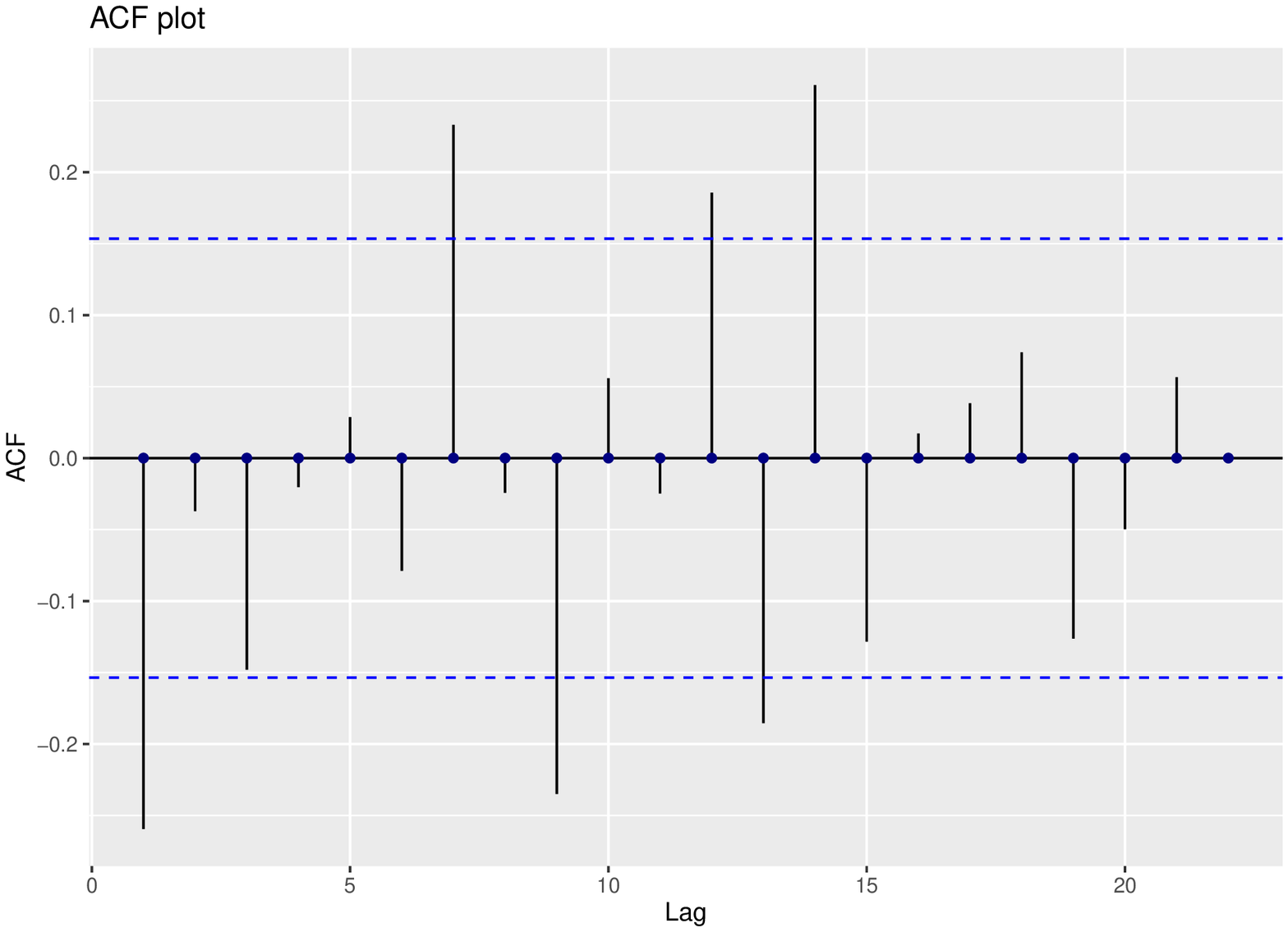}
         \end{minipage}
          &
          \begin{minipage}{.3\textwidth}
          \includegraphics[width=40mm, height=30mm]{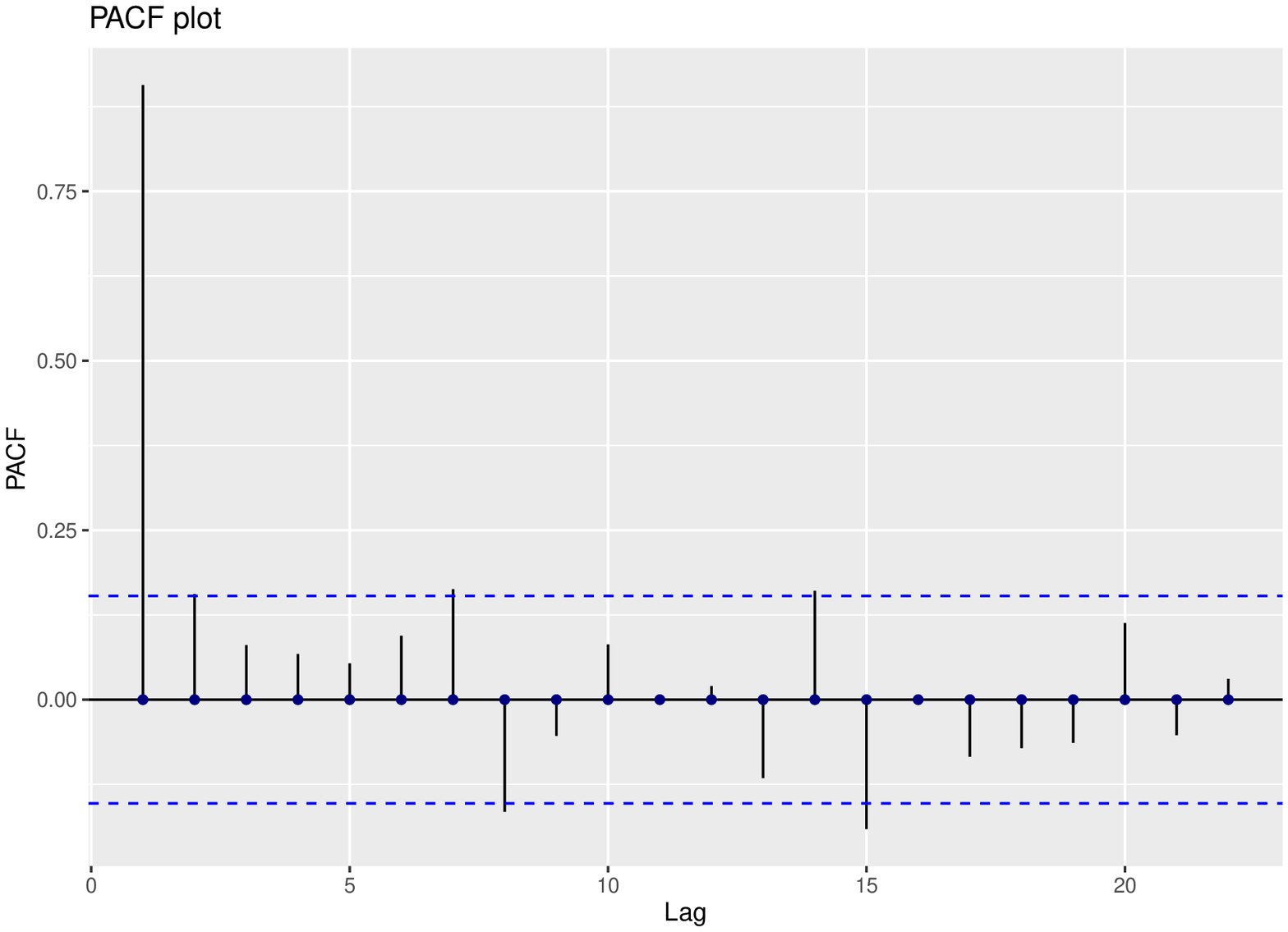}
          \end{minipage}
        \\ \hline
    \end{tabular} 
\end{table}

\begin{figure}[H]
	\includegraphics[width=0.52\textwidth]{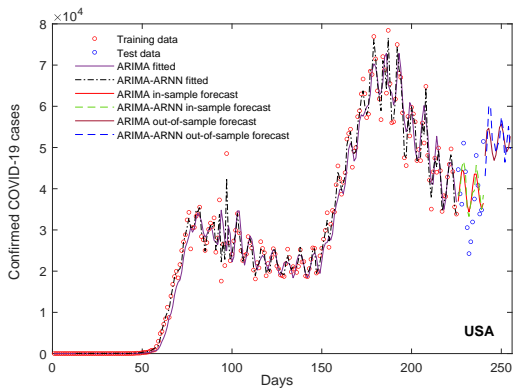}(a)
	\includegraphics[width=0.52\textwidth]{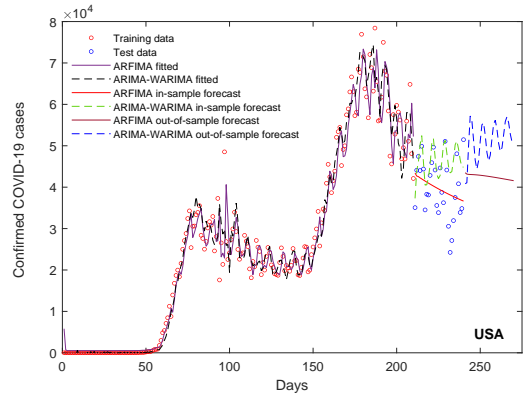}(b)
	\caption{Plots of (a) 15-days ahead forecast results for USA COVID-19 data obtained using ARIMA and hybrid ARIMA-ARNN models. (b) 30-days ahead forecast results from ARFIMA and hybrid ARIMA-WARIMA models.}
	\label{Fig:USA_forecasts}
\end{figure}

\paragraph{\textbf{Results for USA COVID-19 data:}}
Among the single models, ARIMA(2,1,4) performs best in terms of accuracy metrics for 15-days ahead forecasts. TBATS and ARNN(16,8) also have competitive accuracy metrics. Hybrid ARIMA-ARNN model improves the earlier ARIMA forecasts and has the best accuracy among all hybrid/ensemble models (see Table \ref{usa_accuracy_table_15_days}). Hybrid ARIMA-WARIMA also does a good job and improves ARIMA model forecasts. In-sample and out-of-sample forecasts obtained from ARIMA and hybrid ARIMA-ARNN models are depicted in Fig. \ref{Fig:USA_forecasts}(a). Out-of-sample forecasts are generated using the whole dataset as training data.

\begin{table}[H]
	\centering \caption{Performance metrics with 15 days-ahead test set for USA}\label{usa_accuracy_table_15_days}
	\begin{tabular}{|c|c|c|c|c|}
		\hline
		\multirow{2}{*}{Model} & \multicolumn{4}{c|}{15-days ahead forecast} 
		\\ \cline{2-5} & RMSE & MAE & MAPE & SMAPE \\ \hline
		ARIMA(2,1,4) & \textbf{7187.02} & \textbf{6094.95} & \textbf{16.89} & \textbf{16.07} \\ \hline
		ETS(A,N,N) & 8318.73 & 6759.65 & 17.82 & 17.86  \\ \hline
		SETAR & 8203.21 & 6725.96 & 18.19 & 17.77 \\ \hline
		TBATS & 7351.04 & 6367.46 & 17.86 & 16.73 \\ \hline
		Theta & 8112.22 & 6791.52 & 18.51 & 17.95 \\ \hline
		ANN & 9677.105 & 8386.223 & 25.15 & 21.69 \\ \hline
		ARNN(16,8) & 7633.92 & 6647.18 & 19.75 & 17.42 \\ \hline
		WARIMA & 9631.98 & 8182.84 & 21.09 & 22.21 \\ \hline
		BSTS & 10666.15 & 8527.72 & 20.91 & 23.26 \\ \hline
		ARFIMA(1,0.14,1) & 8413.33 & 6696.09 & 17.48 & 17.68 \\ \hline
		Hybrid ARIMA-ANN & 7113.72 & 6058.29 & 16.90 & 15.99 \\ \hline
		Hybrid ARIMA-ARNN & \textbf{5978.04} & \textbf{4650.89} & \textbf{13.22} & \textbf{12.45} \\ \hline 
		Hybrid ARIMA-WARIMA & 6582.93 & 5217.023 & 14.33 & 13.80 \\ \hline
		Hybrid WARIMA-ANN & 10633.97 & 8729.11 & 21.85 & 24.22 \\ \hline
		Hybrid WARIMA-ARNN & 9558.34 & 8138.71 & 21.05 & 22.05 \\ \hline
		Ensemble ARIMA-ETS-Theta & 7602.06 & 6388.96 & 17.32 & 16.89 \\ \hline
		Ensemble ARIMA-ETS-ARNN & 7012.95 & 6184.23 & 18.09 & 16.45 \\ \hline
		Ensemble ARIMA-Theta-ARNN & 6933.88 & 6054.97 & 17.42 & 16.07 \\ \hline
		Ensemble ETS-Theta-ARNN &  7044.20 & 5950.40 & 16.97 & 15.82 \\ \hline
		Ensemble ANN-ARNN-WARIMA & 7437.21 & 6465.18 & 18.66 & 17.11 \\ \hline
	\end{tabular}
\end{table}

ARFIMA(2,0,0) is found to have the best accuracy metrics for 30-days ahead forecasts among single forecasting models. BSTS and SETAR also have good agreement with the test data in terms of accuracy metrics. Hybrid ARIMA-WARIMA model and has the best accuracy among all hybrid/ensemble models (see Table \ref{usa_accuracy_table_30_days}). In-sample and out-of-sample forecasts obtained from ARFIMA and hybrid ARIMA-WARIMA models are depicted in Fig. \ref{Fig:USA_forecasts}(b). 

\begin{table}[H]
	\centering \caption{Performance metrics with 30 days-ahead test set for USA}\label{usa_accuracy_table_30_days}
	\begin{tabular}{|c|c|c|c|c|}
		\hline
		\multirow{2}{*}{Model} & \multicolumn{4}{c|}{30-days ahead forecast} 
		\\ \cline{2-5} & RMSE & MAE & MAPE & SMAPE \\ \hline
		ARIMA(2,1,4) with drift & 12370.18 & 10499.44 & 29.87 & 24.26 \\ \hline
		ETS(A,Ad,N) & 11929.897 & 9951.090 & 28.95 & 23.49  \\ \hline
		SETAR & 8593.527 & 6904.605 & 20.18 & 17.25 \\ \hline
		TBATS & 10314.23 & 8587.83 & 24.95 & 20.73 \\ \hline
		Theta & 12234.16 & 9858.115 & 29.03 & 23.24 \\ \hline
		ANN & 15241.65 & 12973.2 & 37.11 & 28.86 \\ \hline
		ARNN(16,8) & 19000.09 & 17311.86 & 46.95 & 36.01 \\ \hline
		WARIMA & 12455.31 & 9501.018 & 22.55 & 27.45 \\ \hline
		BSTS & 8459.763 & 6444.994 & 15.94 & 16.87 \\ \hline
		ARFIMA(2,0,0) & \textbf{6847.32} & \textbf{5651.33} & \textbf{14.83} & \textbf{14.40} \\ \hline
		Hybrid ARIMA-ANN & 12269.99 & 10339.18 & 29.46 & 23.92 \\ \hline
		Hybrid ARIMA-ARNN & 12584.03 & 10566.16 & 30.14 & 24.32 \\ \hline
		Hybrid ARIMA-WARIMA & \textbf{8514.36} & \textbf{6702.07} & \textbf{19.52} & \textbf{16.59} \\ \hline
		Hybrid WARIMA-ANN & 14983.09 & 11918.16 & 28.55 & 36.52 \\ \hline
		Hybrid WARIMA-ARNN &  12294.48 & 9330.15 & 22.14 & 26.88 \\ \hline
		Ensemble ARIMA-ETS-Theta & 12014.39 & 9978.22 & 29.04 & 23.49 \\ \hline
		Ensemble ARIMA-ETS-ARNN & 11484.49 & 10035.78 & 28.35 & 23.49 \\ \hline
		Ensemble ARIMA-Theta-ARNN & 13596.9 & 12000.69 & 33.86 & 27.21 \\ \hline
		Ensemble ETS-Theta-ARNN & 13074.13 & 11429.5 & 32.52 & 26.26 \\ \hline
		Ensemble ANN-ARNN-WARIMA & 11652.2 & 9947.16 & 30.60 & 24.23 \\ \hline
	\end{tabular}
\end{table}

\begin{figure}[H]
	\includegraphics[width=0.52\textwidth]{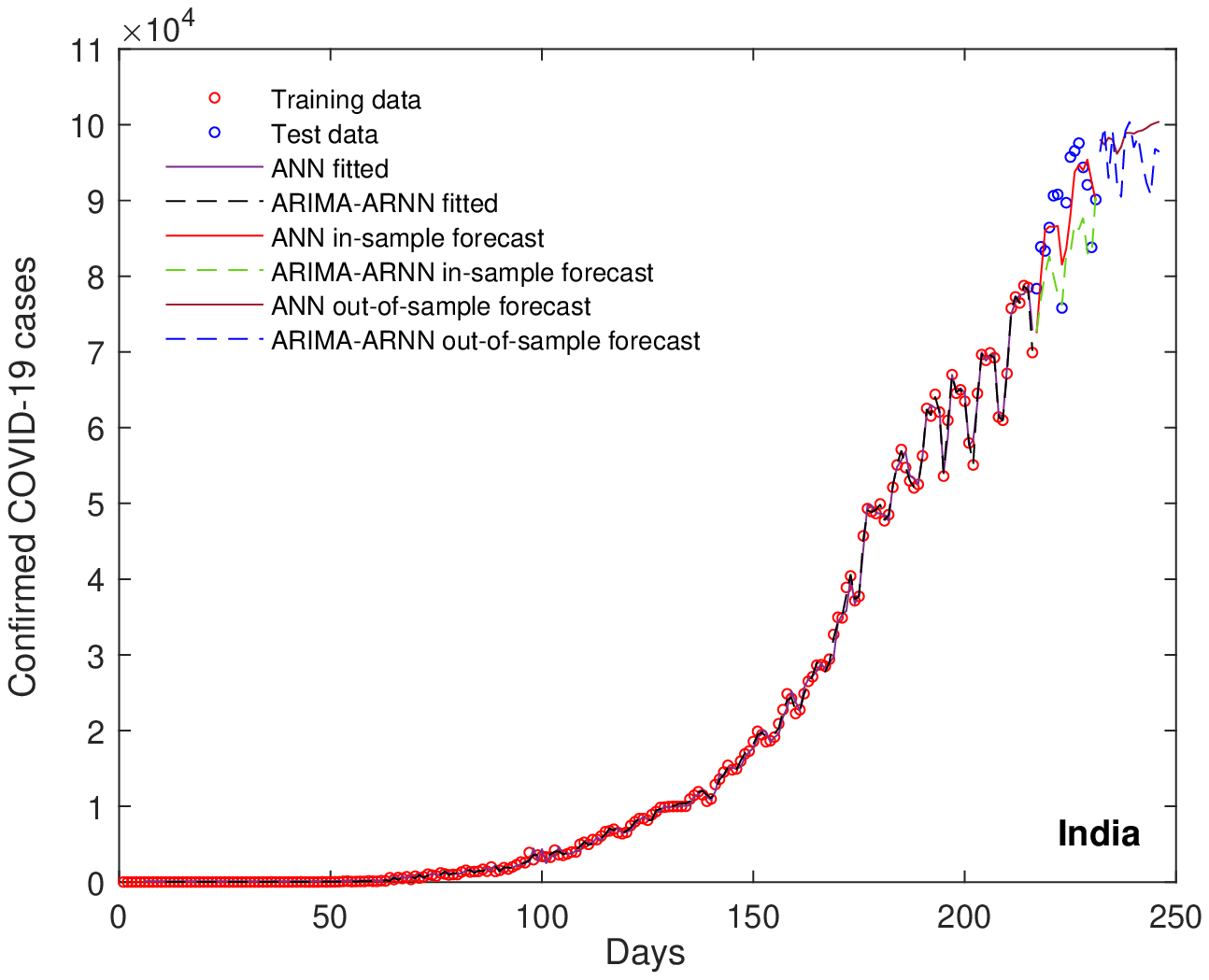}(a)
	\includegraphics[width=0.52\textwidth]{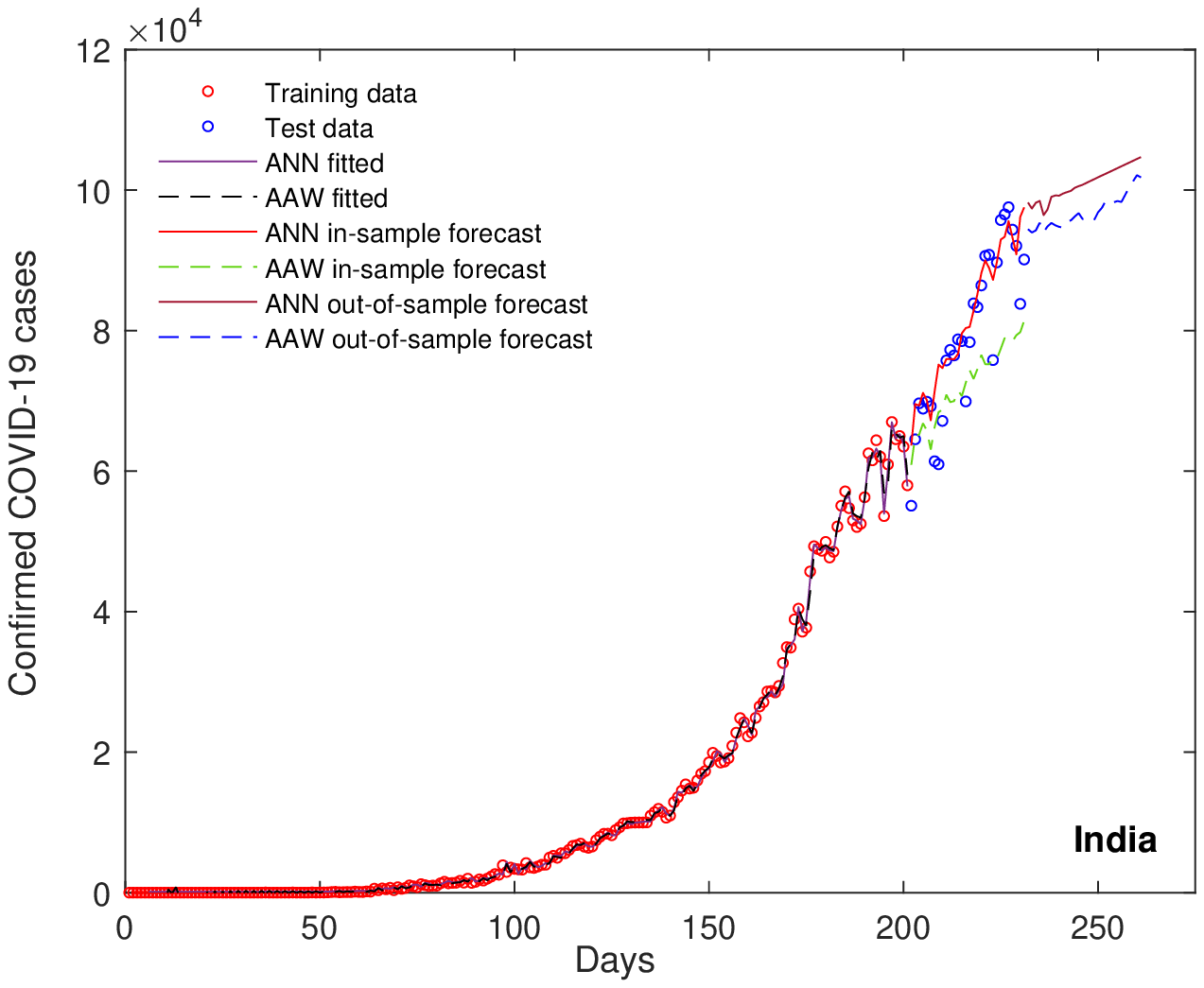}(b)
	\caption{Plots of (a) 15-days ahead forecast results for India COVID-19 data obtained using ANN and hybrid ARIMA-ARNN models and (b) 30-days ahead forecast results from ANN and ANN-ARNN-WARIMA (AAW) models.}
	\label{Fig:India_forecasts}
\end{figure}

\paragraph{\textbf{Results for India COVID-19 data:}}
Among the single models, ANN performs best in terms of accuracy metrics for 15-days ahead forecasts. ARIMA(1,2,5) also has competitive accuracy metrics in the test period. Hybrid ARIMA-ARNN model improves the ARIMA(1,2,5) forecasts and has the best accuracy among all hybrid/ensemble models (see Table \ref{india_accuracy_table_15_days}). Hybrid ARIMA-ANN and hybrid ARIMA-WARIMA also do a good job and improves ARIMA model forecasts. In-sample and out-of-sample forecasts obtained from ANN and hybrid ARIMA-ARNN models are depicted in Fig. \ref{Fig:India_forecasts}(a). Out-of-sample forecasts are generated using the whole dataset as training data (see Fig. \ref{Fig:India_forecasts}). 

\begin{table}[H]
	\centering \caption{Performance metrics with 15 days-ahead test set for India}\label{india_accuracy_table_15_days}
	\begin{tabular}{|c|c|c|c|c|}
		\hline
		\multirow{2}{*}{Model} & \multicolumn{4}{c|}{15-days ahead forecast} 
		\\ \cline{2-5} & RMSE & MAE & MAPE & SMAPE \\ \hline
		ARIMA(1,2,5) & 8141.76 & 7479.43 & 8.36 & 8.72 \\ \hline
		ETS(A,A,N) & 15431 & 14415.73 & 15.92 & 17.18  \\ \hline
		SETAR & 22835.95 & 21851.45 & 24.24 & 27.84 \\ \hline
		TBATS & 11764.61 & 10837.68 & 12.00 & 12.89 \\ \hline
		Theta & 18405.29 & 17403.03 & 19.24 & 21.50 \\ \hline
		ANN & \textbf{6663.10} & \textbf{5891.94} & \textbf{6.54} & \textbf{6.81} \\ \hline
		ARNN(2,2) & 25617.9 & 24539.67 & 27.23 & 31.86 \\ \hline
		WARIMA & 12201.48 & 11103.41 & 12.25 & 13.18 \\ \hline
		BSTS & 13535.1 & 12402.34 & 13.65 & 14.84 \\ \hline
		ARFIMA(0,0.49,4) & 34848.86 & 33323.88 & 37.03 & 46.25  \\ \hline
		Hybrid ARIMA-ANN & 8080.862 & 7399.7 & 8.28 & 8.64 \\ \hline
		Hybrid ARIMA-ARNN & \textbf{7762.32} & \textbf{6560.26} & \textbf{7.20} & \textbf{7.67} \\ \hline
		Hybrid ARIMA-WARIMA & 8144.77 & 7455.34 & 8.32 & 8.68 \\ \hline
		Hybrid WARIMA-ANN & 11883.45 & 10697.21 & 11.79 & 12.65 \\ \hline
		Hybrid WARIMA-ARNN & 11623.15 & 10339.16 & 11.33 & 12.17 \\ \hline
		Ensemble ARIMA-ETS-Theta & 13734.28 & 12641.35 & 13.93 & 15.14 \\ \hline
		Ensemble ARIMA-ETS-ARNN & 15940.9 & 14941.65 & 16.50 & 18.17 \\ \hline
		Ensemble ARIMA-Theta-ARNN & 16883.48 & 15897.06 & 17.57 & 19.45 \\ \hline
		Ensemble ETS-Theta-ARNN & 19750.31 & 18780.61 & 20.79 & 23.42 \\ \hline
		Ensemble ANN-ARNN-WARIMA & 14512.1 & 13496.63 & 14.88 & 16.25 \\ \hline
	\end{tabular}
\end{table}

ANN is found to have the best accuracy metrics for 30-days ahead forecasts among single forecasting models for India COVID-19 data. Ensemble ANN-ARNN-WARIMA model and has the best accuracy among all hybrid/ensemble models (see Table \ref{india_accuracy_table_30_days}). In-sample and out-of-sample forecasts obtained from ANN and ensemble ANN-ARNN-WARIMA models are depicted in Fig. \ref{Fig:India_forecasts}(b). 

\begin{table}[H]
	\centering \caption{Performance metrics with 30 days-ahead test set for India}\label{india_accuracy_table_30_days}
	\begin{tabular}{|c|c|c|c|c|}
		\hline
		\multirow{2}{*}{Model} & \multicolumn{4}{c|}{30-days ahead forecast} 
		\\ \cline{2-5} & RMSE & MAE & MAPE & SMAPE \\ \hline
		ARIMA(1,2,5) & 17755.52 &  15657.27 & 18.67 & 21.01 \\ \hline
		ETS(A,A,N) & 14873.78 & 13051.98 & 15.57 & 17.18  \\ \hline
		SETAR & 21527.58 & 18609.71 & 21.98 & 25.49 \\ \hline
		TBATS & 24849.07 & 21843.15 & 25.96 & 30.82 \\ \hline
		Theta & 21713.03 & 19191.21 & 22.84 & 26.47 \\ \hline
		ANN & \textbf{6379.91} & \textbf{4800.13} & \textbf{6.48} & \textbf{6.13} \\ \hline
		ARNN(8,4) & 13225.43 & 10287.29 & 11.90 & 13.06 \\ \hline
		WARIMA & 14720.81 & 12738.66 & 15.15 & 16.72 \\ \hline
		BSTS & 14332.3 & 12493.74 & 14.88 & 16.34 \\ \hline
		ARFIMA(0,0.5,4) & 40115.62 & 36452.33 & 43.87 & 58.73  \\ \hline
		Hybrid ARIMA-ANN & 17640.51 & 15535.58 & 18.53 & 20.83 \\ \hline
		Hybrid ARIMA-ARNN & 17580.41 & 15507.04 & 18.51 & 20.80 \\ \hline
		Hybrid ARIMA-WARIMA & 17869.14 & 15771.05 & 18.78 & 21.19 \\ \hline
		Hybrid WARIMA-ANN & 14616.89 & 12613.57 & 15 & 16.53 \\ \hline
		Hybrid WARIMA-ARNN & 16052.8 & 14067.29 & 16.83 & 18.74 \\ \hline
		Ensemble ARIMA-ETS-Theta & 18081.97 & 15928.84 & 18.96 & 21.40 \\ \hline
		Ensemble ARIMA-ETS-ARNN & 15615.2 & 13419.82 & 15.86 & 17.61 \\ \hline
		Ensemble ARIMA-Theta-ARNN & 17933.14 & 15330.84 & 18.07 & 20.41 \\ \hline
		Ensemble ETS-Theta-ARNN & 16442.65 & 14160.56 & 16.74 & 18.69 \\ \hline
		Ensemble ANN-ARNN-WARIMA & \textbf{9090.214} & \textbf{7427.787} & \textbf{8.83} & \textbf{9.32} \\ \hline
	\end{tabular}
\end{table}

\begin{figure}[H]
	\includegraphics[width=0.52\textwidth]{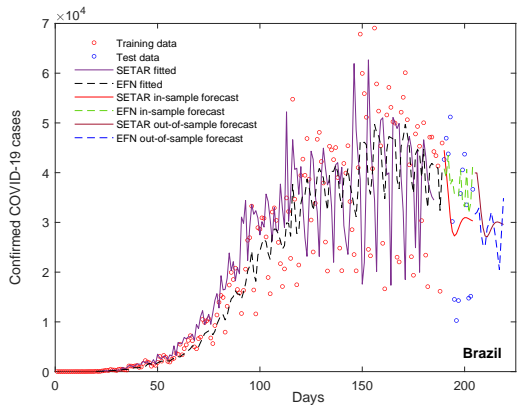}(a)
	\includegraphics[width=0.52\textwidth]{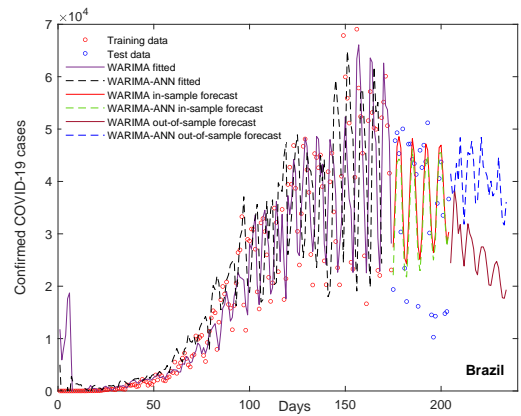}(b)
	\caption{Plots of (a) 15-days ahead forecast results for Brazil COVID-19 data obtained using SETAR and ETS-Theta-ARNN (EFN) models and (b) 30-days ahead forecast results from WARIMA and hybrid WARIMA-ANN models.}
	\label{Fig:Brazil_forecasts}
\end{figure}

\paragraph{\textbf{Results for Brazil COVID-19 data:}}
Among the single models, SETAR performs best in terms of accuracy metrics for 15-days ahead forecasts. Ensemble ETS-Theta-ARNN (EFN) model has the best accuracy among all hybrid/ensemble models (see Table \ref{brazil_accuracy_table_15_days}). In-sample and out-of-sample forecasts obtained from SETAR and ensemble EFN models are depicted in Fig. \ref{Fig:Brazil_forecasts}(a). 

\begin{table}[H]
	\centering \caption{Performance metrics with 15 days-ahead test set for Brazil}\label{brazil_accuracy_table_15_days}
	\begin{tabular}{|c|c|c|c|c|}
		\hline
		\multirow{2}{*}{Model} & \multicolumn{4}{c|}{15-days ahead forecast} 
		\\ \cline{2-5} & RMSE & MAE & MAPE & SMAPE \\ \hline
		ARIMA(3,1,2) & 16553.75 & 12530.04 & 76.62 & 41.66 \\ \hline
		ETS(A,A,N) & 13793.618 & 11038.765 & 63.41 & 38.99  \\ \hline
		SETAR & \textbf{11645.6} & \textbf{10148.91} & \textbf{49.77} & \textbf{37.35} \\ \hline
		TBATS & 15842.01 & 11803.72 & 72.67 & 40.05 \\ \hline
		Theta & 16263.93 & 12614.74 & 65.71 & 42.21 \\ \hline
		ANN & 19622.3 & 16536.91 & 83.45 & 53.78 \\ \hline
		ARNN((19,10)) & 13733.19 & 11951.27 & 57.59 & 40.36 \\ \hline
		WARIMA & 17167.66 & 13487.76 & 80.45 & 43.85\\ \hline
		BSTS & 21154.89 & 16702.38 & 98.97 & 49.62 \\ \hline
		ARFIMA(2,0.5,1) & 14023.22 & 11109.03 & 63.94 & 39.03 \\ \hline
		Hybrid ARIMA-ANN & 17541.86 & 13436.8 & 81.47 & 42.93 \\ \hline
		Hybrid ARIMA-ARNN & 18151.56 & 15254.77 & 79.64 & 46.73 \\ \hline
		Hybrid ARIMA-WARIMA & 16596.75 & 12704.16 & 77.16 & 41.94 \\ \hline
		Hybrid WARIMA-ANN & 16797.05 & 13378.25 & 78.94 & 43.96 \\ \hline
		Hybrid WARIMA-ARNN & 19211.01 & 16043.31 & 83.34 & 48.11 \\ \hline
		Ensemble ARIMA-ETS-Theta & 15271.82 & 11497.86 & 70.54 & 39.68 \\ \hline
		Ensemble ARIMA-ETS-ARNN & 13517.19 & 11260.21 & 62.81 & 39.61 \\ \hline
		Ensemble ARIMA-Theta-ARNN & 14546.36 & 11975.91 & 66.79 & 41.13 \\ \hline
		Ensemble ETS-Theta-ARNN & \textbf{13431.11} & \textbf{11324.4} & \textbf{62.67} & \textbf{39.83} \\ \hline
		Ensemble ANN-ARNN-WARIMA & 15565.1 & 13201.37 & 71.83 & 44.10 \\ \hline
	\end{tabular}
\end{table}

WARIMA is found to have the best accuracy metrics for 30-days ahead forecasts among single forecasting models for Brazil COVID-19 data. Hybrid WARIMA-ANN model has the best accuracy among all hybrid/ensemble models (see Table \ref{brazil_accuracy_table_30_days}). In-sample and out-of-sample forecasts obtained from WARIMA and hybrid WARIMA-ANN models are depicted in Fig. \ref{Fig:Brazil_forecasts}(b). 

\begin{table}[H]
	\centering \caption{Performance metrics with 30 days-ahead test set for Brazil}\label{brazil_accuracy_table_30_days}
	\begin{tabular}{|c|c|c|c|c|}
		\hline
		\multirow{2}{*}{Model} & \multicolumn{4}{c|}{30-days ahead forecast} 
		\\ \cline{2-5} & RMSE & MAE & MAPE & SMAPE \\ \hline
		ARIMA(5,1,1) with drift & 17647.13 & 14924.74 & 69.57 & 41.85 \\ \hline
		ETS(A,A,N) & 20270.82 & 15186.14 & 81.30 & 42.45  \\ \hline
		SETAR & 16136.69 & 15085.91 & 52.75 & 49.03 \\ \hline
		TBATS & 14166.74 & 10629.13 & 56.19 & 33.78 \\ \hline
		Theta & 17662.39 & 12880.03 & 70.55 & 38.38 \\ \hline
		ANN & 22403 & 18241.79 & 90.86 & 47.29 \\ \hline
		ARNN(9,5) & 13458.51 & 10884.02 & 40.10 & 30.92 \\ \hline
		WARIMA & \textbf{10628.51} & \textbf{9075.32} & \textbf{38.24} & \textbf{30.41} \\ \hline
		BSTS & 16876.78 & 15314.18 &  45.58 & 50.17 \\ \hline
		ARFIMA(2,0.5,1) & 12647.79 & 11616.15 & 47.49 & 37.56 \\ \hline
		Hybrid ARIMA-ANN & 17559.43 & 14810.82 & 69.11 & 41.58 \\ \hline
		Hybrid ARIMA-ARNN & 17274.78 & 14511.77 & 67.87 & 41.00 \\ \hline
		Hybrid ARIMA-WARIMA & 17464.81 & 14724.52 & 68.89 & 41.49 \\ \hline
		Hybrid WARIMA-ANN & \textbf{10841.65} & \textbf{8886.71} & \textbf{35.56} & \textbf{29.76} \\ \hline
		Hybrid WARIMA-ARNN & 10649.35 & 9104.54 & 38.39 & 30.48 \\ \hline
		Ensemble ARIMA-ETS-Theta & 18096.57 & 13854.34 & 72.82 & 40.27 \\ \hline
		Ensemble ARIMA-ETS-ARNN  & 16186 & 13705.63 & 64.26 & 39.84 \\ \hline
		Ensemble ARIMA-Theta-ARNN  & 15406.54 & 12793.94  & 60.87 & 38.06 \\ \hline
		Ensemble ETS-Theta-ARNN & 15737.01 & 12512.65 & 63.26 & 37.89 \\ \hline
		Ensemble ANN-ARNN-WARIMA  & 13543.31 & 11230.96 & 52.96847 & 34.57 \\ \hline
	\end{tabular}
\end{table}

\begin{figure}[H]
	\includegraphics[width=0.52\textwidth]{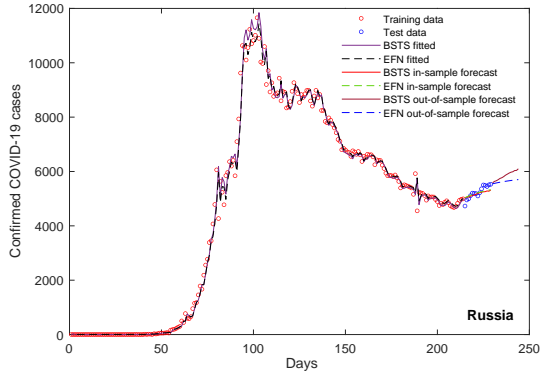}(a)
	\includegraphics[width=0.52\textwidth]{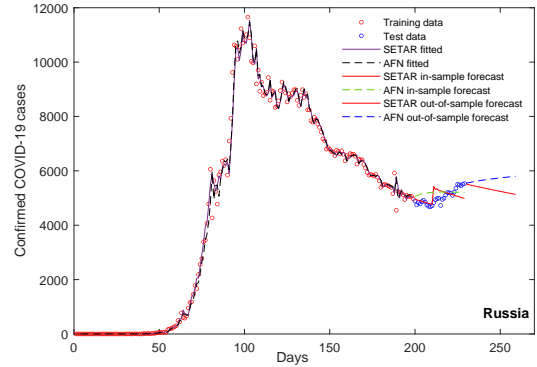}(b)
	\caption{Plots of (a) 15-days ahead forecast results for Russia COVID-19 data obtained using BSTS and ETS-Theta-ARNN (EFN) models and (b) 30-days ahead forecast results from SETAR and ARIMA-Theta-ARNN (AFN) models.}
	\label{Fig:Russia_forecasts}
\end{figure}

\paragraph{\textbf{Results for Russia COVID-19 data:}}
BSTS performs best in terms of accuracy metrics for a 15-days ahead forecast in the case of Russia COVID-19 data among single models. Theta and ARNN(3,2) also show competitive accuracy measures. Ensemble ETS-Theta-ARNN (EFN) model has the best accuracy among all hybrid/ensemble models (see Table \ref{Russia_accuracy_table_15_days}). Ensemble ARIMA-ETS-ARNN and ensemble ARIMA-Theta-ARNN also performs well in the test period. In-sample and out-of-sample forecasts obtained from BSTS and ensemble EFN models are depicted in Fig. \ref{Fig:Russia_forecasts}(a). 

\begin{table}[H]
	\centering \caption{Performance metrics with 15 days-ahead test set for Russia}\label{Russia_accuracy_table_15_days}
	\begin{tabular}{|c|c|c|c|c|}
		\hline
		\multirow{2}{*}{Model} & \multicolumn{4}{c|}{15-days ahead forecast} 
		\\ \cline{2-5} & RMSE & MAE & MAPE & SMAPE \\ \hline
		ARIMA(0,2,3) & 307.34 & 260.18 & 4.87 & 5.02 \\ \hline
		ETS(A,Ad,N) & 215.43 & 178.64 & 3.36 & 3.42  \\ \hline
		SETAR & 436.81 & 383.72 & 7.19 & 7.52 \\ \hline
		TBATS & 215.61 & 178.79 & 3.36 & 3.42 \\ \hline
		Theta & 186.06 & 157.30 & 2.97 & 3.01 \\ \hline
		ANN & 367.19 & 313.66 & 5.87 & 6.09 \\ \hline
		ARNN(3,2) & 208.58 & 184.74 & 3.61 & 3.52 \\ \hline
		WARIMA & 568.44 & 499.58 & 9.35 & 9.92\\ \hline
		BSTS & \textbf{160.18} & \textbf{132.28} & \textbf{2.51} & \textbf{2.53} \\ \hline
		ARFIMA(1,0.1,0) & 351.12 & 297.92 & 5.57 & 5.77 \\ \hline
		Hybrid ARIMA-ANN & 308.49 & 261.17 & 4.89 & 5.03 \\ \hline
		Hybrid ARIMA-ARNN & 245.84 & 207.72 & 3.92 & 3.99 \\ \hline
		Hybrid ARIMA-WARIMA & 299.14 & 251.59 & 4.72 & 4.85 \\ \hline
		Hybrid WARIMA-ANN & 489.38 & 425.98 & 7.98 & 8.38 \\ \hline
		Hybrid WARIMA-ARNN & 542.01 & 473.94 & 8.87 & 9.38 \\ \hline
		Ensemble ARIMA-ETS-Theta & 234.64 & 195.71 & 3.68 & 3.75 \\ \hline
		Ensemble ARIMA-ETS-ARNN & 168.14 & 135.34 & 2.57 & 2.59 \\ \hline
		Ensemble ARIMA-Theta-ARNN & 192.28 & 158.52 & 2.99 & 3.03 \\ \hline
		Ensemble ETS-Theta-ARNN & \textbf{157.25} & \textbf{127.98} & \textbf{2.44} & \textbf{2.45} \\ \hline
		Ensemble ANN-ARNN-WARIMA & 288.26 & 243.69 & 4.57 & 4.69 \\ \hline
	\end{tabular}
\end{table}

SETAR is found to have the best accuracy metrics for 30-days ahead forecasts among single forecasting models for Russia COVID-19 data. Ensemble ARIMA-Theta-ARNN (AFN) model has the best accuracy among all hybrid/ensemble models (see Table \ref{Russia_accuracy_table_30_days}). All five ensemble models show promising results for this dataset. In-sample and out-of-sample forecasts obtained from SETAR and ensemble AFN models are depicted in Fig. \ref{Fig:Russia_forecasts}(b). 

\begin{table}[H]
	\centering \caption{Performance metrics with 30 days-ahead test set for Russia}\label{Russia_accuracy_table_30_days}
	\begin{tabular}{|c|c|c|c|c|}
		\hline
		\multirow{2}{*}{Model} & \multicolumn{4}{c|}{30-days ahead forecast} 
		\\ \cline{2-5} & RMSE & MAE & MAPE & SMAPE \\ \hline
		ARIMA(1,2,1) & 732.12 & 546.87 & 10.40 & 11.44 \\ \hline
		ETS(A,Ad,N) & 337.44 & 264.40 & 5.08 & 5.25  \\ \hline
		SETAR & \textbf{285.41} & \textbf{217.23} & \textbf{4.25} & \textbf{4.24} \\ \hline
		TBATS & 337.78 & 264.62 & 5.08 & 5.25 \\ \hline
		Theta & 327.46 & 297.91 & 6.04 & 5.82 \\ \hline
		ANN & 460 & 340.96 & 6.48 & 6.86 \\ \hline
		ARNN(3,2) & 727.63 & 693.61 & 13.98 & 12.97 \\ \hline
		WARIMA & 961.24 & 727.34 & 13.86 & 15.73 \\ \hline
		BSTS & 686.06 & 509.87 & 9.79 & 10.59 \\ \hline
		ARFIMA(1,0.01,0) & 303.35 & 239.76 & 4.63 & 4.74 \\ \hline
		Hybrid ARIMA-ANN &  734.05 & 548.49 & 10.43 & 11.48 \\ \hline
		Hybrid ARIMA-ARNN & 715.58 & 536.69 & 10.22 & 11.19 \\ \hline
		Hybrid ARIMA-WARIMA & 729.96 & 549.97 & 10.47 & 11.5 \\ \hline
		Hybrid WARIMA-ANN & 1012.61 & 772.11 & 14.73 & 16.82 \\ \hline
		Hybrid WARIMA-ARNN & 939.26 & 715.72 & 13.65 & 15.41 \\ \hline
		Ensemble ARIMA-ETS-Theta & 324.95 & 257.24 & 4.96 & 5.10 \\ \hline
		Ensemble ARIMA-ETS-ARNN & 330.79 & 280.85 & 5.51 & 5.56 \\ \hline
		Ensemble ARIMA-Theta-ARNN & \textbf{299.50} & \textbf{264.55} & \textbf{5.36} & \textbf{5.22} \\ \hline
		Ensemble ETS-Theta-ARNN & 337.63 & 293.23 & 6 & 5.77\\ \hline
		Ensemble ANN-ARNN-WARIMA & 399.84 & 324.34 & 6.29 & 6.46 \\ \hline
	\end{tabular}
\end{table}

\begin{figure}[H]
	\includegraphics[width=0.52\textwidth]{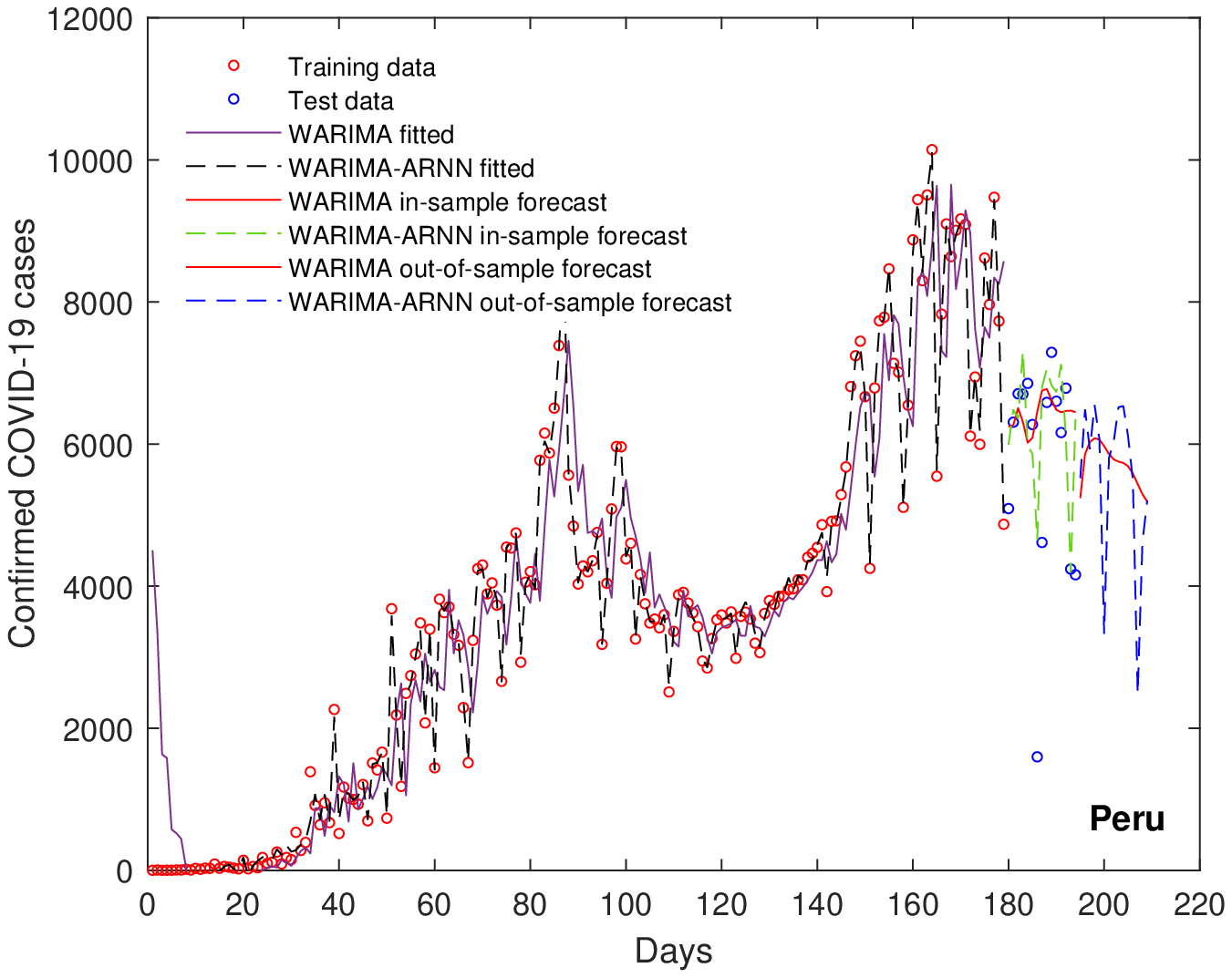}(a)
	\includegraphics[width=0.52\textwidth]{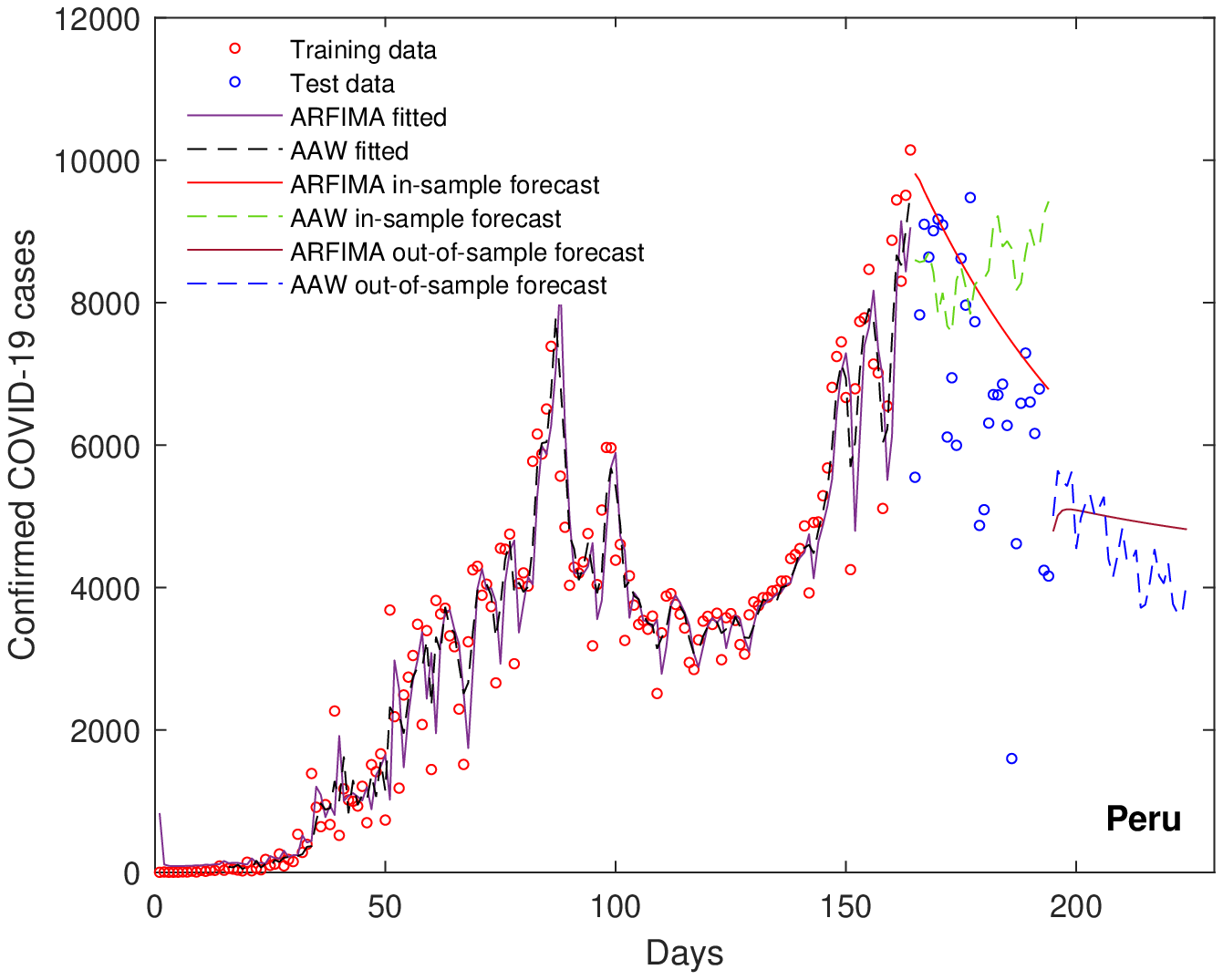}(b)
	\caption{Plots of (a) 15-days ahead forecast results for Peru COVID-19 data obtained using WARIMA and hybrid WARIMA-ARNN models and (b) 30-days ahead forecast results from ARFIMA and ensemble ANN-ARNN-WARIMA (AAW) models.}
	\label{Fig:Peru_forecasts}
\end{figure}

\paragraph{\textbf{Results for Peru COVID-19 data:}}
WARIMA and ARFIMA(2,0.09,1) perform better than other single models for 15-days ahead forecasts in Peru. Hybrid WARIMA-ARNN model improves the WARIMA forecasts and has the best accuracy among all hybrid/ensemble models (see Table \ref{Peru_accuracy_table_15_days}). In-sample and out-of-sample forecasts obtained from WARIMA and hybrid WARIMA-ARNN models are depicted in Fig. \ref{Fig:Peru_forecasts}(a). 

\begin{table}[H]
	\centering \caption{Performance metrics with 15 days-ahead test set for Peru}\label{Peru_accuracy_table_15_days}
	\begin{tabular}{|c|c|c|c|c|}
		\hline
		\multirow{2}{*}{Model} & \multicolumn{4}{c|}{15-days ahead forecast} 
		\\ \cline{2-5} & RMSE & MAE & MAPE & SMAPE \\ \hline
		ARIMA(1,1,1) with drift & 2275.49 & 1686.84 & 49.97 & 28.99 \\ \hline
		ETS(M,A,N) & 1689.96 & 1189.05 & 31.89 & 23.15  \\ \hline
		SETAR & 1935.78 & 1286.56 & 41.57 & 23.71 \\ \hline
		TBATS & 1944.26 & 1301.07 & 41.72 & 24.06 \\ \hline
		Theta & 1831.88 & 1146.27 & 38.37 & 21.92 \\ \hline
		ANN & 1771.59 & 1211.24 & 38.89 & 22.75 \\ \hline
		ARNN(15,8) & 2564.65 & 2244.78 & 57.13 & 35.78 \\ \hline
		WARIMA & \textbf{1659.24} & 1060.67 & \textbf{35.22} & 20.85\\ \hline
		BSTS & 1740.18 & 1082.16 & 36.48 & 21.07 \\ \hline
		ARFIMA(2,0.09,1) & 1712.47 & \textbf{1022.55} & 35.65 & \textbf{20.13} \\ \hline 
		Hybrid ARIMA-ANN & 2189.18 & 1596.80 & 47.93 & 27.96 \\ \hline
		Hybrid ARIMA-ARNN & 1646.88 & 1244.03 & 34.95 & 23.43 \\ \hline
		Hybrid ARIMA-WARIMA & 2082.15 & 1385.87 & 43.93 & 24.99 \\ \hline
		Hybrid WARIMA-ANN & 1560.68 & 1206.92 & 34.11 & 23.43 \\ \hline
		Hybrid WARIMA-ARNN & \textbf{1121.10} & \textbf{827.90} & \textbf{23.33} & \textbf{17.46} \\ \hline
		Ensemble ARIMA-ETS-Theta & 1677.24 & 1040.93 & 35.50 & 20.46 \\ \hline
		Ensemble ARIMA-ETS-ARNN & 1748.39 & 1185.23 & 38.18 & 22.48 \\ \hline
		Ensemble ARIMA-Theta-ARNN & 1801.56 & 1324.73 & 39.97 & 24.39 \\ \hline
		Ensemble ETS-Theta-ARNN & 1613.15 & 1048.04 & 34.76 & 20.62 \\ \hline
		Ensemble ANN-ARNN-WARIMA & 1864.99 & 1329.83 & 41.16 & 24.43 \\ \hline
	\end{tabular}
\end{table}

ARFIMA(2,0,0) and ANN depict competitive accuracy metrics for 30-days ahead forecasts among single forecasting models for Peru COVID-19 data. Ensemble ANN-ARNN-WARIMA (AAW) model has the best accuracy among all hybrid/ensemble models (see Table \ref{Peru_accuracy_table_30_days}). In-sample and out-of-sample forecasts obtained from ARFIMA(2,0,0) and ensemble AAW models are depicted in Fig. \ref{Fig:Peru_forecasts}(b). 

\begin{table}[H]
	\centering \caption{Performance metrics with 30 days-ahead test set for Peru}\label{Peru_accuracy_table_30_days}
	\begin{tabular}{|c|c|c|c|c|}
		\hline
		\multirow{2}{*}{Model} & \multicolumn{4}{c|}{30-days ahead forecast} 
		\\ \cline{2-5} & RMSE & MAE & MAPE & SMAPE \\ \hline
		ARIMA(1,1,1) with drift & 3889.85 & 3288.04 & 70.17 & 41.92 \\ \hline
		ETS(M,A,N) & 7881.14 & 6892.41 & 81.37 & 66.91  \\ \hline
		SETAR & 4598.98 & 4077.59 & 83.67 & 48.90 \\ \hline
		TBATS & 2924.92 & 2366.84 & 52.90 & 33.13 \\ \hline
		Theta & 3862.84 & 3374.84  & 70.68 & 42.93 \\ \hline
		ANN & 2183.98 & 1818.07 & \textbf{30.57} & 32.12 \\ \hline
		ARNN(15,8) & 2833.39 & 2339.49 & 49.10 & 32.92 \\ \hline
		WARIMA & 5579.69 & 4840.75 & 89.04 & 54.14 \\ \hline
		BSTS & 5422.13 & 4851.34 &  87.98 & 54.82 \\ \hline
		ARFIMA(2,0,0) & \textbf{2052.01} & \textbf{1513.62} & 35.37 & \textbf{23.27} \\ \hline
		Hybrid ARIMA-ANN & 3756.5 & 3131.88& 67.50 & 40.46 \\ \hline
		Hybrid ARIMA-ARNN & 4137.45 & 3619.54& 74.50 & 44.93 \\ \hline
		Hybrid ARIMA-WARIMA & 4164.69 & 3602.27& 75.52 & 44.78 \\ \hline
		Hybrid WARIMA-ANN & 6372.936 & 5722.291 & 95.95 & 60.80 \\ \hline
		Hybrid WARIMA-ARNN & 5563.043 & 4819.09 & 93.16 & 53.97 \\ \hline
		Ensemble ARIMA-ETS-Theta & 5176.14 & 4518.43 & 92.73 & 51.99 \\ \hline
		Ensemble ARIMA-ETS-ARNN & 4908.85 & 4153.58 & 87.26 & 48.69 \\ \hline
		Ensemble ARIMA-Theta-ARNN & 3410.39 & 2785.71 & 61.39 & 37.11 \\ \hline
		Ensemble ETS-Theta-ARNN & 4826.01 & 4048.24 & 85.09 & 47.82 \\ \hline
		Ensemble ANN-ARNN-WARIMA & \textbf{2626.8} & \textbf{2003.06} & \textbf{47.02} & \textbf{29.02} \\ \hline
	\end{tabular}
\end{table}

Results from all the five datasets reveal that none of the forecasting models performs uniformly, and therefore, one should be carefully select the appropriate forecasting model while dealing with COVID-19 datasets. 

\section{Discussions}
In this study, we assessed several statistical, machine learning, and composite models on the confirmed cases of COVID-19 data sets for the five countries with the highest number of cases. Thus, COVID-19 cases in the USA, followed by India, Brazil, Russia, and Peru, are considered. The datasets mostly exhibit nonlinear and nonstationary behavior. Twenty forecasting models were applied to five datasets, and an empirical study is presented here. The empirical findings suggest no universal method exists that can outperform every other model for all the datasets in COVID-19 nowcasting. Still, the future forecasts obtained from models with the best accuracy will be useful in decision and policy makings for government officials and policymakers to allocate adequate health care resources for the coming days in responding to the crisis. However, we recommend updating the datasets regularly and comparing the accuracy metrics to obtain the best model. As this is evident from this empirical study that no model can perform consistently as the best forecasting model, one must update the datasets regularly to generate useful forecasts. Time series of epidemics can oscillate heavily due to various epidemiological factors, and these fluctuations are challenging to be captured adequately for precise forecasting. 

All five different countries, except Brazil and Peru, will face a diminishing trend in the number of new confirmed cases of COVID-19 pandemic. Followed by the both short-term out of sample forecasts reported in this study, the lockdown and shutdown periods can be adjusted accordingly to handle the uncertain and vulnerable situations of the COVID-19 pandemic. Authorities and health care can modify their planning in stockpile and hospital-beds, depending on these COVID-19 pandemic forecasts. 

Models are constrained by what we know and what we assume but used appropriately, and with an understanding of these limitations, they can and should help guide us through this pandemic. Since purely statistical approaches do not account for how transmission occurs, they are generally not well suited for long-term predictions about epidemiological dynamics (such as when the peak will occur and whether resurgence will happen) or inference about intervention efficacy. Several forecasting models, therefore, limit their projections to two weeks or a month ahead. 

\section{Conclusion and Future Challenges}
In this research, we have focused on analyzing the nature of the COVID-19 time series data and understanding the data characteristics of the time series. This empirical work studied a wide range of statistical forecasting methods and machine learning algorithms. We have also presented more systematic representations of single, ensemble, and hybrid approaches available for epidemic forecasting. This quantitative study could be used to assess and forecast COVID-19 confirmed cases, which will benefit epidemiologists and modelers in their real-world applications. 

Considering this scope of the study, we can present a list of challenges of pandemic forecasting (short-term) with the forecasting tools presented in this chapter:
\begin{itemize}
    \item Collect more data on the factors that contribute to daily confirmed cases of COVID-19.
    \item model the entire predictive distribution, with particular focus on accurately quantifying uncertainty \cite{holmdahl2020wrong}.
    \item There is no universal model that can generate `best' short-term forecasts of COVID-19 confirmed cases. 
    \item Continuously monitor the performance of any model against real data and either re-adjust or discard models based on accruing evidence. 
    \item Developing models in real-time for a novel virus, with poor quality data, is a formidable task and real challenge for epidemic forecasters. 
    \item Epidemiological estimates and compartmental models can be useful for long-term pandemic trajectory prediction, but they often assume some unrealistic assumptions \cite{ioannidis2020forecasting}.
    \item Future research is needed to collect, clean, and curate data and develop a coherent approach to evaluate the suitability of models with regard to COVID-19 predictions and forecast uncertainties. 
\end{itemize}

\section*{Data and codes}
For the sake of repeatability and reproducibility of this study, all codes and data sets are made available at \url{https://github.com/indrajitg-r/Forecasting-COVID-19-cases}.

\bibliographystyle{plain}
\bibliography{COVID_bib}



\end{document}